\documentclass[pra,twocolumn,superscriptaddress,floatfix,nofootinbib,amsmath,amssymb]{revtex4-1}

\usepackage[utf8]{inputenc}
\usepackage[T1]{fontenc}
\usepackage[english]{babel}
\usepackage{upquote}
\usepackage{float}
\usepackage{graphicx}
\usepackage{bm}
\usepackage{braket}
\usepackage{xcolor}
\usepackage{amsfonts}
\usepackage{comment}
\usepackage{dirtytalk}
\usepackage{url}
\usepackage{bbold}
\usepackage{realboxes}
\usepackage{my_macro}
\usepackage{enumitem}
\usepackage{booktabs}
\usepackage{algpseudocode}
\usepackage{subcaption}
\usepackage{caption}
\usepackage{hyperref}
\usepackage{cleveref}
\usepackage{tikz}
\usetikzlibrary{positioning}
\usepackage{xcolor}
\hypersetup{
    bookmarksopen=true,
    colorlinks=true,
    linkcolor=blue,
    citecolor=green,
    urlcolor=blue
}

\newcommand{\tr}{\text{tr}}

\usepackage{listings}
\UseRawInputEncoding
\usepackage{titlesec} 
\titleformat{\paragraph}[block]{\small\bfseries}{}{}{}[]

\numberwithin{equation}{section}
\numberwithin{figure}{section}
\numberwithin{table}{section}

\definecolor{bkgd}{RGB}{240,242,246}
\definecolor{ceruleanblue}{rgb}{0.16, 0.32, 0.75}
\definecolor{orange-red}{rgb}{1.0, 0.27, 0.0}
\definecolor{anotherblue}{RGB}{37,92,243}
\definecolor{blackblue}{RGB}{46,60,85}
\definecolor{goldyellow}{RGB}{199,146,12}

\lstdefinestyle{altstyle2}{
    backgroundcolor=\color{bkgd},
    basicstyle=\ttfamily\footnotesize\color{blackblue},
    breakatwhitespace=false,
    breaklines=true,
    captionpos=b,
    commentstyle=\color{goldyellow},
    keepspaces=true,
    keywordstyle=\color{orange-red},
    language=Python,
    numbersep=5pt,
    numberstyle=\tiny\color{ceruleanblue},
    showspaces=false,
    showstringspaces=false,
    showtabs=false,
    stringstyle=\color{anotherblue},
    tabsize=2,
    numbers=left
}

\definecolor{codegreen}{rgb}{0,0.6,0}
\definecolor{codegray}{rgb}{0.5,0.5,0.5}
\definecolor{codepurple}{rgb}{0.58,0,0.82}
\definecolor{backcolour}{rgb}{0.95,0.95,0.92}
\lstdefinestyle{mystyle}{
    backgroundcolor=\color{backcolour},
    commentstyle=\color{codegreen},
    keywordstyle=\color{magenta},
    numberstyle=\tiny\color{codegray},
    stringstyle=\color{codepurple},
    basicstyle=\ttfamily\footnotesize,
    breakatwhitespace=false,
    breaklines=true,
    captionpos=b,
    keepspaces=true,
    language=Python,
    numbers=left,
    numbersep=5pt,
    showspaces=false,
    showstringspaces=false,
    showtabs=false,
    tabsize=2
}
\begin{document}

\title{MindSpore Quantum: A User-Friendly, High-Performance, and AI-Compatible Quantum Computing Framework
}

\author{Xusheng Xu}
\email{thuxuxs@163.com}
\affiliation{MindSpore Quantum Special Interest Group}
\author{Jiangyu Cui}
\affiliation{MindSpore Quantum Special Interest Group}
\author{Zidong Cui}
\affiliation{Institute of Fundamental and Frontier Sciences, University of Electronic Science and Technology of China, Chengdu 610051, China}
\author{Runhong He}
\affiliation{State Key Laboratory of Computer Science, Institute of Software Chinese Academy of Sciences, Beijing 100190}
\author{Qingyu Li}
\affiliation{Institute of Fundamental and Frontier Sciences, University of Electronic Science and Technology of China, Chengdu 610051, China}
\author{Xiaowei Li}
\affiliation{Institute for Quantum Science and Engineering, Southern University of Science and Technology, Shenzhen 518055, China}
\author{Yanling Lin}
\affiliation{School of Computer Science and Technology,
    University of Science and Technology of China, Hefei 230027, China}
\author{Jiale Liu}
\affiliation{School of Computer Science and Technology,
    University of Science and Technology of China, Hefei 230027, China}
\author{Wuxin Liu}
\affiliation{MindSpore Quantum Special Interest Group}
\author{Jiale Lu}
\affiliation{MindSpore Quantum Special Interest Group}
\author{Maolin Luo}
\affiliation{MindSpore Quantum Special Interest Group}
\author{Chufan Lyu}
\affiliation{Institute of Fundamental and Frontier Sciences, University of Electronic Science and Technology of China, Chengdu 610051, China}
\author{Shijie Pan}
\affiliation{MindSpore Quantum Special Interest Group}
\author{Mosharev Pavel}
\affiliation{MindSpore Quantum Special Interest Group}
\author{Runqiu Shu}
\affiliation{School of Automation Science and Engineering, South China University of Technology, Guangzhou 510641, China}
\author{Jialiang Tang}
\affiliation{Department of Physical Chemistry, University of the Basque Country UPV/EHU, Apartado 644, 48080 Bilbao, Spain}
\author{Ruoqian Xu}
\affiliation{Department of Physical Chemistry, University of the Basque Country UPV/EHU, Apartado 644, 48080 Bilbao, Spain}
\author{Shu Xu}
\affiliation{MindSpore Quantum Special Interest Group}
\author{Kang Yang}
\affiliation{MindSpore Quantum Special Interest Group}
\author{Fan Yu}
\affiliation{MindSpore Quantum Special Interest Group}
\author{Qingguo Zeng}
\affiliation{Institute for Quantum Science and Engineering, Southern University of Science and Technology, Shenzhen 518055, China}
\author{Haiying Zhao}
\affiliation{MindSpore Quantum Special Interest Group}
\author{Qiang Zheng}
\affiliation{School of Computer Science and Technology,
    University of Science and Technology of China, Hefei 230027, China}
\author{Junyuan Zhou}
\email{jedzhou99@163.com}
\affiliation{MindSpore Quantum Special Interest Group}
\author{Xu Zhou}
\affiliation{School of Physics and Astronomy, Sun Yat-sen University, Zhuhai 519082, China}
\author{Yikang Zhu}
\affiliation{School of Computer Science and Technology,
    University of Science and Technology of China, Hefei 230027, China}
\author{Zuoheng Zou}
\email{zouzuoheng@pku.edu.cn}
\affiliation{MindSpore Quantum Special Interest Group}

\author{Abolfazl Bayat}
\affiliation{Institute of Fundamental and Frontier Sciences, University of Electronic Science and Technology of China, Chengdu 610051, China}
\affiliation{Key Laboratory of Quantum Physics and Photonic Quantum Information, Ministry of Education,
    University of Electronic Science and Technology of China, Chengdu 611731, China}
\author{Xi Cao}
\affiliation{Department of Automation, Tsinghua University, Beijing 100084, China}
\affiliation{Key Laboratory of Quantum Physics and Photonic Quantum Information, Ministry of Education,
    University of Electronic Science and Technology of China, Chengdu 611731, China}
\author{Wei Cui}
\affiliation{School of Automation Science and Engineering, South China University of Technology, Guangzhou 510641, China}
\author{Zhendong Li}
\affiliation{Key Laboratory of Theoretical and Computational Photochemistry, Ministry of Education,
    College of Chemistry, Beijing Normal University, Beijing 100875, China}
\author{Guilu Long}
\affiliation{Beijing Academy of Quantum Information Sciences, Beijing 100193, People's Republic of China}
\affiliation{State Key Laboratory of Low-Dimensional Quantum Physics and Department of Physics, Tsinghua University, Beijing 100084, People's Republic of China}
\author{Zhaofeng Su}
\affiliation{School of Computer Science and Technology,
    University of Science and Technology of China, Hefei 230027, China}
\author{Xiaoting Wang}
\affiliation{Institute of Fundamental and Frontier Sciences, University of Electronic Science and Technology of China, Chengdu 610051, China}
\affiliation{Key Laboratory of Quantum Physics and Photonic Quantum Information, Ministry of Education,
    University of Electronic Science and Technology of China, Chengdu 611731, China}
\author{Zizhu Wang}
\affiliation{Institute of Fundamental and Frontier Sciences, University of Electronic Science and Technology of China, Chengdu 610051, China}
\affiliation{Key Laboratory of Quantum Physics and Photonic Quantum Information, Ministry of Education,
    University of Electronic Science and Technology of China, Chengdu 611731, China}
\author{Shijie Wei}
\affiliation{Beijing Academy of Quantum Information Sciences, Beijing 100193, People's Republic of China}
\author{Re-Bing Wu}
\affiliation{Department of Automation, Tsinghua University, Beijing 100084, China}
\author{Pan Zhang}
\affiliation{CAS Key Laboratory for Theoretical Physics, Institute of Theoretical Physics,
    Chinese Academy of Sciences, Beijing 100190, China}

\author{Man-Hong Yung}
\email{yung@sustech.edu.cn}
\affiliation{MindSpore Quantum Special Interest Group}
\affiliation{Institute for Quantum Science and Engineering, Southern University of Science and Technology, Shenzhen 518055, China}
\affiliation{Shenzhen International Quantum Academy, Shenzhen 518048, China}
\affiliation{Guangdong Provincial Key Laboratory of Quantum Science and Engineering, Southern University of Science and Technology, Shenzhen 518055, China}
\affiliation{Shenzhen Key Laboratory of Quantum Science and Engineering, Southern University of Science and Technology, Shenzhen 518055, China}

\date{\today}


\begin{abstract}
    We introduce MindSpore Quantum, a pioneering hybrid quantum-classical framework with a primary focus on the design and implementation of noisy intermediate-scale quantum (NISQ) algorithms. Leveraging the robust support of MindSpore, an advanced open-source deep learning training/inference framework, MindSpore Quantum exhibits exceptional efficiency in the design and training of variational quantum algorithms on both CPU and GPU platforms, delivering remarkable performance. Furthermore, this framework places a strong emphasis on enhancing the operational efficiency of quantum algorithms when executed on real quantum hardware. This encompasses the development of algorithms for quantum circuit compilation and qubit mapping, crucial components for achieving optimal performance on quantum processors. In addition to the core framework, we introduce QuPack---a meticulously crafted quantum computing acceleration engine. QuPack significantly accelerates the simulation speed of MindSpore Quantum, particularly in variational quantum eigensolver (VQE), quantum approximate optimization algorithm (QAOA), and tensor network simulations, providing astonishing speed. This combination of cutting-edge technologies empowers researchers and practitioners to explore the frontiers of quantum computing with unprecedented efficiency and performance.
\end{abstract}
\maketitle
\tableofcontents
\section{Introduction}

Quantum computing is a rapidly evolving field that has the potential to revolutionize the way we process information. Unlike classical computers, which use bits to represent information, quantum computers use quantum bits or qubits. Due to quantum superposition, qubits can encode multiple computational bases simultaneously, allowing quantum computers to perform certain calculations much faster than classical computers. These algorithms include Grover's search algorithm \cite{grover1996fast, long2001grover}, Shor's factoring algorithm \cite{shor1994algorithms} and HHL linear equations solving algorithm \cite{harrow2009quantum}. In recent years, the development of quantum computing chips has been relatively rapid. IBM has successively launched its 433-bit Osprey quantum chip \cite{ibm2022osprey} and 1121-bit Condor quantum chip \cite{castelvecchi2023ibm}. Google also demonstrate the quantum supremacy on 53-bit Sycamore quantum chip \cite{arute2019quantum}. The 66-bit programmable quantum computing chip Zuchongzhi-2 \cite{wu2021strong} has also been released in China. How to use these quantum chips efficiently has become an increasingly important research topic.

Currently, we are still in the NISQ quantum computing stage \cite{preskill2018quantum}. The number of qubits in quantum computers is still relatively small, and the quantum gates are noisy. It is important to use classical hardware, such as CPUs, GPUs, and Ascend, to perform quantum simulation. With the quantum simulator, we can quickly develop and verify quantum algorithms. Furthermore, we can also use the quantum simulator to verify the correctness of the quantum chip or assist in completing the error mitigation of the current noisy quantum chip.

In the NISQ stage, a type of variational quantum algorithm \cite{Peruzzo2014Peruzzo2014,cerezo2021variational,PhysRevA.109.052414} called quantum-classical hybrid variational algorithm may provide superior practical value. These algorithms have a certain resistance to noise, including the variational quantum eigensolver for simulating chemical molecules \cite{mcardle2020quantum,cao2019quantum,yeter2020practical,fan2023circuit}, quantum approximate optimization algorithm \cite{farhi2014quantum,PhysRevResearch.4.013141} for solving combinatorial optimization problems, and quantum machine learning algorithm \cite{benedetti2019parameterized,biamonte2017quantum,das2019machine,PhysRevResearch.6.013027} for processing classification and pattern generation. According to the principle of variational algorithms, variational quantum algorithms rely on classical optimizers to adjust the parameters in the variational quantum circuit to make the quantum system output as close as possible to the expected target, thereby achieving the purpose of optimization learning.

Here we present a hybrid quantum-classical programming framework called MindSpore Quantum (\MindQuantum). Compared to other quantum framework such as QuEST \cite{jones2019quest}, QPanda \cite{dou2022qpanda}, TensorCircuit \cite{zhang2023tensorcircuit}, Qulacs \cite{suzuki2021qulacs}, TensorFlow Quantum \cite{broughton2020tensorflow}, Yao \cite{luo2020yao}, Intel Quantum Simulator \cite{guerreschi2020intel}, and PennyLane \cite{bergholm2018pennylane}, \MindQuantum\ provides a simple, user-friendly, and efficient quantum programming framework that supports the construction, simulation and real chip based execution of quantum circuits. With MindSpore, one can achieve fast development and training of hybrid quantum-classical algorithms. Fig.~\ref{fig:framework} illustrates the overall architecture of \MindQuantum.

\definecolor{c8adacd}{RGB}{138,218,205}
\definecolor{ce8f8f3}{RGB}{232,248,243}
\definecolor{cffc46d}{RGB}{255,196,109}
\definecolor{cf2f2f2}{RGB}{220,220,220}
\definecolor{c6de1ff}{RGB}{109,225,255}
\definecolor{caffaeb}{RGB}{175,250,235}
\definecolor{c0157b4}{RGB}{1,87,180}
\def \globalscale {1.000000}
\begin{figure*}
    \begin{tikzpicture}[scale=0.6, every node/.append style={scale=\globalscale}, inner sep=0pt, outer sep=0pt]
        \path[draw=black,line width=0.0794cm] (0.0397, 29.1174) rectangle (29.2232, 17.8991);
        \node[align=center, font=\bfseries] at (14.63145, 28.5174) {Basic Component};

        \path[fill=c8adacd, draw] (0.4895, 28.0061) rectangle (7.3686, 27.0272);
        \node[align=center] at (3.92905, 27.51665) {Quantum Gate};

        \path[fill=ce8f8f3] (0.4763, 26.8023) rectangle (7.3819, 18.2563);
        \node[draw=none] at (4.2, 25) {
            \begin{minipage}{0.3\textwidth}
                \begin{itemize}
                    \item Fixed gate
                    \item Parameterized gate
                    \item Any control on any gate
                    \item Noise channel
                \end{itemize}
            \end{minipage}
        };
        \path[fill=white,rounded corners=0.6879cm] (0.926, 22.8865) rectangle (6.9321, 18.6796);

        \path[fill=c8adacd, draw] (7.6332, 28.0061) rectangle (14.4859, 27.0272);
        \node[align=center] at (11.05955, 27.51665) {Quantum Circuit};

        \path[fill=ce8f8f3] (7.62, 26.8023) rectangle (14.5256, 18.2563);
        \node[draw=none] at (11.3437, 25.5) {
            \begin{minipage}{0.3\textwidth}
                \begin{itemize}
                    \item Append operator
                    \item $+=$ operator
                    \item Chain rule
                \end{itemize}
            \end{minipage}
        };
        \path[fill=white,rounded corners=0.6879cm] (8.0698, 22.8865) rectangle (14.0758, 18.6796);

        \path[fill=c8adacd, draw] (14.777, 28.0061) rectangle (21.6561, 27.0272);
        \node[align=center] at (18.21655, 27.51665) {Parameter Resolver};
        \path[fill=ce8f8f3] (14.7638, 26.8023) rectangle (21.6694, 18.2563);
        \node[draw=none] at (18.4875, 25.9) {
            \begin{minipage}{0.3\textwidth}
                \begin{itemize}
                    \item Fine tuning
                    \item Arithmetic calculation
                \end{itemize}
            \end{minipage}
        };
        \path[fill=white,rounded corners=0.6879cm] (15.2135, 22.8865) rectangle (21.2196, 18.6796);

        \path[fill=c8adacd, draw] (21.9207, 28.0061) rectangle (28.7734, 27.0272);
        \node[align=center] at (25.34705, 27.51665) {Observable};
        \path[fill=ce8f8f3] (21.9075, 26.8023) rectangle (28.8131, 18.2563);
        \node[draw=none] at (25.6312, 25) {
            \begin{minipage}{0.3\textwidth}
                \begin{itemize}
                    \item Hamiltonian
                    \item QubitOperator
                    \item FermionOperator
                    \item Transform
                \end{itemize}
            \end{minipage}
        };
        \path[fill=white,rounded corners=0.6879cm] (22.3573, 22.8865) rectangle (28.3633, 18.6796);
        \node at (3.92905, 20.78305) {\includegraphics[scale=0.4]{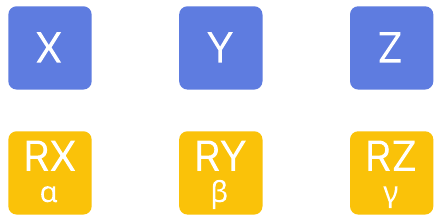}};
        \node at (11.07285, 20.78305) {\includegraphics[scale=0.35]{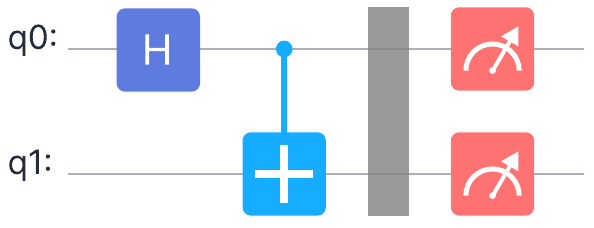}};
        \node at (18.21665, 21.18305){\begin{lstlisting}[numbers=none, backgroundcolor=]
a, b = np.pi, 0.5
ParameterResolver({
 'a': a, 'b': b
})\end{lstlisting}};

        \node[align=center] at (18.21665, 19.68305) {\textcolor{blue}{$\pi * a+ 1 / 2 * b$}};
        \node at (25.36045, 21.18305-0.2){\begin{lstlisting}[numbers=none, backgroundcolor=]
o = QubitOperator(
 'X0 Y1', 'a'
) + QubitOperator(
 'Z2', np.sqrt(2)
)
        \end{lstlisting}};
        \node[align=center] at (25.36045, 19.68305-0.2) {\textcolor{blue}{$aX_0Y_1 + \sqrt{2}Z_2$}};

        \path[draw=black,line width=0.0794cm] (0.0397, 17.608) rectangle (14.4595, 10.2791);
        \node[align=center, font=\bfseries] at (7.2496, 17.008) {NISQ Algorithm};

        \path[fill=cffc46d,draw] (0.5159, 16.047) rectangle (6.9453, 14.7505);
        \node[align=center] at (3.7306, 15.39875) {Ansatz Library};

        \path[draw=black] (0.5159, 14.7241) rectangle (6.9453, 10.5436);

        \path[fill=cffc46d,draw] (7.62, 16.0602) rectangle (14.0758, 14.7373);
        \node[align=center] at (10.8479, 15.39875) {VQE};

        \path[fill=cffc46d, draw] (7.62, 13.9435) rectangle (14.0758, 12.6206);
        \node[align=center] at (10.8479, 13.28205) {QAOA};

        \path[fill=cffc46d, draw] (7.62, 11.8269) rectangle (14.0758, 10.504);
        \node[align=center] at (10.8479, 11.16545) {QNN};
        \node at (3.7306, 12.58205) {\includegraphics[scale=0.065]{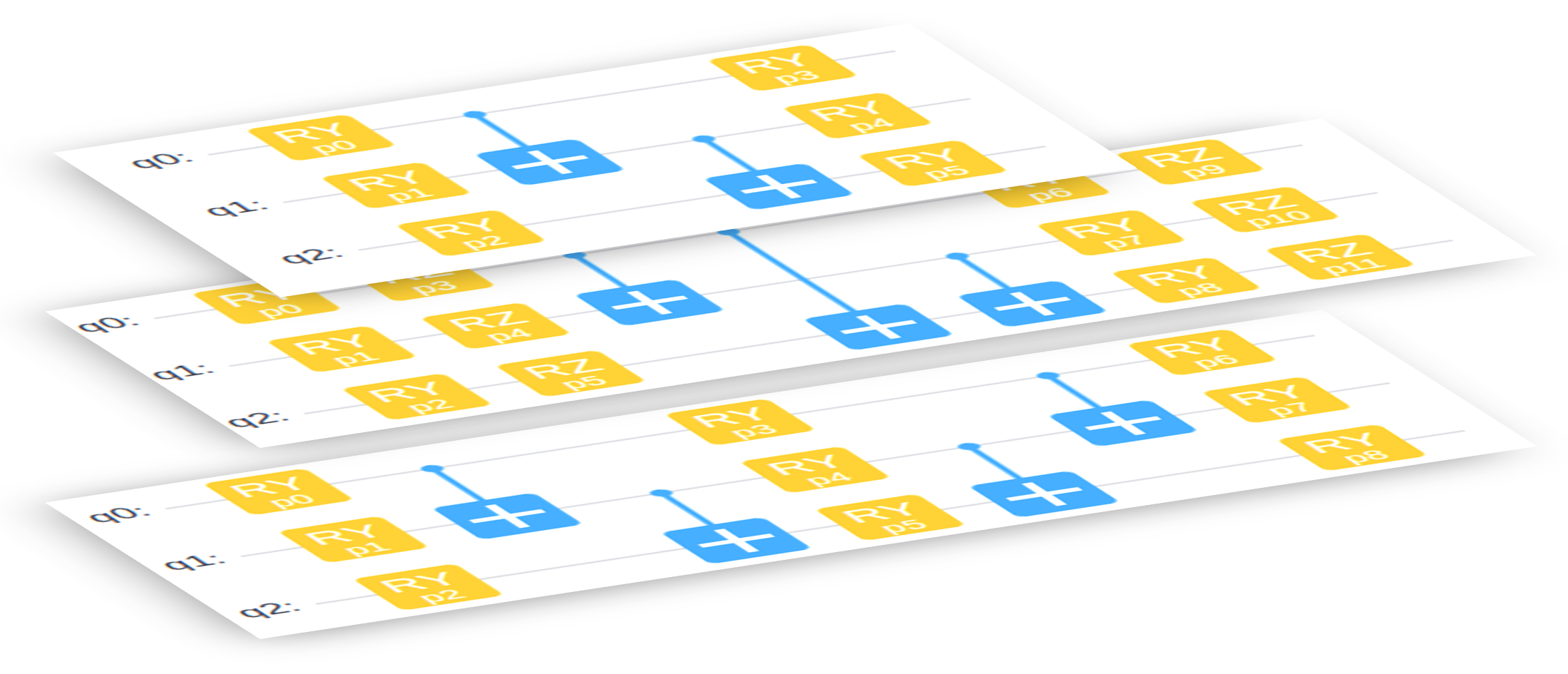}};

        \path[draw=black,line width=0.0794cm] (14.8034, 17.608) rectangle (29.2232, 10.2791);
        \node[align=center, font=\bfseries] at (22.0133, 17.008) {Compiler Algorithm};

        \path[fill=cffc46d,draw] (19.7379, 15.2929) rectangle (28.3633, 13.97);
        \node[align=center] at (24.0506, 14.63145) {DAG based circuit compiler};

        \path[fill=cffc46d,draw] (19.7379, 12.4883) rectangle (28.3633, 11.1654);
        \node[align=center] at (24.0506, 11.82685) {Qubit Mapping};
        \node at (17.2506, 13.22915) {\includegraphics[scale=0.5]{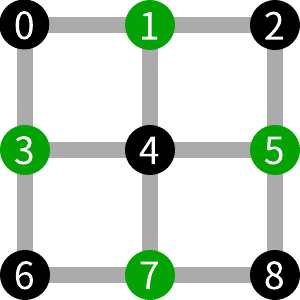}};

        \path[draw=black,line width=0.0794cm] (0.0397, 9.8822) rectangle (14.4595, 3.2941);
        \node[align=center, font=\bfseries] at (7.2496, 9.2822) {Quantum Simulator};

        \path[fill=c6de1ff] (0.4895, 8.5328) rectangle (14.0626, 7.2364);
        \node[align=center] at (7.2496, 7.8846) {Core Simulation Algorithm};

        \path[fill=caffaeb] (0.4763, 6.9585) rectangle (14.0758, 3.5719);
        \node[draw=none] at (5.7, 5.2415) {
            \begin{minipage}{0.4\textwidth}
                \begin{itemize}
                    \item Evolution of Circuit
                    \item Sampling Measurement
                    \item Expectation of Observable
                    \item Gradient calculation in batch mode
                \end{itemize}
            \end{minipage}
        };

        \path[draw=black,line width=0.0794cm] (0.0661, 2.8972) rectangle (14.4859, 0.0397);
        \node[align=center, font=\bfseries] at (7.276,2.2972) {Policy Detail};

        \path[fill=c0157b4] (0.2646, 1.7992) rectangle (4.6038, 0.2117);
        \node[align=center,text=white] at (2.4342,1.00545) {X86(AVX)};

        \path[fill=c0157b4] (5.0271, 1.7992) rectangle (9.3663, 0.2117);
        \node[align=center,text=white] at (7.1967,1.00545) {GPU(CUDA)};

        \path[fill=c0157b4] (9.7896, 1.7992) rectangle (14.1288, 0.2117);
        \node[align=center,text=white] at (11.9592,1.00545) {Ascend(NENO)};

        \path[fill=cf2f2f2] (14.8034, 9.8822) rectangle (29.2232, 3.2941);
        \path[draw=black,line width=0.0794cm,dash pattern=on 0.1588cm off 0.1588cm] (14.8034, 9.8822) rectangle (29.2232, 3.2941);
        \node[align=center, font=\bfseries] at (22.0133, 6.58815) {Backend Provider};

        \path[fill=cf2f2f2] (14.8034, 2.8972) rectangle (29.2232, 0.0397);
        \path[draw=black,line width=0.0794cm,dash pattern=on 0.1588cm off 0.1588cm] (14.8034, 2.8972) rectangle (29.2232, 0.0397);
        \node[align=center, font=\bfseries] at (22.0133, 1.46845) {QPU};

    \end{tikzpicture}
    \caption{The framework of \MindQuantum}
    \label{fig:framework}
\end{figure*}

\begin{description}
    \item[First layer] This layer represents the fundamental building blocks of the framework. We provide various quantum gates and convenient ways to construct quantum circuits. To support variational quantum algorithms, we also offer a parameter resolver that converts certain quantum gates into variational ones. Additionally, we provide the capability to describe various observables.
    \item[Second layer] This layer is the algorithm layer, which includes general quantum algorithms and variational quantum algorithms. Furthermore, this layer encompasses quantum compilation algorithms that compile and map quantum circuits to quantum chips.
    \item[Third layer] This layer is the execution layer, which is divided into a quantum simulator and a quantum chip backend provider depending on the hardware used for execution. In the quantum simulator, we design high-performance simulation logic tailored for CPU, GPU, and Ascend architectures, allowing for flexible switching between single-precision and double-precision simulation modes.
\end{description}

The arrangement of this whitepaper is as follows: In Chapter \hyperref[sec:elements]{\ref*{sec:elements}}, we will introduce the basic elements in \MindQuantum\ for constructing quantum circuits and observables. Chapter \hyperref[sec:backend]{\ref*{sec:backend}} will provide a comprehensive demonstration of the usage methods for the quantum simulator. In Chapter \hyperref[sec:toolbox]{\ref*{sec:toolbox}}, we will discuss the modules in \MindQuantum\ that are relevant to variational quantum algorithms, including various types of variational quantum circuits and methods for solving gradients of variational quantum circuits. Chapter \hyperref[sec:applications]{\ref*{sec:applications}} will showcase the examples of recent research results in the current academic community using \MindQuantum. Chapter \hyperref[sec:qupack]{\ref*{sec:qupack}} will introduce the quantum acceleration engine, QuPack, which can significantly enhance quantum simulation efficiency in certain scenarios. In Chapter \hyperref[sec:benchmark]{\ref*{sec:benchmark}}, we will benchmark the performance of \MindQuantum\ against some alternative quantum computing frameworks in different scenarios. Chapter \hyperref[sec:chip]{\ref*{sec:chip}} will explain how to utilize \MindQuantum\ to run quantum algorithms on real quantum chips. We will summarize the whole paper in Chapter \hyperref[sec:summary]{\ref*{sec:summary}}. Finally, in Chapter \hyperref[sec:acknowledgement]{\ref*{sec:acknowledgement}}, we will express our gratitude to all those who have contributed to \MindQuantum\ and those who use \MindQuantum\ in their research.

\section{Elements of MindSpore Quantum}
\label{sec:elements}

View demo code of this chapter: \demonotebook{02}{chapt02} \ \demonotebookgithub{02}{chapt02}

\subsection{Data Type}

In quantum computing, the complex number is essential for drawing quantum information. A complex number is an element of a number system that extends the real number with a specific element denoted $i$, called the imaginary unit, and satisfying the equation $i^2=-1$.
The complex number can be expressed in the form $c=a+bi$, where $a$ and $b$ are real numbers, $a$ is called the real part, and $b$ is called the imaginary part.

In computer digital format, the data type of complex number $c$ is determined by the type of $a$ and $b$.
In general, if $a$ and $b$ are Single, the type of $c$ is complex64 (used 64 bits to store $c$). If $a$ and $b$ are Double, the type of $c$ is complex128 (used 128 bits).

In different programming languages, the symbol of the imaginary unit may be different.
For example, In C, $i$ is denoted as $I$, in Python, it is denoted as $1j$.
In \MindQuantum, single- and double-precision floating-point formats are supported, in addition, their complex version formats are also supported which are indispensable to simulating the quantum evolution.
These data types can be easily converted with the same data type of the NumPy \cite{harris2020array} package. They have the same mathematics property but more easily implement in \MindQuantum\ than declaring the data as NumPy's data type.

\begin{lstlisting}
import numpy as np
import mindquantum as mq

# We support for different types
all_types = [mq.float32, mq.float64, mq.complex64, mq.complex128]

# convert from numpy and to numpy
mq_float64 = mq.to_mq_type(np.float64)
numpy_float64 = mq.to_np_type(mq.float64)
\end{lstlisting}

The exponential wall is the main obstacle to simulating a quantum system in a classical computer.
As the size of the quantum system increases, the required classical resources exponentially increase.
A quantum state with $N$ qubits can be described by $2^N$ amplitudes, all of which are complex numbers.

If we want to store an $N=20$ quantum state with $2^{20}$ amplitudes and all amplitudes are represented as single precision (Complex64), the classical memory is
\begin{equation}
    G=2^{20}\times 64 \mathrm{bits} = 8 \mathrm{Mb}.
\end{equation}

If we use complex double to represent the amplitude, the memory requirement is $G=2^{20}\times 128 bits=16 Mb$. Here we give a table that shows the memory consuming for storage full amplitudes quantum state with different data accuracy.

\begin{table}[ht]
    \begin{tabular}{ccc}
        \toprule
        Qubit Number & Complex128 & Complex64 \\
        \midrule
        6            & 1kB        & 0.5kB     \\
        16           & 1MB        & 0.5MB     \\
        26           & 1GB        & 0.5GB     \\
        30           & 16GB       & 8GB       \\
        36           & 1TB        & 0.5TB     \\
        40           & 16TB       & 8TB       \\
        \bottomrule
    \end{tabular}
    \caption{Memory consuming for storage full amplitudes quantum state.}
    \label{tab:mem_consume}
\end{table}

It is not hard to see, double precision requires twice as much storage as single precision because every amplitude is $128$ bits in Double but $64$ bits in Single.
Although double precision requires more space and usually more computing time (which depends on the specific processing unit), we can get higher precision which is very important for some takes in quantum computing, such as finding the eigenvalue of quantum Hamiltonians.
In fact, using single-precision to represent the amplitude can only simulate one more qubit than double-precision at most in the same memory resources.
Then we strongly recommend that complex double as the first choice if you have enough memory resources.

\subsection{Parameter Resolver}
\subsubsection{Parameters in MindQuantum}
The parameterized quantum gates or circuits are key ingredients for quantum computing, especially for hybrid quantum-classical algorithms.
In \MindQuantum, we provide a Python Class \code{mindquantum.core.ParameterResolver} to efficiently manage these classical parameters.
Suppose we would like to initialize a parameterized gate, where the value of the parameter may change during an optimization process.
\ParameterResolver solves this problem by
denoting each parameter as a unique symbol, such as $\theta=0.1$, where $\theta$ is the symbol of the parameter and $0.1$ is the corresponding value.
Then we can use some symbols to denote the parameters instead of the real value when we write the code in \MindQuantum.
When we declare an instance of \ParameterResolver, we only need to give the symbols and the values as a Python dict, where the symbol is the key of the dict and the values as the value of the dict,

\begin{lstlisting}
import numpy as np
from mindquantum.core.parameterresolver import ParameterResolver

# Construct ParameterResolver with a dict
pr1 = ParameterResolver({'theta': np.pi})

# Construct ParameterResolver with a symbol 'a'
pr2 = ParameterResolver('a')

# This is equal to
pr3 = ParameterResolver({'a': 1.0})

# Declare a constant
pr4 = ParameterResolver(3)

# Arithmetic operator is also supported
pr5 = pr1 + pr2 * np.sqrt(2)
print(pr5.expression())
\end{lstlisting}

The expression of \code{pr5} will be displayed as $\sqrt{2}*a + \pi*\text{theta}$.

When parameterized circuits are simulated by \MindQuantum, the \ParameterResolver replaces the symbol in circuits as the corresponding real value declared in the \ParameterResolver.

In a hybrid quantum-classical algorithm, the parameters are usually optimized by some gradient descent methods such as SGD, ADAM, and BFGS. When we declare an instance of \ParameterResolver, by default, all parameters need to be differentiable.
However, we can explicitly state whether a parameter needs to be differentiated as follows,

\begin{lstlisting}
params.no_grad_part('theta')
\end{lstlisting}

Here, \code{params} is an instance of \ParameterResolver. The \code{no_grad_part('theta')} method is invoked to specify that the parameter named \code{theta} should be non-differentiable. This means that in any subsequent optimization steps or gradient calculations, the \code{theta} parameter will be excluded.

In a hybrid algorithm, parameterized quantum gates or circuits are usually divided into two parts, encoder, and ansatz. The encoder encodes the classical data into the quantum state. And the ansatz is an assumption about the form of an unknown function, and we can approximate the solution to the problem by optimizing it. By default, all parameters in \ParameterResolver are initialized as ansatz's parameters. We can convert these parameters between encoder's and ansatz's parameters in the following way

\begin{lstlisting}
#declare all parameters as encoder's parameters
params.as_encoder()

#declare all parameters as ansatz's parameters
params.as_ansatz()

#declare a given symbol as encoder's parameter
params.encoder_part('theta')

#declare a given symbol as ansatz's parameter
params.ansatz_part('theta')
\end{lstlisting}

For more specific examples, please refer to section \ref{sec:rotate_three_states}.

We can also change the data type of parameters in a \ParameterResolver instance in the following way
\begin{lstlisting}
# reset the data type of params as mq.complex128
params.astype(mq.complex128)
\end{lstlisting}

\subsubsection{Parameter Generator}

Besides, \MindQuantum\ also provides a powerful tool to generate parameters quickly, \PRGenerator.
\begin{lstlisting}
from mindquantum.core.parameterresolver import PRGenerator

prg = PRGenerator(suffix='a')
a = prg.new()
b = prg.new()
print(a)
print(b)
print(len(prg.all_pr))
\end{lstlisting}
The output is:
\begin{lstlisting}
p0_a
p1_a
2
\end{lstlisting}

\subsection{Quantum Gate}
In this section, we introduce quantum gates and demonstrate their use in \MindQuantum. We categorize quantum gates into three types: fixed quantum gates, parameterized quantum gates, and custom quantum gates. Additionally, we utilize the \code{on()} method to set target and control qubits and the \code{controlled()} function for adding control qubits to multiple gates.

\subsubsection{Quantum Gates Overview}

In quantum computing, quantum gates are fundamental operations that manipulate qubits, the basic units of quantum information. Analogous to classical logic gates that operate on bits, quantum gates act on qubits and are represented mathematically by unitary matrices, preserving the quantum state's normalization and reversibility.

Quantum gates are essential for constructing quantum circuits and implementing quantum algorithms by performing operations such as rotations, flips, and entanglement creation. These operations enable unique computational capabilities distinct to quantum computers.

Quantum gates are categorized based on the number of qubits they act upon: single-qubit gates modify individual qubit states, while multi-qubit gates facilitate interactions between qubits, enabling complex quantum operations and entanglement. Gates can also be classified as non-parameterized (e.g., \X, \Y, \Z, \H) or parameterized (e.g., \RX, \RY, \RZ), the latter providing additional computational flexibility. In \MindQuantum, these gates---including parameterized ones using \ParameterResolver---are easily constructed.

\subsubsection{Fixed Quantum Gates}
Fixed quantum gates perform specific operations on quantum states, represented by unitary matrices. Here are some examples for a qubit state $\ket{\phi}=a\ket{0}+b\ket{1}$:

\begin{align*}
    I\ket{\phi} & =
    \begin{bmatrix}
        1 &  & 0 \\
        0 &  & 1
    \end{bmatrix}
    \begin{bmatrix}
        a \\
        b
    \end{bmatrix}=
    \begin{bmatrix}
        a \\
        b
    \end{bmatrix},  \\
    X\ket{\phi} & =
    \begin{bmatrix}
        0 &  & 1 \\
        1 &  & 0
    \end{bmatrix}
    \begin{bmatrix}
        a \\
        b
    \end{bmatrix}=
    \begin{bmatrix}
        b \\
        a
    \end{bmatrix},  \\
    Y\ket{\phi} & =
    \begin{bmatrix}
        0 & -i \\
        i & 0
    \end{bmatrix}
    \begin{bmatrix}
        a \\
        b
    \end{bmatrix}=
    \begin{bmatrix}
        -bi \\
        ai
    \end{bmatrix},  \\
    Z\ket{\phi} & =
    \begin{bmatrix}
        1 & 0 \\
        0 & -1
    \end{bmatrix}
    \begin{bmatrix}
        a \\
        b
    \end{bmatrix}=
    \begin{bmatrix}
        a \\
        -b
    \end{bmatrix}, \\
    H\ket{\phi} & =
    \frac{1}{\sqrt{2}}
    \begin{bmatrix}
        1 & 1  \\
        1 & -1
    \end{bmatrix}
    \begin{bmatrix}
        a \\
        b
    \end{bmatrix}=
    \frac{1}{\sqrt{2}}
    \begin{bmatrix}
        a+b \\
        a-b
    \end{bmatrix}.
\end{align*}

Example usage in \MindQuantum:
\begin{lstlisting}
from mindquantum.core.gates import H, SWAP

h = H.on(0)
swap = SWAP.on([0, 1])
print(swap)
\end{lstlisting}
The output is:
\begin{lstlisting}
SWAP(0 1)
\end{lstlisting}

\subsubsection{Parameterized Quantum Gate}
Pauli matrices are very important matrices, and they form a set of bases for spatial operators. When the Pauli matrix appears on the exponents, three classes of useful unitary operators are produced, namely rotation operators with respect to $\hat{x}, \hat{y}, \hat{z}$, defined by the following equations.

\begin{align*}
    Rx(\theta) & =
    e^{-\frac{i\theta X}{2}}=
    \cos{\frac{\theta}{2}}I-i\sin{\frac{\theta}{2}}X \\
               & =\begin{bmatrix}
        \cos{\frac{\theta}{2}}   &  & -i\sin{\frac{\theta}{2}} \\
        -i\sin{\frac{\theta}{2}} &  & \cos{\frac{\theta}{2}}
    \end{bmatrix},        \\
    Ry(\theta) & =
    e^{-\frac{i\theta Y}{2}}=
    \cos{\frac{\theta}{2}}I-i\sin{\frac{\theta}{2}}Y \\
               & =    \begin{bmatrix}
        \cos{\frac{\theta}{2}} &  & -\sin{\frac{\theta}{2}} \\
        \sin{\frac{\theta}{2}} &  & \cos{\frac{\theta}{2}}
    \end{bmatrix},    \\
    Rz(\theta) & =
    e^{-\frac{i\theta Z}{2}}=
    \cos{\frac{\theta}{2}}I-i\sin{\frac{\theta}{2}}Z \\
               & =    \begin{bmatrix}
        e^{\frac{-i\theta}{2}} &  & 0                     \\
        0                      &  & e^{\frac{i\theta}{2}}
    \end{bmatrix}.
\end{align*}

Since there is an infinite number of $2*2$ matrices, the number of quantum gates is also infinite. According to the Z-Y decomposition, a unitary matrix $U$ over any single qubit can be represented as $U=e^{i\alpha}Rz(\beta)Ry(\gamma)Rz(\delta)$. Therefore, in order to construct general quantum gates, it is necessary to use parameters to construct rotating gates. There are three initialization methods provided in \MindQuantum, because \ParameterResolver has three initialization methods. Take the RX gate as an example:

\begin{lstlisting}
from mindquantum.core.gates import RX
from mindquantum.core.parameterresolver import ParameterResolver as PR
import numpy as np

rx1 = RX(0.5)
rx2 = RX('a')
rx3 = RX({'a': 0.2, 'b': 0.5})

a, b = PR('a'), PR('b')
rx4 = RX(0.2 * a + 0.5 * b)

mat_rx1 = rx1.matrix()
mat_rx4 = rx4.matrix(pr={'a': 1, 'b': 2})
\end{lstlisting}

Note that in the above demo, \code{rx1} is actually a non-parameterized gate, since we already set the rotation angle $\theta = 0.5$.

\subsubsection{Custom Quantum Gate}
Creating arbitrary quantum gates is sometimes necessary for specific applications. In \MindQuantum, two methods are available for constructing custom gates:

\textit{Universal Math Gate} -- If the matrix representation of a gate is known, it is convenient to construct the gate in \MindQuantum. Two parameters are required to initialize \UnivMathGate, which require the gate name and the matrix value. If the matrix is not unitary, the state vector cannot be normalized.
Example:
\begin{lstlisting}
from mindquantum.core.gates import UnivMathGate
import numpy as np

x_mat = np.array([[0,1],[1,0]])
custom_X_gate = UnivMathGate('custom_X', x_mat).on(0, 1)
print(custom_X_gate)
\end{lstlisting}

Output:
\begin{lstlisting}
custom_X(0 <-: 1)
\end{lstlisting}

\textit{Universal Parameterized Gate} -- For some applications we need to construct a parameterized gate in which the parameter is varying. In \MindQuantum, we can easily construct a customized parameterized gate by \geneunivparameterizedgate, and its usage is basically as same as that of \RX gate. Two parameters are required to initialize such a gate. One is a function or method to use only one parameter (similar to theta in \RX) to generate a unitary matrix, noting that no error is reported if the resulting matrix is not unitary. The other is the function or method that produces the derivative of this matrix, which is used to calculate the gradient. This method supports the generation of arbitrary qubit operators and can accelerate the performance by numba.JIT.

Example:
\begin{lstlisting}
from mindquantum.core.gates import gene_univ_parameterized_gate

def matrix(theta):
    return np.array([[np.exp(1j * theta), 0],
                     [0, np.exp(-1j * theta)]])

def diff_matrix(theta):
    return 1j * np.array([[np.exp(1j * theta), 0],
                          [0, -np.exp(-1j * theta)]])

TestGate = gene_univ_parameterized_gate('Test', matrix, diff_matrix)

# Non-parameterized usage
test1 = TestGate(0.5).on(0)

# Parameterized usage
test2 = TestGate('a').on(0)
\end{lstlisting}

\subsubsection{``On'' Method}
For some controlled quantum gates, we need to specify the target qubits and control qubits of the quantum gate. In \MindQuantum, we can implement this function through the \code{on()} method. This method takes two parameters. One is the target qubits and the other is the control qubits, both of which can be single qubit or multiple qubits. It is worth emphasizing that any gate can add arbitrary control operations.

Example:
\begin{lstlisting}
from mindquantum.core.gates import X

x = X.on(0, [1, 2])
print('Target qubit:{}, Control qubits:{}'.format(x.obj_qubits, x.ctrl_qubits))
\end{lstlisting}

Output:
\begin{lstlisting}
Target qubit:[0], Control qubits:[1, 2]
\end{lstlisting}

\subsubsection{``Controlled'' Function}
\begin{figure}[h]
    \centering
    \includegraphics[width=0.9\linewidth]{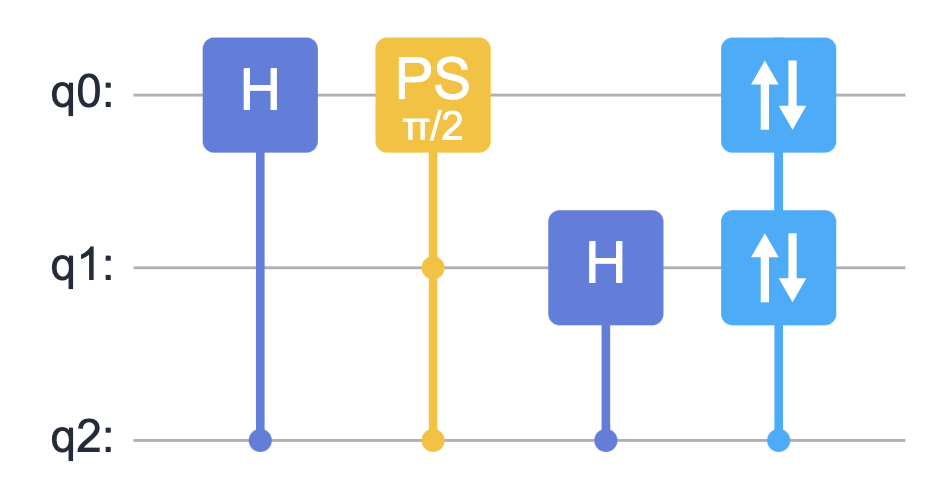}
    \caption{Add control qubits}
    \label{fig:controlled-method}
\end{figure}

In addition to the \code{on()} method, we can also add control bits via the \code{controlled()} function. The \code{controlled()} function is used to add control qubits (which can be multiple) to any quantum circuit or quantum operator. For example, we build a quantum circuit containing only two qubits and add a control qubit q2 to it by \code{controlled()} method:
\begin{lstlisting}
from mindquantum.algorithm.library import qft
from mindquantum.core.circuit import controlled

u1 = qft(range(2))
u2 = controlled(u1)
u2 = u2(2)
u2.svg()
\end{lstlisting}

As shown in Fig.~\ref{fig:controlled-method}, the QFT circuit now is controlled by q2.

\begin{figure}[h]
    \centering
    \includegraphics[width=0.9\linewidth]{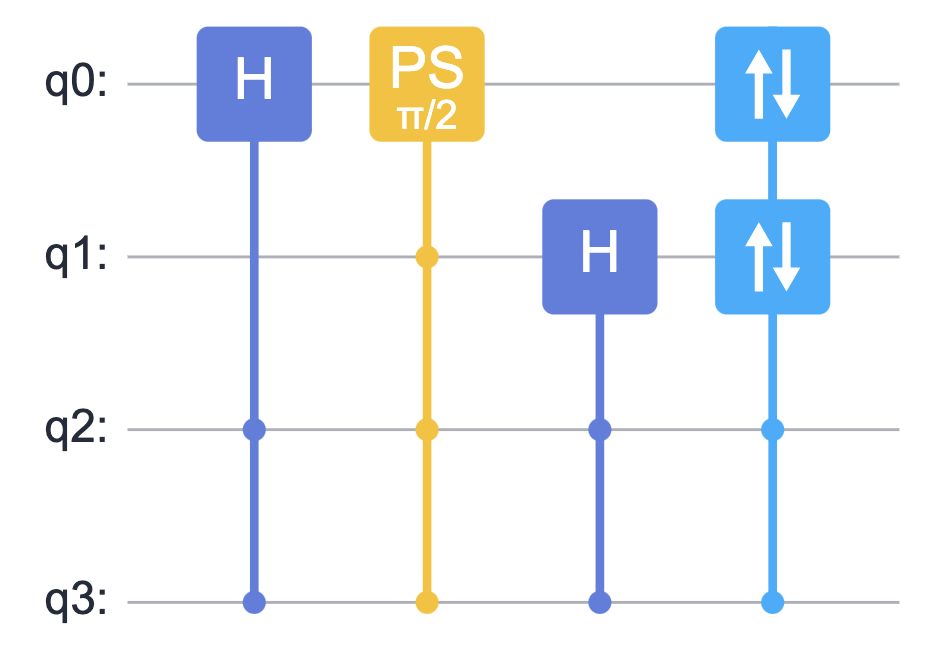}
    \caption{Add control qubits in batches}
    \label{fig:batch-control}
\end{figure}

In addition, we can add control bits to quantum circuits in batches. In the following example, the target qubits are q0 and q1, with control qubits q2 and q3, respectively:
\begin{lstlisting}
u = controlled(qft)
u = u([2, 3], [0, 1])
u.svg()
\end{lstlisting}

The result is shown in Fig.~\ref{fig:batch-control}.

\subsection{Quantum Circuit}
The quantum circuit is a graphical representation of the sequence of quantum gates applied to qubits in a quantum computation or quantum algorithm. Similar to classical digital circuits made up of logic gates that manipulate classical bits, quantum circuits are composed of quantum gates that manipulate qubits.

\subsubsection{Construct a quantum circuit}
\begin{figure}[h]
    \begin{center}
        \includegraphics[width=0.9\linewidth]{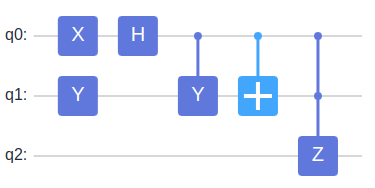}
    \end{center}
    \caption{Quantum Circuit.}
    \label{fig:Quantum-Circuit}
\end{figure}

In \MindQuantum\ we use \Circuit to represent quantum circuit:

\begin{lstlisting}
from mindquantum.core.circuit import Circuit
circ = Circuit()
\end{lstlisting}

There are several basic operations for adding a quantum gate into \Circuit
\begin{itemize}
    \item list: directly construct a quantum gate with gate list.
    \item \code{+=}: use \code{+=} to add a quantum gate to the circuit.
    \item \code{Circuit.x}: add an X gate to the circuit.
\end{itemize}
\begin{lstlisting}
from mindquantum.core.gates import H, Y, X, Z
circ = Circuit([X.on(0), Y.on(1)])
circ += H.on(0)
circ += Y.on(1, 0)
circ.x(1, 0)
circ.z(2, [0, 1])
\end{lstlisting}

the result is shown in Fig.~\ref{fig:Quantum-Circuit}.

\subsubsection{Display a quantum circuit}
SVG(Scalable Vector Graphics) is based on the XML markup language and is used to describe vector graphics in two dimensions. \MindQuantum\ provides a function to export quantum circuit to SVG format.
\begin{lstlisting}
circ.svg().to_file(filename='circuit.svg')
\end{lstlisting}
Please note that if you are in jupyter notebook environment, you can directly get SVG image by \code{circ.svg()}.

\subsubsection{Common interfaces}

Quantum Circuit is the basic element of quantum algorithm, \MindQuantum\ provide a lot of useful methods for \Circuit.

\begin{itemize}
    \item \propnqubits: get the number of qubits of quantum circuit.
    \item \propparamsname: get the parameter names of circuit.
    \item \prophasmeasuregate: get whether to be measured.
    \item \methodmatrix: get the circuit matrix.
    \item \methodgetqs: get the final quantum state.
    \item \methodsummary: get information about the current circuit, including the number of blocks, gates, gates without parameters, gates with parameters and parameters.
\end{itemize}

\subsubsection{Advanced operator on circuit}

Constructing a large size quantum circuit is not straightforward. Here in \MindQuantum, we provide some pre-defined methods to make \Circuit more user-friendly. For example, let the origin quantum circuit be:

\begin{lstlisting}
from mindquantum.core.circuit import *

circ = Circuit().h(0).rx('a', 2, 0)
\end{lstlisting}

\begin{figure}[H]
    \begin{center}
        \includegraphics[width=0.4\linewidth]{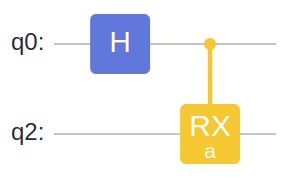}
    \end{center}
    \caption{Quantum Circuit need to manipulate.}
\end{figure}
We now show how to manipulate on this quantum circuit:

\begin{enumerate}
    \item \methodcompress{circ} : compress all qubit to first $n$ qubits.\par
          \begin{minipage}{\linewidth}
              \centering
              \includegraphics[width=0.4\linewidth]{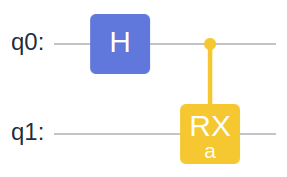}
          \end{minipage}
    \item \methodcontrol{circ}{1} : add control qubits on this circuit\par
          \begin{minipage}{\linewidth}
              \centering
              \includegraphics[width=0.4\linewidth]{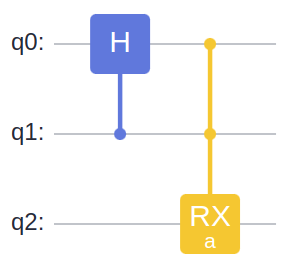}
          \end{minipage}
    \item \methoddagger{circ} : get hermitian conjugate version\par
          \begin{minipage}{\linewidth}
              \centering
              \includegraphics[width=0.4\linewidth]{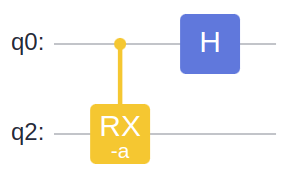}
          \end{minipage}
    \item \methodapply{circ}{[2, 1]} : apply circuit to other qubits\par
          \begin{minipage}{\linewidth}
              \centering
              \includegraphics[width=0.4\linewidth]{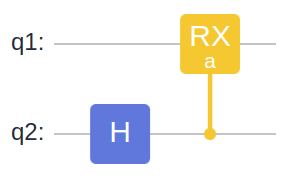}
          \end{minipage}
    \item \methodreverse{circ} : reverse the qubit order in circuit\par
          \begin{minipage}{\linewidth}
              \centering
              \includegraphics[width=0.4\linewidth]{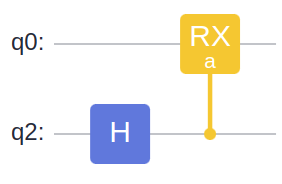}
          \end{minipage}
    \item \methodshift{circ}{2} : shift the qubit range with given step\par
          \begin{minipage}{\linewidth}
              \centering
              \includegraphics[width=0.4\linewidth]{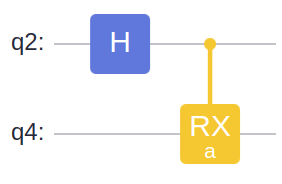}
          \end{minipage}
\end{enumerate}

\subsection{Observable and Hamiltonian}
Simulating many-body physical systems was proposed by Richard Feynman as a main application for quantum computation. Among the various many-body physics models, we focus on the Heisenberg model and the Fermi-Hubbard model. The Hamiltonians for these models can be expressed as follows:

\begin{align*}
    H_\text{Heisenberg}    & = -J\sum_{<i,j>}\sigma_i\otimes \sigma_j-h\sum_i\sigma_i,                                            \\
    H_\text{Fermi-Hubbard} & = -\sum_{<i,j>,\sigma}\left(a_{i\sigma}^\dagger a_{j\sigma} + a_{j\sigma}^\dagger a_{i\sigma}\right) \\
                           & + U\sum_i n_{i\uparrow}n_{i\downarrow}.
\end{align*}

In \MindQuantum, it is easy to construct those Hamiltonian with the help of \QubitOperator and \FermionOperator.

\subsubsection{Qubit Operators}

In Heisenberg model, $\sigma_i$ represents the Pauli operator. The matrix form of Pauli operators are:

\begin{align*}
    \sigma_X & = \begin{pmatrix}
        0 & 1 \\
        1 & 0
    \end{pmatrix}, \sigma_Y = \begin{pmatrix}
        0 & -i \\
        i & 0
    \end{pmatrix}, \\
    \sigma_I & =\begin{pmatrix}
        1 & 0 \\
        0 & 1
    \end{pmatrix}, \sigma_Z = \begin{pmatrix}
        1 & 0  \\
        0 & -1
    \end{pmatrix}.
\end{align*}
In \MindQuantum, \QubitOperator is used to build this kind of operator. Given a Pauli word $\sigma_{X,3}\otimes \sigma_{Y,1}\otimes \sigma_{Z,0}=X_3 Y_1 Z_0$, applying Pauli $Z$, Pauli $Y$ and Pauli $X$ on qubit 3, 1 and 0,  we can easily construct it with:

\begin{lstlisting}
from mindquantum.core.operators import QubitOperator

ops = QubitOperator('Z0 Y1 X3')
\end{lstlisting}
Please note that since $[Z_0, Y_1]= [Z_0,X_3] = [Y_1, X_3]=0$, so the order of Pauli word in the Pauli string does not matter.

\QubitOperator also support arithmetic operations, in order to build more complex operators:

\begin{lstlisting}
from mindquantum.core.operators import QubitOperator
from mindquantum.core.parameterresolver import ParameterResolver as PR

op1 = QubitOperator('X0')
op2 = QubitOperator('Z1', 'a')
op3 = QubitOperator('Y1')
op4 = 2 * op1 * op2 + op3 * PR('b')
print(op4)
print(op4.subs({'a':1, 'b':2}).matrix().toarray())
\end{lstlisting}
The output is:
\begin{lstlisting}
2*a [X0 Z1] +
  b [Y1]

[[ 0.+0.j  2.+0.j  0.-2.j  0.+0.j]
 [ 2.+0.j  0.+0.j  0.+0.j  0.-2.j]
 [ 0.+2.j  0.+0.j  0.+0.j -2.+0.j]
 [ 0.+0.j  0.+2.j -2.+0.j  0.+0.j]]
\end{lstlisting}
In the last line, we used \code{.subs} to set the value of parameters and obtained the CSR format sparse matrix with \code{.matrix}.

\subsubsection{Fermion Operators}

In Fermi-Hubbard model, $a_i^\dagger$ and $a_i$ are creation and annihilation operators for fermionic particles. Different from Pauli operators, the fermion operator follows anti-commutation relation:
\begin{align*}
    \{a_i, a_j^\dagger\} & = a_ia_j^\dagger + a_j^\dagger a_i = \delta_{ij}, \\
    \{a_i, a_j\}         & = \{a_i^\dagger, a_j^\dagger\}=0.
\end{align*}
In qubit system, the creation and annihilation operators acting on state $\ket{0}$ and $\ket{1}$ obey the following rules:
\begin{align*}
    a\ket{0}          & =0,        & a\ket{1}          & =\ket{0}, \\
    a^\dagger \ket{0} & = \ket{1}, & a^\dagger \ket{1} & =0.
\end{align*}

The matrix forms of the creation and annihilation operators are:

\begin{align*}
    a=\begin{pmatrix}
        0 & 1 \\
        0 & 0
    \end{pmatrix},
    a^\dagger=\begin{pmatrix}
        0 & 0 \\
        1 & 0
    \end{pmatrix}.
\end{align*}
In \MindQuantum, we can use \FermionOperator to construct fermion operator. Suppose we first create a state on qubit 1 and annihilate a state on qubit 0, or $a_1^\dagger\otimes a_0$, one can build it with:
\begin{lstlisting}
from mindquantum.core.operators import FermionOperator
op1 = FermionOperator('1^ 0')
\end{lstlisting}
In the fermion string, we use \verb|^| to represent $\dagger$ and the number to label the corresponding qubit. The arithmetic operation of \FermionOperator is very similar to \QubitOperator.

\subsubsection{Operator Functions}

\MindQuantum\ also supplies a number of advanced functions for operators. Here are some examples:

- \methodcommutator{op1}{op2} : Output the commutator of two operators.

\begin{lstlisting}
from mindquantum.core.operators import QubitOperator, FermionOperator, commutator
qub_op1 = QubitOperator("X1 Y2")
qub_op2 = QubitOperator("X1 Z2")
commutator(qub_op1, qub_op1)  # 0
commutator(qub_op1, qub_op2)  # (2j) [X2]
\end{lstlisting}

- \methodcountqubits{op1} : Count the number of qubits before deleting unused qubits.

\begin{lstlisting}
from mindquantum.core.operators import QubitOperator,FermionOperator, count_qubits
qubit_op = QubitOperator("X1 Y2")
count_qubits(qubit_op)  # 3
fer_op = FermionOperator("1^")
count_qubits(fer_op)  # 2
\end{lstlisting}

- \methodhermitianconj{op1} : Get the hermitian conjugation of given operator.

\begin{lstlisting}
from mindquantum.core.operators import FermionOperator, hermitian_conjugated
fer_op = FermionOperator("1^ 3")
hermitian_conjugated(fer_op)
\end{lstlisting}

\subsubsection{Transformation}
Quantum simulation of fermionic systems is challenging. To effectively reduce the resources used to simulate fermionic Hamiltonians on quantum hardware, we can simulate fermions with qubits, which involves the conversion between the fermionic Hamiltonian and the qubit Hamiltonian. \MindQuantum\ supplies the module \code{mindquantum.algorithm.nisq.Transform} for transformation between fermions (\FermionOperator) and bosons (\QubitOperator). The following functions are provided:
\begin{itemize}
    \item \methodjordanwigner: Apply the Jordan-Wigner transformation, which maps fermionic annihilation operators (fermions) to qubits (bosons) via:
          \begin{equation}
              \begin{split}
                  a_j^{\dagger} \to {\sigma}_j^{-} X \prod_{i=0}^{j-1} {\sigma}^Z_i \\
                  a_j \to {\sigma}_j^{+} X \prod_{i=0}^{j-1} {\sigma}^Z_i,
              \end{split}
          \end{equation}
          where ${\sigma}_j^{+} = {\sigma}_j^X+i{\sigma}_j^Y$, ${\sigma}_j^{-} = {\sigma}_j^X-i{\sigma}_j^Y$, are the spin-up operator and spin-down operator respectively. The operator $\prod_{i=0}^{p-1} {{\sigma}^Z_i}$ is called a ``parity operator'' with eigenvalues $\pm 1$, because it measures whether the number of qubits in the down state (or ``$-1$'' state) from index 0 to $p-1$ is even or odd. If the number is even, the operator yields $+1$; if odd, it yields $-1$. This transformation preserves the locality of the initial occupation number. The problem with this method is that as a consequence of the non-locality of the parity operator, the number of extra qubit operations required to simulate a single fermionic operator scales as $O(n)$.
    \item \methodparity: Apply the parity transformation, which uses qubit $j$ to store the $parity$ of all occupied orbitals up to orbital $j$. That is, we could let qubit $j$ store $p_j = (\sum_{i=0}^j f_i) mod 2$. Here, $f_i$ represents the occupation number of the $i$-th orbital, which can be either 0 (unoccupied) or 1 (occupied). This encoding of fermionic states in qubit states is called the $parity$ basis. The occupation number basis states, which can be represented as vectors (e.g. $(j_7,...,j_1,j_0)^T$), can be mapped to the parity basis as follows:
          \begin{equation}
              p_i = \sum_j {[\pi_n]_{ij} f_j},
          \end{equation}
          where $n$ is the number of orbitals, $\pi_n$ is the ($n \times n$) matrix defined below:
          \begin{equation}
              [\pi_n]_{ij} =
              \begin{cases}
                  1 & i<j     \\
                  0 & i\geq j \\
              \end{cases}.
          \end{equation}
          The representations of the creation and annihilation operators in the parity basis are then:
          \begin{equation}
              \begin{split}
                  a_j^{\dagger} \to \frac{1}{2} (\prod_{i=j+1}^n ({\sigma}^X_i X))({\sigma}^X_j - i{\sigma}^Y_j) X {\sigma}^Z_{j-1} \\
                  a_j \to \frac{1}{2} (\prod_{i=j+1}^n ({\sigma}^X_i X))({\sigma}^X_j + i{\sigma}^Y_j) X {\sigma}^Z_{j-1}.
              \end{split}
          \end{equation}
    \item \methodbravyikitaev \cite{bravyi2002fermionic, fenwick1994new}: The previous two approaches are dual in the sense that they encode the information required to represent fermionic operators with qubits. With the occupation number basis and its associated Jordan-Wigner transformation, the occupation information is stored locally but the parity information is non-local, whereas in the parity basis method and its corresponding operator transformation, the parity information is stored locally but the occupation information is non-local.
          The Bravyi-Kitaev transformation is a middle ground, it balances the locality of occupation and parity information for improved simulation efficiency. In this scheme, qubits store the parity of a set of $2^x$ orbitals, where $x \geq 0$. A qubit of index $j$ always stores orbital $j$. For even values of $j$, this is the only orbital that it stores, but for odd values of $j$, it also stores a certain set of adjacent orbitals with index less than $j$. The map from the occupation number basis to the Bravyi-Kitaev basis is:
          \begin{equation}
              b_i = \sum_j {[\beta_n]_{ij} f_j},
          \end{equation}
          where $n$ is the number of orbitals, $\beta_n$ is an ($n \times n$) square matrix. See \cite{seeley2012bravyi} for a detailed explanation.
    \item \methodbravyikitaevsuperfast: A fast version of $bravyi\_kitaev()$, which can perform the Bravyi-Kitaev transformation at a faster speed \cite{setia2018bravyi}. Note that only Hermitian operators such as the following can be transformed:
          \begin{equation}
              C + \sum_{p,q} h_{p,q} a_p^{\dagger} a_q + \sum_{p,q,r,s} h_{p,q,r,s} a_p^{\dagger} a_q^{\dagger} a_r a_s,
          \end{equation}
          where $C$ is a constant.
    \item \methodreversedjordanwigner: Apply inverse transformation of Jordan-Wigner, which will transform $QubitOperator$ to $FermionOperator$.
    \item \methodternarytree: Apply the Ternary-Tree transformation. This function is based on \cite{jiang2020optimal}.
\end{itemize}
In the code below, we use the function $Transform()$ to transform $FermionOperator$ to $QubitOperator$.
\begin{lstlisting}
from mindquantum.core.operators import FermionOperator
from mindquantum.algorithm.nisq import Transform
op1 = FermionOperator('1^')
op_transform = Transform(op1)
op_transform.jordan_wigner()
# 0.5 [Z0 X1] +
# -0.5j [Z0 Y1]
op_transform.parity()
# 0.5 [Z0 X1] +
# -0.5j [Y1]
op_transform.bravyi_kitaev()
# 0.5 [Z0 X1] +
# -0.5j [Y1]
op2 = FermionOperator('1^', 'a')
Transform(op2).jordan_wigner()
# 0.5*a [Z0 X1] +
# -0.5*I*a [Z0 Y1]
\end{lstlisting}

\subsection{Basic Usage of Simulator}
\label{sec:sim_basic_usage}
There are many methods to simulate a quantum system. For example, you can use a full state vector to simulate a pure quantum system and use a density matrix to simulate a mixed state quantum system. In \MindQuantum, both of these methods are supported.

\begin{lstlisting}
from mindquantum.core.gates import H
from mindquantum.simulator import Simulator

sim = Simulator('mqvector', 2)
sim.apply_gate(H.on(0))
print(sim)
\end{lstlisting}
The output is:

\begin{lstlisting}[escapeinside={\[}{\]}]
mqvector simulator with 2 qubits (little endian), dtype: mindquantum.complex128.
Current quantum state:
[$\surd{}$]2/2|00>
[$\surd{}$]2/2|01>
\end{lstlisting}

In the above demo, we initialized a full state vector simulator named \code{"mqvector"}, which is a full state quantum simulator. Up to version 0.9.0, all simulator that \MindQuantum\ supported is shown in the following table:
\begin{table}[ht]
    \begin{tabular}{ccccc}
        \toprule
        Name          & CPU & GPU & Complex64 & Complex128 \\
        \midrule
        mqvector      & Yes & No  & yes       & yes        \\
        mqvector\_gpu & No  & Yes & yes       & yes        \\
        mqmatrix      & Yes & No  & yes       & yes        \\
        \bottomrule
    \end{tabular}
    \caption{Supported quantum simulator.}
    \label{tab:simulator_supported}
\end{table}
After applying a Hadamard gate, we gate a quantum state $\left(\ket{00}+\ket{01}\right)/\sqrt{2}$. What we need to emphasize is that in the whole framework, we use little endian notation, that is, for base $\ket{01}$, the first qubit is in state $\ket{1}$ end the second qubit is in state $\ket{0}$.

Every simulator in \MindQuantum\ maintains a quantum state, and we can use methods start with ``apply'' to change the quantum state and use methods start with ``get'' to extracting information from the quantum state.

\subsubsection{``apply'' method}

``apply'' method is going to apply an operator on the maintained quantum state and change the state. We can apply a single gate or a quantum circuit, as well as a Hamiltonian, to manipulate the quantum state. We should notice that after applying a Hamiltonian, the quantum state will not be a will defined quantum state.

\begin{lstlisting}
from mindquantum.core.circuit import Circuit
from mindquantum.core.operators import Hamiltonian, QubitOperator
from mindquantum.simulator import Simulator

circ = Circuit().h(0).x(1, 0)
ham = Hamiltonian(QubitOperator('X0 Y1'))
sim = Simulator('mqvector', 2)
sim.apply_hamiltonian(ham)
sim.apply_circuit(circ)
\end{lstlisting}
The output is:
\begin{lstlisting}[escapeinside={\[}{\]}]
mqvector simulator with 2 qubits (little endian), dtype: mindquantum.complex128.
Current quantum state:
-[$\surd{}$]2/2j|01[$\rangle$]
[$\surd{}$]2/2j|10[$\rangle$]
\end{lstlisting}

\subsubsection{``get'' method}

``get'' method will not change the maintained quantum state but extract useful information.

- \getqs: You can get the quantum matrix or a ket string expression (with argument \code{ket=True}) from a simulator with this method.
\begin{lstlisting}
from mindquantum.algorithm.library import qft
from mindquantum.simulator import Simulator

sim = Simulator('mqvector', 3)
sim.apply_circuit(qft(range(3)))
qs = sim.get_qs()
print(sim.get_qs(ket=True))
\end{lstlisting}

- \getexpectation: Calculating the expectation of a given observable is a very common requirement. In \MindQuantum, we expand the definition of observable to below:

\begin{equation}
    E = \bra{\varphi}U_l^\dagger H U_r \ket{\psi}.
\end{equation}

Following this definition, we are easily to get the physical observable if we set $\ket{\varphi}=\ket{\psi}$ and $U_l = U_r$ and we cal also to calculate the inner product of two quantum state if we set $H=I$

\begin{lstlisting}
from mindquantum.core.circuit import Circuit
from mindquantum.core.operators import Hamiltonian, QubitOperator

# Get expectation
psi = Simulator('mqvector', 2)
H = Hamiltonian(QubitOperator("Z0 Z1"))
U_r = Circuit().rx(2.5, 0)
print("Expectation:", psi.get_expectation(H, U_r))

# Get inner product
phi = Simulator("mqvector", 2)
H = Hamiltonian(QubitOperator(""))
U_l = Circuit().ry(1.5, 0)
print("Inner product:")
print(psi.get_expectation(H, U_r, U_l, phi))
\end{lstlisting}
The output is:

\begin{lstlisting}
Expectation: (-0.8011436155469336+0j)
Inner product:
(0.23071786267161518-0.6468646992187576j)
\end{lstlisting}

\subsubsection{sampling}

Quantum sampling is a fundamental aspect of quantum computing that plays a pivotal role in solving certain complex problems efficiently. It leverages the principles of quantum superposition and entanglement to explore multiple possibilities simultaneously, enabling quantum computers to outperform classical counterparts in specific scenarios.

In \MindQuantum, we can use a simulator to sample the measurement result of a quantum circuit.
\begin{lstlisting}
from mindquantum.utils import random_circuit
from mindquantum.simulator import Simulator

circ = random_circuit(3, 5, seed=42)
circ.barrier().measure_all()
sim = Simulator('mqvector', circ.n_qubits)
res = sim.sampling(circ, shots=1000, seed=42)
\end{lstlisting}

\begin{figure}[h]
    \begin{center}
        \includegraphics[width=0.9\linewidth]{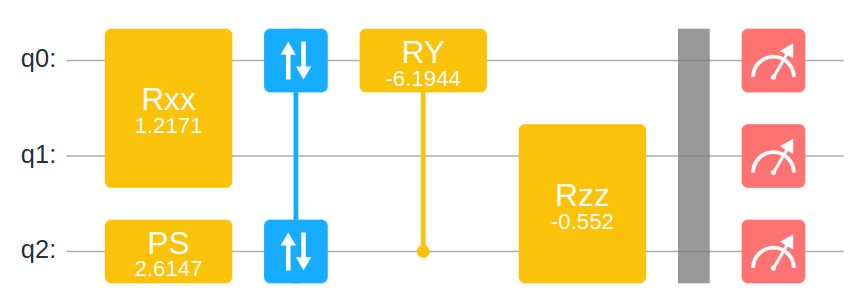}
    \end{center}
    \caption{Sampling circuit}
\end{figure}

\begin{figure}[h]
    \begin{center}
        \includegraphics[width=0.9\linewidth]{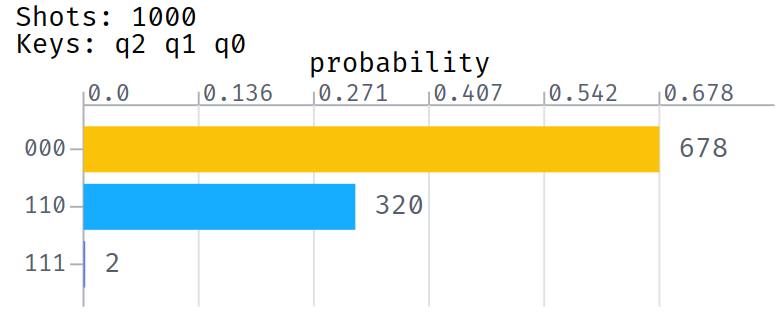}
    \end{center}
    \caption{Sampling result}
\end{figure}

\section{Circuit Simulation Backend}
\label{sec:backend}

\subsection{State Vector Simulator}
\label{sec_state_vec}
\subsubsection{Policy Based Design pattern}
In \MindQuantum, we use \Simulator to simulate quantum states. A \Simulator maintains a quantum state, as mentioned in section \ref{sec:sim_basic_usage}, methods start with ``apply'' will change this maintained quantum state, and methods start with ``get'' will not change anything but only extrude information from the quantum state.

Simulating quantum systems imposes high performance requirements on simulators. In \MindQuantum, we utilize C++ or CUDA as the underlying simulator implementation language, accelerating quantum simulations based on different hardware instruction sets. We adopt a policy-based development approach, optimizing the fundamental simulator interface for different instruction sets and encapsulating them into policy types. Building upon a unified quantum simulation framework, we generate different quantum simulators using various simulator policies, such as AVX-supported simulators, neon-supported simulators, and GPU-supported simulators.

\begin{figure}[ht]
    \centering
    \includegraphics[scale=0.6]{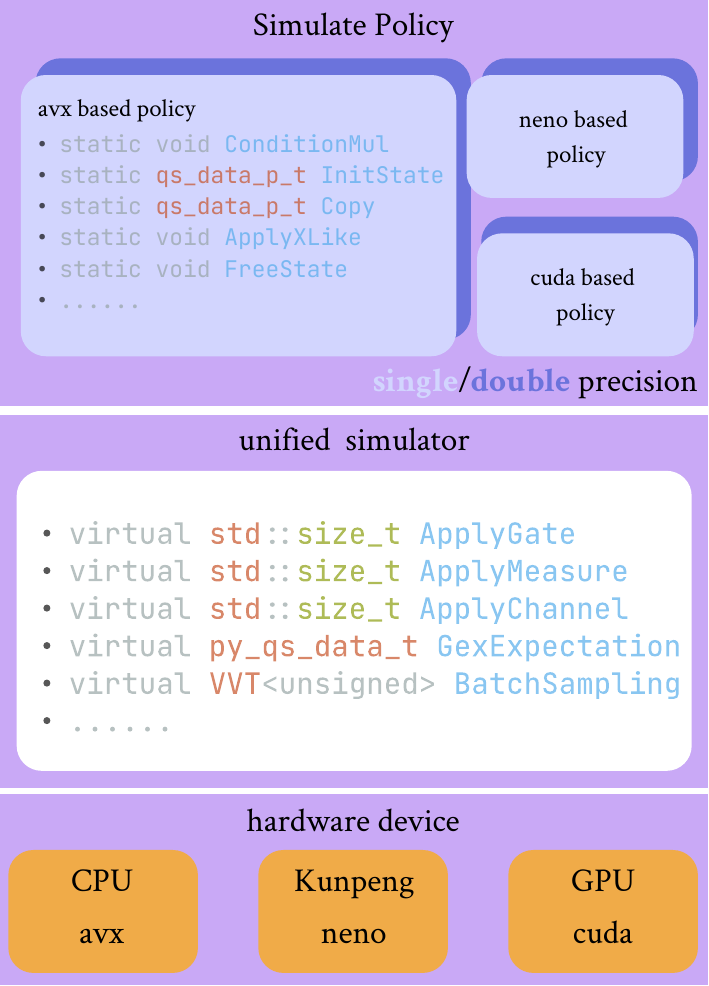}
    \captionsetup{justification=raggedright,singlelinecheck=false}
    \caption{\label{3_1_sim_str} The layer structure of the simulator. The \textit{simulate policy} layer applies quantum gates to quantum states based on the current simulation hardware. The \textit{unified simulator} layer uses different policies to complete the circuit simulation and expectation and gradient calculation. The \textit{hardware device} layer shows that we support CPU, GPU and Kunpeng platform.}
\end{figure}

Depending on the specific quantum system being simulated, we have the option to choose between a full amplitude quantum simulator or a density matrix simulator. In \MindQuantum, these two simulators are referred to as \code{"mqvector"} and \code{"mqmatrix"}, respectively. In the current \version version, the \code{"mqvector"} simulator also supports GPU acceleration and is named \code{"mqvector_gpu"}. Both \code{"mqvector"} and \code{"mqmatrix"} use the same design pattern as in Fig.~\ref{3_1_sim_str}. Based on these simulators, noise simulation is also supported with the help of quantum channel, which will be talked in section ~\ref{sec:noise_simulation}.

\subsubsection{Gate optimization with SIMD}
SIMD (Single Instruction Multiple Data) is a computer architecture and parallel processing technique designed to accelerate the execution of tasks that can be parallelized by applying a single operation to multiple data elements simultaneously. The fundamental building block of a quantum simulator is the interaction between quantum gate and quantum state, involving multiple parallel operations, so we can utilize SIMD to accelerate the speed of quantum simulations. In this section, we will show how to optimize gate evolution based on AVX instruction with \code{complex128} data type.

\begin{figure}
    \begin{tikzpicture}[scale=0.5]
        \node[left] at (0,0.5) {256 bits reg};
        \draw[line width=1.2] (0,0) rectangle (1, 1);
        \draw[line width=1.2] (1,0) rectangle (2, 1);
        \draw[line width=1.2] (2,0) rectangle (3, 1);
        \draw[fill=black] (3.6, 0.5) circle (0.1);
        \draw[fill=black] (4.0, 0.5) circle (0.1);
        \draw[fill=black] (4.4, 0.5) circle (0.1);
        \draw[line width=1.2] (5,0) rectangle (6, 1);
        \draw[line width=1.2] (6,0) rectangle (7, 1);
        \draw[line width=1.2] (7,0) rectangle (8, 1);
        \node[left] at (0,-1) {reg a};
        \draw[line width=1.2] (0,-1.5) rectangle (2, -0.5);
        \draw[line width=1.2] (2,-1.5) rectangle (4, -0.5);
        \draw[line width=1.2] (4,-1.5) rectangle (6, -0.5);
        \draw[line width=1.2] (6,-1.5) rectangle (8, -0.5);
        \node at (1,-1) {$b_i$};
        \node at (3,-1) {$b_r$};
        \node at (5,-1) {$a_i$};
        \node at (7,-1) {$a_r$};
        \node[left] at (0,-2.5) {reg b};
        \draw[line width=1.2] (0,-3) rectangle (2, -2);
        \draw[line width=1.2] (2,-3) rectangle (4, -2);
        \draw[line width=1.2] (4,-3) rectangle (6, -2);
        \draw[line width=1.2] (6,-3) rectangle (8, -2);
        \node at (1,-2.5) {$d_i$};
        \node at (3,-2.5) {$d_r$};
        \node at (5,-2.5) {$c_i$};
        \node at (7,-2.5) {$c_r$};
        \draw[dashed, line width=1.2, fill=red!50] (8, -0.5) -- (8, -1.5) -- (10, -1.4) -- (12, -1.4) -- (12, 0.5) -- (10, 0.5) -- cycle;
        \draw[dashed, line width=1.2, fill=blue!50] (8, -2) -- (8, -3) -- (10, -3.4) -- (12, -3.4) -- (12, -1.6) -- (10, -1.6) -- cycle;
        \node[right] at (9, 1) {$\ket{\psi}=$};
        \node[right] at (9.5, -0) {$a\ket{00}+$};
        \node[right] at (9.5, -1) {$b\ket{01}+$};
        \node[right] at (9.5, -2) {$c\ket{10}+$};
        \node[right] at (9.5, -3) {$d\ket{11}$};
    \end{tikzpicture}
    \caption{Store a 2-qubit quantum state into two 256-bit registers.}
    \label{fig:avx}
\end{figure}
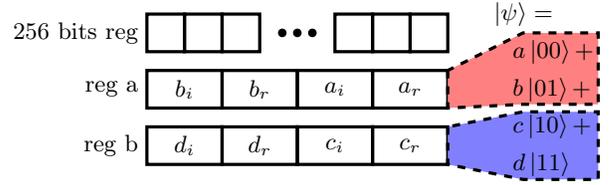

AVX instruction provide us with a 256-bit width vector register, which can store 4 numbers with double precision or two complex numbers with double precision. In Fig.~\ref{fig:avx}, we take a two-qubit full state vector for example, probability amplitudes $a$ and $b$ can be loaded into a 256-bit register simultaneously, and $c$ and $d$ can do the same.

We split basic quantum gates into different types, such as X-like, Z-like or matrix gate. For X-like gate, the elements are only in anti-diagonal:
\begin{equation}
    \begin{pmatrix}
        0 & a \\
        b & 0
    \end{pmatrix}.
\end{equation}
When X-like gate is act on first qubit, we need to swap $(b*a_r, b*a_i)$ and $(a*b_r, a*b_i)$ in a single AVX register. And if this gate is act on other qubit, we need to swap $b*\text{reg\_a}$ and $a*\text{reg\_b}$. For Z-like gate, the elements are only in diagonal:
\begin{equation}
    \begin{pmatrix}
        a & 0 \\
        0 & b
    \end{pmatrix}.
\end{equation}
In this situation, we just need to change $(a_r, a_i)$ to $(a*a_r, a*a_i)$ and $(b_r, b_i)$ to $(b*b_r, b*b_i)$ if the gate act on first qubit, or just multiply $a$ to reg a and multiply $b$ to reg b if acting on other qubit.

\subsection{Density Matrix Simulator}
Different from the state vector simulator, density matrix simulator is more suitable for mixed states or open quantum system simulation, but much more memory consuming. The density operator of a quantum system is described as:
\begin{equation}
    \rho=\sum_ip_i\ket{\psi_i}\bra{\psi_i},
\end{equation}
where $p_i$ is the probability of the system in pure state $\ket{\psi_i}$. The evolution of a quantum gate $U$ on density operator is:
\begin{equation}
    \rho'=U\rho U^\dagger.
\end{equation}
The expectation of an observable $H$ under $\rho$ is:
\begin{equation}
    \left<H\right> = \sum_ip_i\bra{\psi_i}H\ket{\psi_i}=\tr(\rho H).
\end{equation}

The usage, developing strategy and optimization process are very similar with state vector simulator, please refer to section ~\ref{sec_state_vec}.

\subsection{Channel Based Noise Simulator}
\label{sec:noise_simulation}

View demo code of this section: \demonotebook{03}{chapt03} \ \demonotebookgithub{03}{chapt03}

\subsubsection{Quantum Channel}
Due to quantum decoherence and coupling with the environment, errors often occur in current quantum computers during the calculation process. They occur in stages such as state preparation, quantum gate operation, and measurement. This is often called quantum noise. Quantum noise can be characterized by quantum channels. In quantum information theory, quantum channels refer to completely positive trace-preserving (CPTP) maps in the operator space, which can also be regarded as a quantum operation. All quantum channels can be represented by Kraus operators

\begin{equation}
    \Psi(\rho) = \sum_i K_i \rho K_i^\dagger,
\end{equation}

where $\Psi$ is a quantum channel, $\rho$ is a density matrix, and $\{K_i\}$ are Kraus operators of $\Psi$. The Kraus operators satisfy the completeness condition

\begin{equation}
    \sum_i K_i^\dagger K_i = I.
\end{equation}

In density matrix simulator \code{"mqmatrix"}, we support the simulation of quantum channels based on the above mathematical form. In addition to the density matrix method, in the state vector simulator \code{"mqvector"} in \MindQuantum, we also support the Monte Carlo method to simulate quantum channels. The noise gate will affect the qubits with a certain probability. By sampling the circuit multiple times, we can get the noise-containing simulation results of quantum circuits. This process is much closer to how a real quantum computer works.

The following quantum channels are implemented in \MindQuantum.

\PauliChannel: Pauli channel can be seen as Pauli operators randomly apply to a quantum state $\rho$ with respect to a probability distribution.
\begin{equation}
    \Psi(\rho) = (1-p_x-p_y-p_z) \rho + p_x X \rho X + p_y Y \rho Y + p_z Z \rho Z.
\end{equation}
It becomes \BitFlipChannel\ if only $p_x$ is nonzero. Similarly, \PhaseFlipChannel\ has only one nonzero value $p_z$ and \BitPhaseFlipChannel\ is nonzero $p_y$. A special case is, all Pauli operators have same probabilities, which is called \DepolarizingChannel. Depolarizing channel is widely used in the description of quantum noise. Here is the formula in 1-qubit case:
\begin{equation}
    \Psi(\rho) =  (1-p) \rho + \frac{p}{4}(I\rho I+X\rho X+Y\rho Y+Z\rho Z).
\end{equation}
Depolarizing channel in \MindQuantum\ also supports multiple qubits case.
\begin{lstlisting}
from mindquantum.core.gates import DepolarizingChannel
from mindquantum.core.circuit import Circuit
circ = Circuit()
circ += DepolarizingChannel(0.02).on(0)
circ += DepolarizingChannel(0.01, 2).on([0, 1])
\end{lstlisting}

Damping Channel: Common damping channels include \AmplitudeDampingChannel\ and \PhaseDampingChannel. The amplitude damping channel can describe the dissipation of system energy, while the phase damping channel describes the loss of quantum information without exchanging energy with environment.
Amplitude damping channel applies noise as:
\begin{gather*}
    \epsilon(\rho) = E_0 \rho E_0^\dagger + E_1 \rho E_1^\dagger
    \\
    \text{where}\ {E_0}=\begin{bmatrix}1 & 0               \\
        0 & \sqrt{1-\gamma}\end{bmatrix},
    \ {E_1}=\begin{bmatrix}0 & \sqrt{\gamma} \\
        0 & 0\end{bmatrix}.
\end{gather*}
Phase damping channel applies noise as:
\begin{gather*}
    \epsilon(\rho) = E_0 \rho E_0^\dagger + E_1 \rho E_1^\dagger
    \\
    \text{where}\ {E_0}=\begin{bmatrix}1 & 0               \\
        0 & \sqrt{1-\gamma}\end{bmatrix},
    \ {E_1}=\begin{bmatrix}0 & 0             \\
        0 & \sqrt{\gamma}\end{bmatrix}.
\end{gather*}

\KrausChannel: The custom single-bit quantum channel in \MindQuantum. It can be constructed by passing in the Kraus operators.
\begin{lstlisting}
from mindquantum.core.gates import KrausChannel
from mindquantum.core.circuit import Circuit
from cmath import sqrt
gamma = 0.5
kmat0 = [[1, 0], [0, sqrt(1 - gamma)]]
kmat1 = [[0, sqrt(gamma)], [0, 0]]
amplitude_damping = KrausChannel('damping', [kmat0, kmat1])
circ = Circuit()
circ += amplitude_damping.on(0)
\end{lstlisting}

\subsubsection{Channel Adder}
A real quantum chip is comprised of multiple qubits. Due to inherent variabilities in the manufacturing process, each qubit exhibits distinct sources of error. Furthermore, the errors associated with the same qubit vary when different quantum gates are applied. To faithfully replicate the errors inherent in a quantum chip, a precise error model must be thoughtfully engineered. Within \MindQuantum, we have developed a suite of tools, called Channel Adder, which facilitates the rapid construction of noise models. Through the Channel Adder, we can conveniently introduce distinct quantum channels after various quantum gates, each tailored to specific qubits, and utilize noisy simulators to simulate these noise models.

The \ChannelAdder class consists of three main functions: \code{_accepter()}, \code{_excluder()}, and \code{_handler(BasicGate)}. Their functionalities are as follows:
\begin{itemize}
    \item \code{_accepter()}: Returns a list of functions called the accept rule set, where each accept rule function takes a quantum gate as input. When the function returns True, it indicates that we can add a channel after that quantum gate.
    \item \code{_excluder()}: Returns a list of functions called the reject rule set, where each reject rule function takes a quantum gate as input. When the function returns True, it indicates that we reject adding a channel after that quantum gate.
    \item \code{_handler(BasicGate)}: Takes a quantum gate as input and returns a quantum circuit representing a custom channel added after the input quantum gate.
\end{itemize}

\begin{figure}
    \centering
    \begin{subfigure}{0.2\textwidth}
        \centering
        \includegraphics[width=\textwidth]{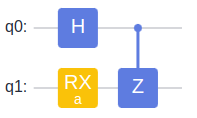}
        \caption{Circuit without noise.}
    \end{subfigure}
    \begin{subfigure}{0.4\textwidth}
        \centering
        \includegraphics[width=0.9\textwidth]{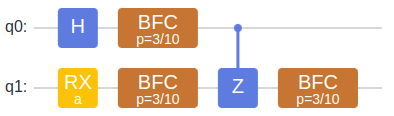}
        \caption{Circuit after a \code{BitFlipAdder}.}
    \end{subfigure}
    \caption{An example of applying a BitFlipAdder to a quantum circuit.}
    \label{fig:bit_flip_adder}
\end{figure}

Take \BitFlipAdder as an example. This adder will add a bit flip channel after every quantum gate.
\begin{lstlisting}
from mindquantum.core.circuit import Circuit, BitFlipAdder

circ = Circuit().h(0).rx('a', 1).z(1, 0)
noise_model = BitFlipAdder(0.3, with_ctrl=False)
noise_circ = noise_model(circ)
print(noise_model)
\end{lstlisting}
The output is:
\begin{lstlisting}
BitFlipAdder<flip_rate=0.3, with_ctrl=False>
\end{lstlisting}
After applying the noise model to a circuit, a bit flip channel will add to each quantum gate, see Fig.~\ref{fig:bit_flip_adder}.

\begin{figure}
    \centering
    \begin{subfigure}{0.32\textwidth}
        \centering
        \includegraphics[width=\textwidth]{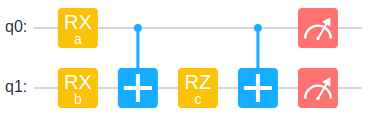}
        \caption{Circuit without noise.}
    \end{subfigure}
    \begin{subfigure}{0.5\textwidth}
        \centering
        \includegraphics[width=\textwidth]{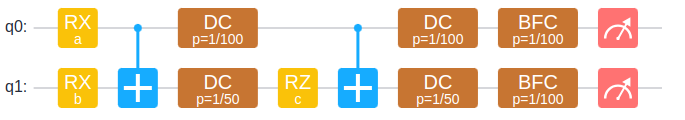}
        \caption{Circuit after a complex Channel Adder.}
    \end{subfigure}
    \caption{An example of applying a complex Channel Adder to a quantum circuit.}
    \label{fig:complex_adder}
\end{figure}

Here we will have a more complex example, so that we can see the power of Channel Adder. Assuming negligible noise in single-qubit gate operations on different qubits of the quantum chip, we consider that the two-qubit gates exhibit distinct depolarizing channel on different qubits. Additionally, the measurement of the circuit is subject to a bit-flip error with a flip probability of 0.01. We assume the depolarizing channels on different qubits to be:
\begin{lstlisting}
from mindquantum.core import *
dc0 = DepolarizingChannel(0.01)
dc1 = DepolarizingChannel(0.02)
dc2 = DepolarizingChannel(0.03)
\end{lstlisting}
Then we are going to build the whole noise model as:
\begin{lstlisting}
adder1 = MixerAdder([
    NoiseExcluder(),
    ReverseAdder(MeasureAccepter()),
    QubitNumberConstrain(2),
    NoiseChannelAdder(dc0, focus_on=0),
])
adder2 = MixerAdder([
    NoiseExcluder(),
    ReverseAdder(MeasureAccepter()),
    QubitNumberConstrain(2),
    NoiseChannelAdder(dc1, focus_on=1),
])
adder3 = MixerAdder([
    NoiseExcluder(),
    ReverseAdder(MeasureAccepter()),
    QubitNumberConstrain(2),
    NoiseChannelAdder(dc2, focus_on=2),
])
adder4 = MixerAdder([
    NoiseExcluder(),
    MeasureAccepter(),
    BitFlipAdder(0.01)
], add_after=False)

noise_model = SequentialAdder([
    adder1,
    adder2,
    adder3,
    adder4
])
print(noise_model)
\end{lstlisting}
The output is:
\begin{lstlisting}
SequentialAdder<
  MixerAdder<
    NoiseExcluder<>
    ReverseAdder<
      MeasureAccepter<>
    >
    QubitNumberConstrain<n_qubits=2, with_ctrl=True>
    NoiseChannelAdder<channel=DC(p=1/100), with_ctrl=True>
  >
  MixerAdder<
    NoiseExcluder<>
    ReverseAdder<
      MeasureAccepter<>
    >
    QubitNumberConstrain<n_qubits=2, with_ctrl=True>
    NoiseChannelAdder<channel=DC(p=1/50), with_ctrl=True>
  >
  MixerAdder<
    NoiseExcluder<>
    ReverseAdder<
      MeasureAccepter<>
    >
    QubitNumberConstrain<n_qubits=2, with_ctrl=True>
    NoiseChannelAdder<channel=DC(p=0.03), with_ctrl=True>
  >
  MixerAdder<
    NoiseExcluder<>
    MeasureAccepter<>
    BitFlipAdder<flip_rate=0.01, with_ctrl=True>
  >
>
\end{lstlisting}
A \MixerAdder is a set of adders, that all the \code{_accepter()} and\code{_excluder()} will be met. All the Channel Adder in \SequentialAdder will be executed one by one. From Fig.~\ref{fig:complex_adder}, we can see that this noise model satisfies our requirement.

After we build a noise model, we are easy to simulate a quantum circuit with this noise model in \MindQuantum.
\begin{lstlisting}
circ = Circuit().rx('a', 0).rx('b', 1).x(1, 0).rz('c', 1).x(1, 0).measure_all()
noise_sim = Simulator(NoiseBackend('mqvector', 2, noise_model))
noise_sim.sampling(circ, pr={'a': 1, 'b': 2, 'c':3}, shots=10000)
\end{lstlisting}
Here we construct a noise simulator using \code{"mqvector"}-based \NoiseBackend and the noise model that we built previously.

\section{ToolBox of MindSpore Quantum}
\label{sec:toolbox}

View demo code of this chapter: \demonotebook{04}{chapt04} \ \demonotebookgithub{04}{chapt04}

\subsection{Gradient of Variational Quantum Algorithm}
Owing to the principles of quantum superposition, quantum entanglement, and quantum parallelism, quantum algorithms have the capacity to efficiently address certain computational problems, such as Shor's algorithm \cite{shor1994algorithms}, Grover's algorithm \cite{grover1996fast,long2001grover} and Harrow-Hassidim-Lloyd algorithm \cite{harrow2009quantum}, achieving exponential or polynomial acceleration when compared to classical algorithms. Nevertheless, the current stage of quantum devices is constrained by limited quantum qubits and significant noise, which is often referred to as Noisy Intermediate-Scale Quantum (NISQ) devices \cite{preskill2018quantum}. It is worth noting that the quantum algorithms mentioned above do not exhibit quantum advantage under NISQ era.

Variational Quantum Algorithms (VQA) \cite{cerezo2021variational,yuan2019theory,xu2021variational} are considered the most promising candidates for achieving quantum advantage in the NISQ era. VQA leverages optimization algorithms from classical machine learning to fine-tune parameterized quantum circuits (PQC). As a result, VQA is a hybrid quantum-classical algorithm. VQA has found extensive applications in various domains, including quantum many body simulations \cite{Peruzzo2014Peruzzo2014,kandala2017hardware,kokail2019self,lyu2023variational,Lyu2023symmetryenhanced,cao2022progress}, quantum approximate optimization algorithm \cite{farhi2014quantum,patti2022variational,chandarana2023digitized}, quantum machine learning \cite{benedetti2019parameterized,wei2022quantum,lloyd2018quantum,biamonte2017quantum,abbas2021power} and entanglement purification \cite{zhang2023variational}. In this section, we will provide a brief overview of the VQA algorithm's workflow and how it can be efficiently implemented in \MindQuantum. More examples of VQA algorithm applications will be present in chapter \ref{sec:applications}.

\begin{figure}[h]
  \begin{center}
    \includegraphics[width=0.9\linewidth]{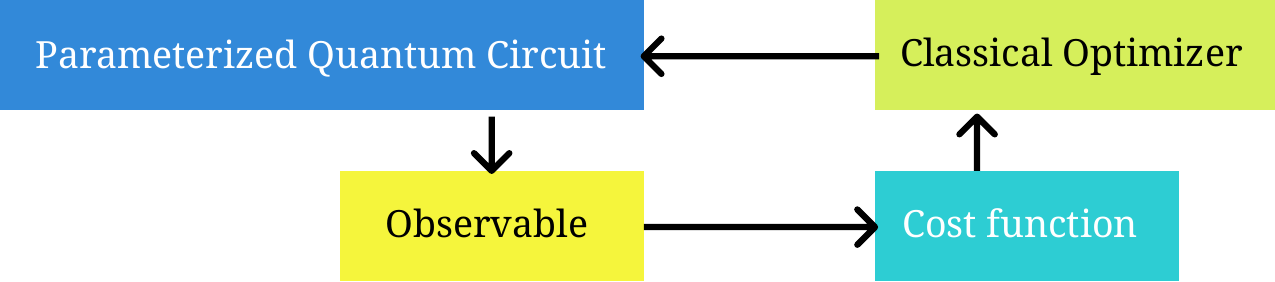}
  \end{center}
  \caption{The work flow of VQA.}
  \label{fig:vqa_work_flow}
\end{figure}

Fig.~\ref{fig:vqa_work_flow} illustrates the primary workflow of the VQA. The central component of VQA is the PQC, which comprises both non-parameterized gates such as Hadamard and Pauli gates, and parameterized gates like the rotation gates around the $x$, $y$, or $z$ axes of the Bloch sphere. Depending on the specific task, the PQC can be divided into two parts: the encoder circuit, responsible for encoding classical data into a quantum state denoted as $U_e(\alpha)$, and the ansatz circuit, serving as the trainable component of VQA, denoted as $U_a(\theta)$.

In VQA, we also require an observable operator denoted as $H$. The expectation value of this observable is calculated as follows:

\begin{equation}
  E(\alpha, \theta) = \bra{0}U_e^\dagger(\alpha)U_a^\dagger(\theta)HU_a(\theta)U_e(\alpha)\ket{0}.
\end{equation}
In quantum many-body simulation and quantum approximate optimization algorithm, we need to convert the problem into an observable and optimize the parameter $\theta$ to find the ground state of this observable (the encoder circuit $U_e$ is always identity in these problems). The task is described as:

\begin{equation}
  \min_\theta E(\theta).
\end{equation}
On the other hand, we need to define a cost function when we treat VQA as a part of machine learning model:

\begin{equation}
  \min_\theta L(E(\alpha, \theta), E_\text{target}),
\end{equation}
where $E_\text{target}$ is the target of the training model and $L$ is a cost function like mean absolute error loss or mean squared error loss.

Optimization algorithms in VQA can be categorized into two types based on their reliance on gradients. The first category comprises gradient-based optimization algorithms such as Adam and BFGS. These algorithms are well-suited for implementation in quantum simulators. The second category consists of non-gradient optimization algorithms such as SPSA and Nelder-Mead. These algorithms find more practical utility in real-world quantum experiments. In \MindQuantum, we are always focusing on gradient based optimizer, and we will show more detail about how we speed up the calculation of gradient of PQC in section \ref{sec:benchmark}.

\subsubsection{Learning a rotation}

In this subsection, we will delve into the implementation of variational quantum algorithms in \MindQuantum\ by exploring a task of learning a rotation quantum gate.

Suppose we have an initial quantum state $\ket{\psi_0}=\ket{0}$, and we are going to find a single qubit unitary operator $U(\theta)$, which can rotate $\ket{\psi_0}$ to $\ket{\psi_f}$ so that to minimum $E_X(\theta) = \bra{\psi_f}X\ket{\psi_f}$:

\begin{equation}
  \min_\theta(\bra{0}U^\dagger(\theta)XU(\theta)\ket{0}).
\end{equation}

The first step of VQA is to build an ansatz. In this task, we choose arbitrary single qubit rotation gate \Uthree to work as ansatz and \Uthree can also decompose into:

\begin{equation}
  U_3(a, b, c) = RZ(b)RX(-\pi/2)RZ(a)RX(\pi/2)RZ(c).
\end{equation}

\begin{lstlisting}
import numpy as np
from mindquantum.core.circuit import Circuit

ansatz = Circuit().rz('c', 0).rx(np.pi/2, 0)
ansatz.rz('a', 0).rx(-np.pi/2, 0).rz('b', 0)
ansatz.svg()
\end{lstlisting}

\begin{figure}[h]
  \begin{center}
    \includegraphics[width=0.9\linewidth]{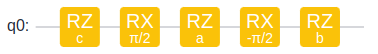}
  \end{center}
  \caption{The ansatz circuit that can be work as an arbitrary single qubit rotation.}
  \label{fig:u3_ansatz}
\end{figure}

The observable is implemented as:

\begin{lstlisting}
from mindquantum.core.operators import QubitOperator
from mindquantum.core.operators import Hamiltonian

obs = QubitOperator('X0')
ham = Hamiltonian(obs)
\end{lstlisting}

In \MindQuantum, the simulator provide a method \getexpectationwithgrad to easily calculate the gradient of $E_X(\theta)$.

\begin{lstlisting}
import numpy as np
from mindquantum.simulator import Simulator

sim = Simulator('mqvector', ansatz.n_qubits)
grad_ops = sim.get_expectation_with_grad(ham, ansatz)

p0 = np.random.uniform(-np.pi, np.pi, len(ansatz.params_name))
f, g = grad_ops(p0)
print(f, g)
\end{lstlisting}
And the output is:
\begin{lstlisting}
[[-0.04602095+0.j]] [[[-1.17340357e-17+0.j -3.47310364e-02+0.j -7.96875999e-01+0.j]]]
\end{lstlisting}

\getexpectationwithgrad receives observable Hamiltonian and ansatz circuit and returns a gradient operator. You can use this operator to calculate gradient in any ansatz parameters. The output of gradient operator is the expectation and its gradient with respect to ansatz parameters. The dimension of expectation value is $[n_\text{batch}, n_\text{ham}]$, where $n_\text{batch}$ is the batch size of encoder data and $n_\text{ham}$ is the number of Hamiltonian. In this task, we do not have encoder and we have only one observable, so $n_\text{batch}=1, n_\text{ham}=1$. The dimension of gradient of ansatz is $[n_\text{batch}, n_\text{ham}, n_\text{ansatz}]$, while $n_\text{ansatz} = 3$ is the number of parameters in ansatz circuit.

The next step is to optimize the PQC with the help of gradient operator.
\begin{lstlisting}
from scipy.optimize import minimize

def fun(p0, grad):
  f, g = grad(p0)
  f = np.real(f)[0, 0]
  g = np.real(g)[0, 0]
  return f, g

res = minimize(fun, p0, method='bfgs', jac=True, args=(grad_ops, ))
print('min E:', res.fun)
print('Best p:', res.x)
\end{lstlisting}
And we get the minimized observable value and corresponded ansatz parameters as:
\begin{lstlisting}
min E: -0.9999999999994561
Best p: [-2.51566965 -4.71238798 -3.14159296]
\end{lstlisting}
So that the rotation that we learned is:
\begin{lstlisting}
ansatz.matrix(pr=dict(zip(ansatz.params_name, res.x)))
\end{lstlisting}
\begin{align*}
  U_a=\frac{1}{\sqrt{2}}\begin{pmatrix}
    e^{-0.313i}  & e^{0.313i} \\
    -e^{-0.313i} & e^{0.313i}
  \end{pmatrix}.
\end{align*}

\subsubsection{Rotate three states}
\label{sec:rotate_three_states}

In previous example, the PQC only contains ansatz circuit, and in this example, we will show how to add encoder circuit into PQC in \MindQuantum. The problem is described as below: find a single qubit rotation gate $U_a(\theta)$, which maximizes the sum of the expectation values of $Z$ for the following three states: $\ket{\psi_0}=\ket{0},\ket{\psi_1}=(\ket{0}+\ket{1})/\sqrt{2}$ and $\ket{\psi_2}=(\ket{0}+i\ket{1})/\sqrt{2}$.

\begin{equation}
  \max_{\theta}\sum_i{\bra{\psi_i}U^\dagger(\theta)ZU(\theta)\ket{\psi_i}}.
\end{equation}

Here, we use an encoder circuit $U_e(a, b) = RX(a)RY(b)$ to prepare state $\ket{\psi_i}=U_e(a, b)\ket{0}$
\begin{lstlisting}
import numpy as np
from mindquantum.core.circuit import Circuit

encoder = Circuit().rx('a', 0).ry('b', 0)
encoder = encoder.as_encoder()

e_data = np.array([[0.0, 0.0],      # psi_0
                   [0.0, np.pi/2],  # psi_1
                   [-np.pi/2, 0.0], # psi_2
                  ])
\end{lstlisting}
In above code, we call \asencoder to convert a quantum circuit to encoder circuit. In this example, we will show how to train the PQC with the help of MindSpore machine learning framework.

\begin{lstlisting}
# Build the arbitrary single qubit rotation ansatz
from mindquantum.framework import MQLayer
import mindspore.nn as nn
import mindspore as ms

ansatz = Circuit().rz('a2', 0).rx(np.pi/2, 0)
ansatz.rz('a0', 0).rx(-np.pi/2, 0).rz('a1', 0)

obs = QubitOperator('Z0')
ham = Hamiltonian(-obs)

sim = Simulator('mqvector', 1)
total_circ = encoder + ansatz
grad_ops = sim.get_expectation_with_grad(ham, total_circ)

e_tensor = ms.Tensor(e_data, ms.float32)

quantum_net = MQLayer(grad_ops)

class Hybrid(nn.Cell):
    def __init__(self, quantum_net):
        super(Hybrid, self).__init__()
        self.quantum_net = quantum_net
    def construct(self, e):
        x = self.quantum_net(e)
        x = x.sum()
        return x

hybrid = Hybrid(quantum_net)
opti = nn.Adam(quantum_net.trainable_params(), learning_rate=0.1)
train_net = nn.TrainOneStepCell(hybrid, opti)
for step in range(100):
    print(-train_net(e_tensor))
\end{lstlisting}
The final sum of expectation that we learned is 1.732. What's more we can easily display these quantum state in \MindQuantum:

\begin{lstlisting}
import matplotlib.pyplot as plt
from mindquantum.io import BlochScene
from mindquantum.core.parameterresolver import ParameterResolver as PR

e_name = encoder.params_name
a_name = ansatz.params_name
a_data = np.array(quantum_net.weight)

init_states = [encoder.get_qs(pr=dict(zip(e_name, i))) for i in e_data]
final_states = []
for i, data in enumerate(e_data):
    pr = PR(dict(zip(e_name, data)))
    pr += PR(dict(zip(a_name, a_data)))
    final_states.append(total_circ.get_qs(pr=pr))

scene = BlochScene()
fig, ax = scene.create_scene()

for i in init_states:
    scene.add_state(ax,
        i,
        linecolor='r',
        with_proj=False)

for i in final_states:
    scene.add_state(ax,
        i,
        linecolor='g',
        with_proj=False)

plt.show()
\end{lstlisting}

\begin{figure}[ht]
  \centering
  \includegraphics[scale=0.48]{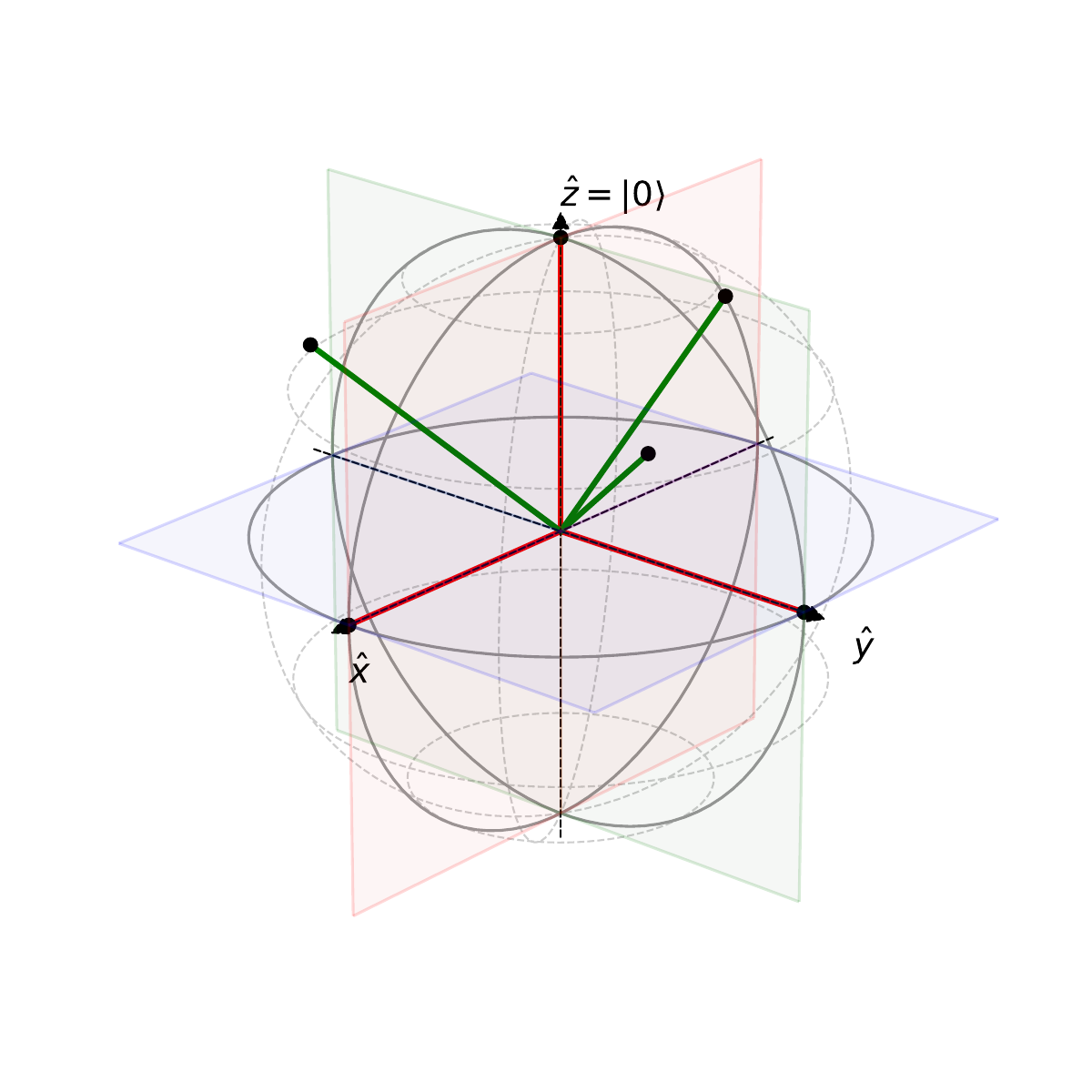}
  \caption{\label{4_1_vqa} Rotation of initial states. The red represent the initial quantum states, while the green represent the final quantum states.}
\end{figure}

\subsubsection{More usage}
In \MindQuantum, except these two basic usages, \getexpectationwithgrad offers a broader range of application scenarios. The definition of equation that we actually calculate gradient is:
\begin{equation}
  E = \left<\varphi\right|U_l^\dagger H U_r \left|\psi\right>.
\end{equation}
You can set $U_l$ to an empty circuit and $H$ to identity \QubitOperator, so that you can get the projection of the new state on $\ket{\varphi}$ and you can also set $\ket{\varphi}=\ket{\psi}$, $U_l = U_r$ and $H$ to a density matrix, such as $H = \ket{0}\bra{0}$, and the result will be
\begin{equation}
  E=\left|\bra{0}U_r\ket{\psi}\right|^2.
\end{equation}
You can also set the observable to a list of Hamiltonian, so that \MindQuantum\ can calculate the expectation and gradient in parallel.

\subsection{Circuit Ansatz Library}
Ansatz is an important term of VQA, which is essentially a guess or an assumption about the form of the quantum state that is suitable for the problem at hand. For example, in variational quantum eigensolver (VQE), the ansatz is chosen to represent the ground state or a low-energy state of a quantum system. The ansatz can be fixed or adaptive, depending on whether the circuit structure is predetermined or dynamically updated during the optimization process. The choice of the ansatz can affect the performance and robustness of the VQA, as well as the resources required for the quantum and classical computations. Therefore, selecting or designing a suitable ansatz is an important part of the success of a VQA.

The \nisq module in \MindQuantum \ contains many commonly used Ansatz, as shown below.

\begin{itemize}
    \item \IQPEncoding: IQP Encoding is a method for encoding classical data into quantum states \cite{Havlíček2019}. It uses rotation gates and controlled gates to construct a quantum circuit, such that the input data can be adjusted by the parameters of the rotation gates, and the output quantum state can reflect the features of the input data. One advantage of IQP Encoding is that it can generate a high-dimensional quantum feature space, which can improve the performance of machine learning.

    \item \HardwareEfficientAnsatz: It is a kind of ansatz that can be easily implemented on quantum chip \cite{Kandala2017,doi:10.1021/acs.jctc.3c00966}. The hardware efficient is constructed by a layer of single qubit rotation gate and a layer of two qubits entanglement gate. The single qubit rotation gate layer is constructed by one or several rotation gates that act on every qubit. The two qubits entanglement gate layer is constructed by CNOT, CZ, Rxx, Ryy, Rzz, etc. acting on entangle mapping.

    \item \StronglyEntangling: It is usually used in quantum machine learning. Its advantage is that it can reach 'wide corners of the Hilbert space' \cite{Schuld_2020}, so it is considered to be very suitable as ansatz of classifier.

    \item \MaxCutAnsatz: \MaxCutAnsatz is a kind of ansatz circuit for QAOA \cite{farhi2014quantum}. It is designed to solve the Max Cut Problem on regular graphs. The Max Cut problem is an NP-complete problem in graph theory, which requires dividing the vertices of a graph into two parts and maximizing the number of edges that are cut. The mathematical representation of \MaxCutAnsatz is as follows:
          $$U(\beta, \gamma) = e^{-i\beta_pH_b}e^{-i\frac{\gamma_p}{2}H_c}
              \cdots e^{-i\beta_0H_b}e^{-i\frac{\gamma_0}{2}H_c}H^{\otimes n}$$,
          where $H_b = \sum_{i\in n}X_{i}, H_c = \sum_{(i,j)\in C}Z_iZ_j$.

    \item \MaxToSATAnsatz: \MaxToSATAnsatz is a kind of ansatz circuit for QAOA \cite{Akshay_2020}. It is designed to solve the Max-2-SAT Problem. Max-2-SAT is a combinatorial optimization problem, where the goal is to find a truth assignment for a given Boolean formula, such that the number of satisfied clauses is maximized. Max-2-SAT is a special case of the Boolean satisfiability problem (SAT), where only clauses with two literals are considered. Max-2-SAT is NP-hard, which means that there is no known polynomial-time algorithm to solve it. It has many applications, such as data clustering, geometric and visualization problems. The mathematical representation of \MaxToSATAnsatz is as follows:
          $$U(\beta, \gamma) = e^{-i\beta_pH_b}e^{-i\frac{\gamma_p}{2}H_c}
              \cdots e^{-i\beta_0H_b}e^{-i\frac{\gamma_0}{2}H_c}H^{\otimes n}$$,
          where $H_b = \sum_{i\in n}X_{i}, H_c = \sum_{l\in m}P(l)$.

    \item \QubitUCCAnsatz: Qubit Unitary Coupled-Cluster (qUCC) ansatz is a variant of unitary coupled-cluster ansatz which uses qubit excitation operators instead of Fermion excitation operators \cite{PhysRevA.102.062612}. The Fock space spanned by qubit excitation operators is equivalent as Fermion operators, therefore the exact wave function can be approximated using qubit excitation operators at the expense of a higher order of Trotterization.

          The greatest advantage of qUCC is that the number of CNOT gates is much smaller than the original version of UCC, even using a 3rd or 4th order Trotterization. Also, the accuracy is greatly improved despite that the number of variational parameters is increased.

    \item \UCCAnsatz: The unitary coupled-cluster ansatz for molecular simulations. The mathematical representation is as follows:
          $$U(\vec{\theta}) = \prod_{j=1}^{N(N\ge1)}{\prod_{i=0}^{N_{j}}{\exp{(\theta_{i}\hat{\tau}_{i})}}}$$,
          where $\hat{\tau}$ are anti-Hermitian operators.
\end{itemize}

\MindQuantum \ also implemented 19 ansatzes for quantum classical hybrid algorithms, all based on this article analyzing the expressibility and entanglement capabilities of various parameterized circuits \cite{Sim_2019}.

\subsubsection{Ansatz for Quantum Chemistry}

In addition, there are some quantum chemistry-related generators in \MindQuantum \ that can generate corresponding VQE ansatz.

\begin{itemize}
    \item \generateuccsd: Generate a UCCSD quantum circuit based on a molecular data generated by \textit{Openfermion} \cite{mcclean2020openfermion}. The basic idea of UCCSD ansatz is to use the unitary coupled cluster (UCC) theory, which expresses the quantum state as a product of a reference state and an exponential form of the coupled cluster operator, that is:
          $$
              |\Psi(\vec{\theta})\rangle = \exp(\hat{T} - \hat{T}^\dagger)|\Psi_0\rangle.
          $$
          Here, $\hat{T}$ is the coupled cluster operator, which contains the single and double excitations, that is:
          $$
              \hat{T} = \sum_{pq}\theta_{pq}a_p^\dagger a_q + \sum_{pqrs}\theta_{pqrs}a_p^\dagger a_q^\dagger a_r a_s.
          $$
          Here, $a_p^\dagger$ and $a_q$ are the creation and annihilation operators of fermions, $\theta_{pq}$ and $\theta_{pqrs}$ are the variational parameters, which can be determined by optimizing the objective function.

    \item \quccsdgenerator: Generate qubit-UCCSD (qUCCSD) ansatz using qubit-excitation operators.
    \item \uccsdzerosingletgenerator: Generate UCCSD operators using CCD0 ansatz for molecular systems.
    \item \uccsdsingletgenerator: Create a singlet UCCSD generator for a system with n electrons.
\end{itemize}

Users can also design their own Ansatz by inheriting the Ansatz base class and implementing the \code{implement()} method.

\begin{lstlisting}
from mindquantum.core.gates import RX, RY
from mindquantum.algorithm.nisq import Ansatz

class MyAnsatz(Ansatz):
    def __init__(self, n_qubits, depth):
        """Initialize my ansatz."""
        Ansatz.__init__(self, 'MyAnsatz', n_qubits, depth)

    def _implement(self, depth):
        """Implement of my ansatz."""
        for i in range(depth):
            for j in range(self.n_qubits):
                self._circuit += RX(f'a_{i}').on(j)
                self._circuit += RY(f'b_{i}').on(j)

ansatz_circuit = MyAnsatz(3, 2).circuit
\end{lstlisting}

\subsection{Quantum Algorithm Subroutine}
In the \library module of \MindQuantum, we implement some common components of quantum algorithms.

\begin{itemize}

    \item \amplitudeencoder: Quantum circuit for amplitude encoding.

    \item \bitphaseflipoperator: Generate a circuit that can flip the sign of any calculation bases.

    \item \generalghzstate: Circuit that prepare a general GHZ state based on zero state. The GHZ state is defined as the equality superposition of three zero states and three one states:

          \begin{equation}
              \left|\text{GHZ}\right> = (\left|000\right> + \left|111\right>)/\sqrt{2}.
          \end{equation}

          Here in this API, we can create a general GHZ state on arbitrary sub qubits of any total qubits.

    \item \generalwstate: General W state. The W state is defined as the equality superposition of bases that only one qubit is in $|1\rangle$
          while others qubits are in $|0\rangle$. For example, a three qubits W state is defined as:

          \begin{equation}
              \left|\rm W\right> = (\left|001\right> + \left|010\right> + \left|100\right>)/\sqrt{3}.
          \end{equation}

          Here in this API, we can define a W state on any sub Hilbert space of any total number qubits.

    \item \qft: Quantum Fourier Transform (QFT). The function of the quantum Fourier transform is similar to that of the classical Fourier transform.
\end{itemize}

Taking \amplitudeencoder as an example, you can use it when building a quantum algorithm like this:
\begin{lstlisting}
from mindquantum.algorithm.library import amplitude_encoder
from mindquantum.simulator import Simulator
sim = Simulator('mqvector', 8)
encoder, parameterResolver = amplitude_encoder([0.5, -0.5, 0.5, 0.5], 8)
sim.apply_circuit(encoder, parameterResolver)
\end{lstlisting}
Their return value is just a circuit, so you can add it into your algorithm wherever needed.

\section{Applications}
\label{sec:applications}
\subsection{Quantum Neural Network}




View demo code of this section: \democode{05}{5.1_Quantum_Neural_Network} \ \democodegithub{05}{5.1_Quantum_Neural_Network}
\subsubsection{Background}
In this subsection, we will go through how to create a quantum neural network to address the issue of iris classification in supervised learning, which is a challenge in classical machine learning.

The iris dataset is a popular dataset in classical machine learning. This dataset comprises 150 samples (split into three subgenus: setosa, versicolor, and virginica, with 50 samples each), and each sample has four features: sepal length, sepal width, petal length, and petal width.

We select the first 100 samples (setosa and versicolor), and 80 samples will be randomly chosen from them as the training set. After that, we built a quantum neural network to train the quantum classifier (ansatz). The remaining 20 samples are then sent to a classification test after learning, with the goal of achieving the greatest prediction accuracy possible.

First, we divide 100 samples into 80 training samples and 20 testing samples. Second, we calculate the parameters required to build an encoder based on the classical data of the training samples. Third, we create an encoder to encode the classical data of the training samples to the quantum state. Fourth, we design an ansatz and train its parameters with the proposed quantum neural network layer and MindSpore operators, which  will enable us to obtain the final classifier. Finally, we classify the remaining 20 testing samples and acquire the accuracy of the predictions.

\subsubsection{Data initialization}
In the first place, we should import the iris dataset.

\begin{lstlisting}
from sklearn import datasets

iris_dataset = datasets.load_iris()
\end{lstlisting}

Since we just need to select the first 100 samples, then execute the following command.

\begin{lstlisting}
import numpy as np

x = iris_dataset.data[:100, :].astype(np.float32)
x_feature_names = iris_dataset.feature_names
y = iris_dataset.target[:100].astype(int)
y_target_names = iris_dataset.target_names[:2]
\end{lstlisting}

In order to have a more intuitive understanding of the dataset consisting of these 100 samples, we could draw a scatterplot of the composition between the different features of all the samples by executing the following command.

\begin{lstlisting}
import matplotlib.pyplot as plt

feature_name = {0: 'sepal length', 1: 'sepal width', 2: 'petal length', 3: 'petal width'}
axes = plt.figure(figsize=(23, 23)).subplots(4, 4)

colormap = {0: 'r', 1: 'g'}
cvalue = [colormap[i] for i in y]

for i in range(4):
    for j in range(4):
        if i != j:
            ax = axes[i][j]
            ax.scatter(x[:, i], x[:, j], c=cvalue)
            ax.set_xlabel(feature_name[i], fontsize=22)
            ax.set_ylabel(feature_name[j], fontsize=22)

plt.show()
\end{lstlisting}



\subsubsection{Encoder and Ansatz}
Following that, we compute the parameters required to build the encoder.  Activate the upcoming command.

\begin{lstlisting}
alpha = x[:, :3] * x[:, 1:]
x = np.append(x, alpha, axis=1)
\end{lstlisting}

Constructing features is a common technique in data preprocessing. Therefore, it can be found that each sample has 7 features at this time, the first four feature values are the original feature values, and the last three feature values are the ones calculated by the above preprocessing. The specific calculation formula is as follows:

\begin{eqnarray}\label{5.1dataprocessing}
    X_{i+4}^{j} = X_{i}^{j} * X_{i+1}^{j},
\end{eqnarray}
where $i = 0, 1, 2$ and $j = 1, 2, \cdots, 100$.

At last, we separate the current data set into a training set and a testing set by running the command below.

\begin{lstlisting}
from sklearn.model_selection import train_test_split

x_train, x_test, y_train, y_test = train_test_split(x, y, test_size=0.2, random_state=0, shuffle=True)
\end{lstlisting}

At this point there are 80 samples in the training set and 20 samples in the testing set, each with 7 features.

Next, we need to build an encoder in MindSpore Quantum to encode classical data into quantum states.

Here, the encoding method we employ is IQP encoding (Instantaneous Quantum Polynomial encoding). Generally speaking, the method of building the encoder is not set in stone, instead, many encoding techniques can be used depending on the demands of the issue at hand, and the encoder may occasionally be modified to improve performance.

The values of the parameters $\alpha_0$, $\alpha_1$, $\cdots$,$\alpha_6$ in the encoder are substituted with the 7 feature values obtained in the data processing of Eq.~\eqref{5.1dataprocessing}.

\begin{figure}[H]
    \centering
    \includegraphics[width=0.49\textwidth]{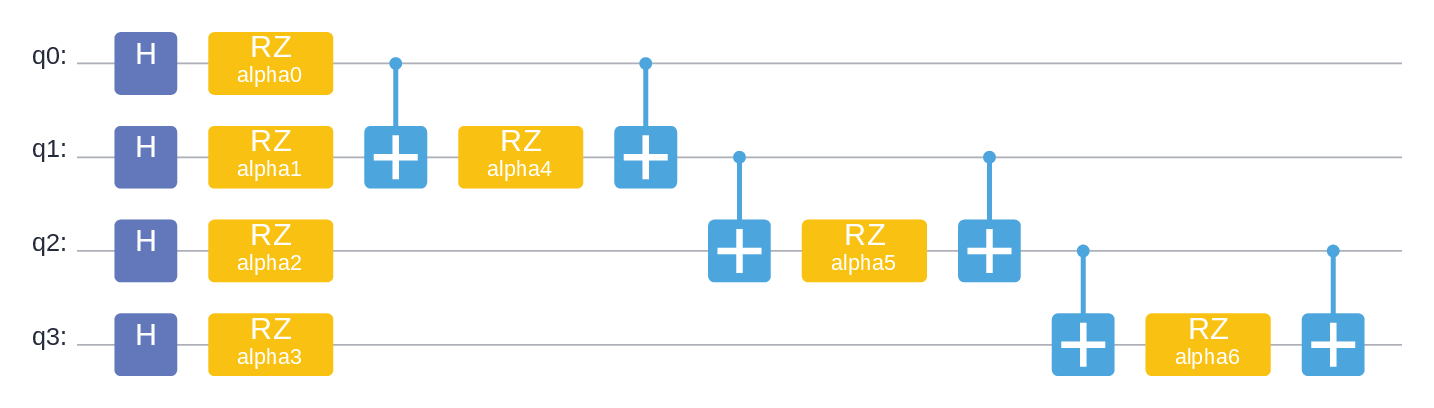}
    \caption{IQP Encoder}
    \label{5.1encoder-circuit}
\end{figure}

Now, we can build the quantum circuit of encoder in \MindQuantum\ (see Fig.~\ref{5.1encoder-circuit}) through the following code.

\begin{lstlisting}
from mindquantum import *

encoder = Circuit()

encoder += UN(H, 4)
for i in range(4):
    encoder += RZ(f'alpha{i}').on(i)
for j in range(3):
    encoder += X.on(j+1, j)
    encoder += RZ(f'alpha{j+4}').on(j+1)
    encoder += X.on(j+1, j)

encoder = encoder.no_grad()
encoder.svg()
\end{lstlisting}

As can be seen from the encoder circuit, the quantum circuit consists of 17 quantum gates, of which 7 are
parameterized quantum gates and the parameters are $\alpha_0$, $\alpha_1$, $\cdots$,$\alpha_6$, respectively. The number of qubits regulated by the encoder circuit is 4.

Furthermore, we need to build an ansatz in MindSpore Quantum.

Like the encoder, the building strategy used by ansatz is flexible, allowing us to examine the effectiveness of various strategies.

\begin{figure}[H]
    \centering
    \includegraphics[width=0.49\textwidth]{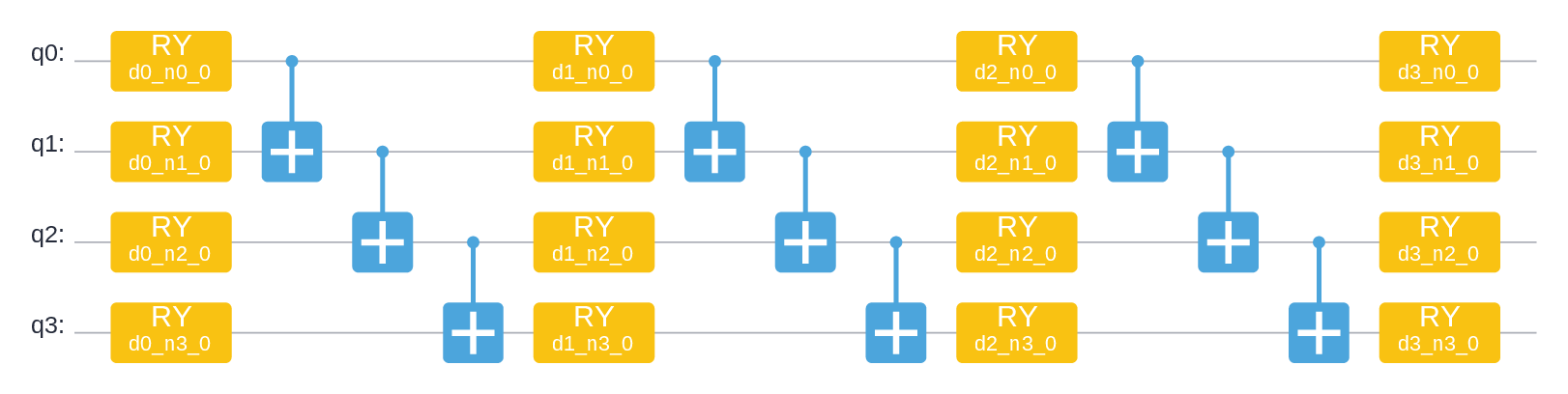}
    \caption{Hardware Efficient Ansatz}
    \label{5.1ansatz-circuit}
\end{figure}

In this case, we apply \HardwareEfficientAnsatz. The encoding scheme is depicted in the following quantum circuit (see Fig.~\ref{5.1ansatz-circuit}) through the following code.

\begin{lstlisting}
ansatz = HardwareEfficientAnsatz(4, single_rot_gate_seq=[RY], entangle_gate=X, depth=3).circuit
ansatz.svg()
\end{lstlisting}

As can be seen from the ansatz circuit, the quantum circuit consists of 25 quantum gates, of which 16 are parameterized quantum gates and the parameters are $d0\_n0\_0$, $d0\_n1\_0$, $d0\_n2\_0$, $\cdots$,  respectively. The number of qubits regulated by the ansatz circuit is 4.

\begin{figure}[H]
    \centering
    \includegraphics[width=0.49\textwidth]{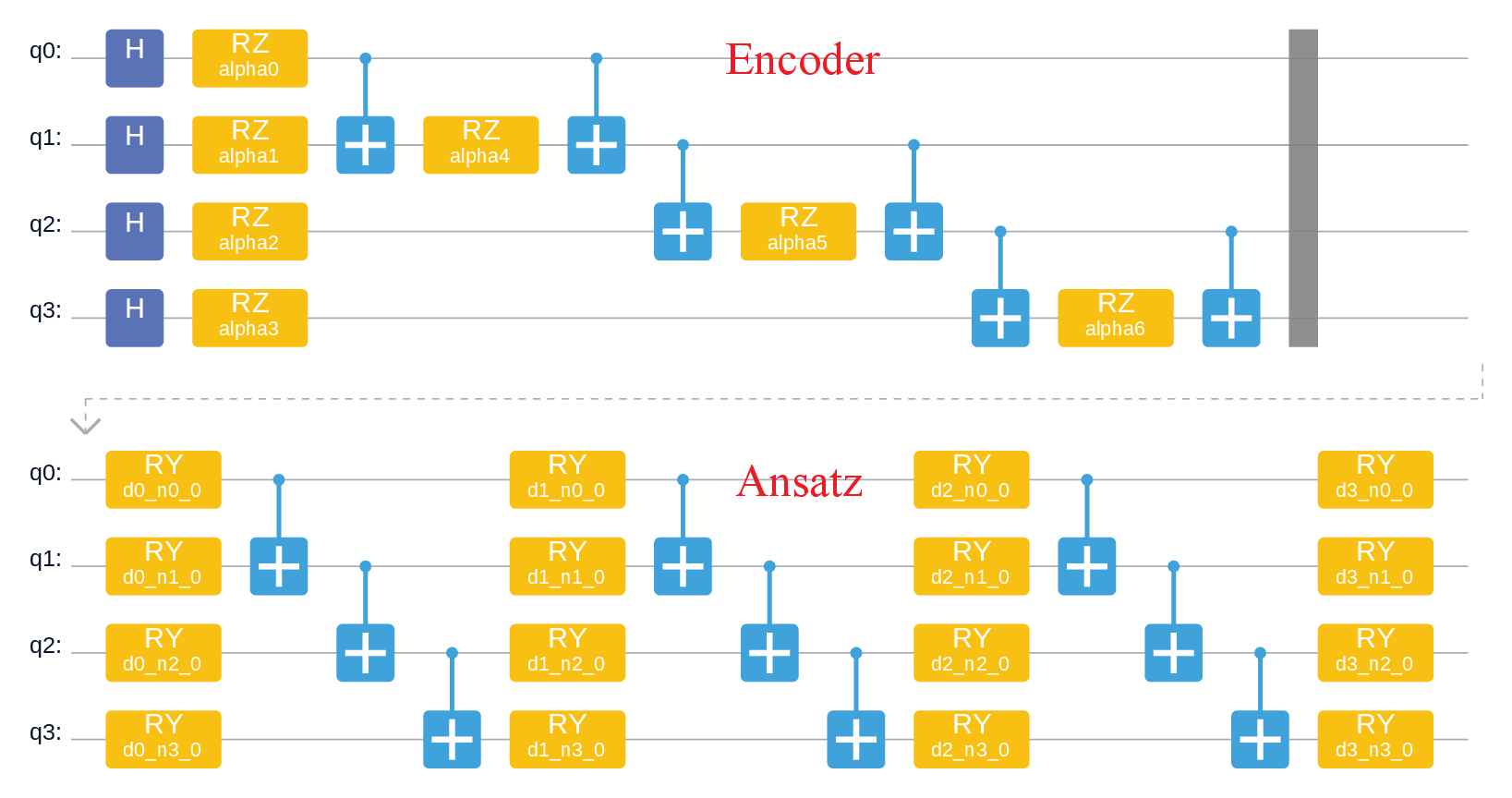}
    \caption{The whole quantum circuit}
    \label{5.1total-circuit}
\end{figure}

In such case, encoder and ansatz might create a whole quantum circuit (see Fig.~\ref{5.1total-circuit}) through the following code.

\begin{lstlisting}
circuit = encoder.as_encoder() + ansatz.as_ansatz()
circuit.svg()
\end{lstlisting}

\subsubsection{Hamiltonian}
We execute Pauli $Z$ operator measurement on the second and third qubits, respectively, to create the corresponding Hamiltonian. Execute the following code.

\begin{lstlisting}
hams = [Hamiltonian(QubitOperator(f'Z{i}')) for i in [2, 3]]
for h in hams:
    print(h)
\end{lstlisting}

\begin{lstlisting}
1 [Z2]
1 [Z3]
\end{lstlisting}

It can be seen from the above print that there are two Hamiltonians constructed at this time, which are to perform the Pauli $Z$ operator on the second and third qubits, respectively, and set the coefficients to 1. We will obtain two Hamiltonian measurement values by the Pauli $Z$ operator measurement.

If the first measurement value is larger, the sample will be classified into the class labeled ``0''.  Similarly, if the second measurement value is larger, this sample will be classified into the class with the label ``1''. By the training of the quantum neural network, it is expected that the first measurement value of the sample labeled ``0''  in the training sample is larger, and the second measurement value of the sample labeled ``1'' is larger. At last, this trained model is applied to predict the classification of new samples.

\subsubsection{Training}
We construct a quantum neural network by the following command.

\begin{lstlisting}
import mindspore as ms
from mindquantum.framework import MQLayer
from mindquantum.simulator import Simulator

ms.set_context(mode=ms.PYNATIVE_MODE, device_target="CPU")
ms.set_seed(1)
sim = Simulator('mqvector', circuit.n_qubits)
grad_ops = sim.get_expectation_with_grad(hams,
                                         circuit,
                                         parallel_worker=5)
QuantumNet = MQLayer(grad_ops)
QuantumNet
\end{lstlisting}

\begin{lstlisting}
MQLayer<
  (evolution): MQOps<4 qubits mqvector VQA Operator>
  >
\end{lstlisting}

As can be seen from the above print, we have successfully built a quantum machine learning layer, which can seamlessly form a larger machine learning network with other operators in MindSpore.

Next, we need to define the loss function and set the parameters to be optimized. After that, we will combine the built quantum machine learning layer and MindSpore operators to form a larger machine learning network. Finally, we will train the model. 

\begin{lstlisting}
from mindspore import LossMonitor, Model
from mindspore.nn import SoftmaxCrossEntropyWithLogits, Adam, Accuracy
from mindspore.dataset import NumpySlicesDataset
from mindspore.train import Callback

loss = SoftmaxCrossEntropyWithLogits(sparse=True, reduction='mean')
opti = Adam(QuantumNet.trainable_params(), learning_rate=0.1)

model = Model(QuantumNet, loss, opti, metrics={'Acc': Accuracy()})

train_loader = NumpySlicesDataset({'features': x_train, 'labels': y_train}, shuffle=False).batch(5)
test_loader = NumpySlicesDataset({'features': x_test, 'labels': y_test}).batch(5)

class StepAcc(ms.Callback):
    def __init__(self, model, test_loader):
        self.model = model
        self.test_loader = test_loader
        self.acc = []

    def step_end(self, run_context):
        self.acc.append(self.model.eval(self.test_loader, dataset_sink_mode=False)['Acc'])

monitor = LossMonitor(16)

acc = StepAcc(model, test_loader)

model.train(20, train_loader, callbacks=[monitor, acc], dataset_sink_mode=False)
\end{lstlisting}

\begin{lstlisting}
epoch: 1 step: 16, loss is 0.6140301823616028
epoch: 2 step: 16, loss is 0.48262983560562134
epoch: 3 step: 16, loss is 0.43457236886024475
epoch: 4 step: 16, loss is 0.4101267457008362
epoch: 5 step: 16, loss is 0.4027639925479889
epoch: 6 step: 16, loss is 0.39859312772750854
epoch: 7 step: 16, loss is 0.39496558904647827
epoch: 8 step: 16, loss is 0.3970319926738739
epoch: 9 step: 16, loss is 0.3954522907733917
epoch: 10 step: 16, loss is 0.39520972967147827
epoch: 11 step: 16, loss is 0.3955090641975403
epoch: 12 step: 16, loss is 0.3953099250793457
epoch: 13 step: 16, loss is 0.39525243639945984
epoch: 14 step: 16, loss is 0.3952508568763733
epoch: 15 step: 16, loss is 0.39521533250808716
epoch: 16 step: 16, loss is 0.39519912004470825
epoch: 17 step: 16, loss is 0.39518338441848755
epoch: 18 step: 16, loss is 0.395169198513031
epoch: 19 step: 16, loss is 0.39515653252601624
epoch: 20 step: 16, loss is 0.3951443135738373
\end{lstlisting}

The loss value keeps dropping during training and tends to stable about 20 iterations before convergent at roughly 0.395.

Based on the above convergence situation, we may present the accuracy of the model's predictions during the training process. 

\begin{lstlisting}
plt.plot(acc.acc)
plt.title('Statistics of accuracy', fontsize=20)
plt.xlabel('Steps', fontsize=20)
plt.ylabel('Accuracy', fontsize=20)

plt.show()
\end{lstlisting}

\begin{figure}[H]
    \centering
    \includegraphics[width=0.4\textwidth]{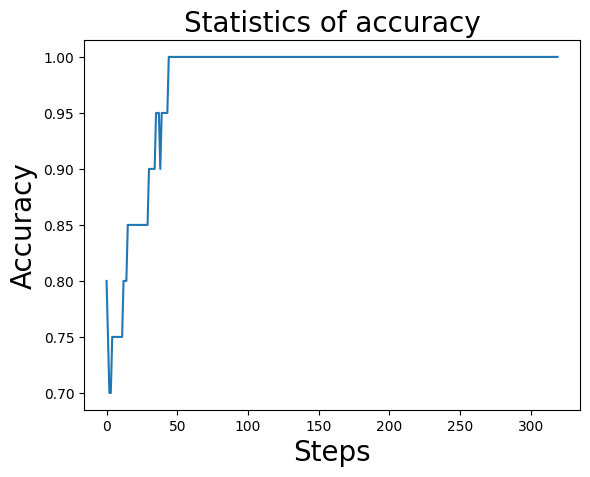}
    \caption{Statistics of accuracy}
    \label{5.1Statistics_of_accuracy}
\end{figure}

As can be seen from the Fig.~\ref{5.1Statistics_of_accuracy}, the prediction accuracy has converged to 1.00 after about 50 steps.

\subsubsection{Prediction}
Finally, we apply the trained model on the testing set.
\begin{lstlisting}
from mindspore import ops

predict = np.argmax(ops.Softmax()(model.predict(ms.Tensor(x_test))), axis=1)
correct = model.eval(test_loader, dataset_sink_mode=False)

print("Predicted classification result:", predict)
print("Actual classification result:", y_test)

print(correct)
\end{lstlisting}

\begin{lstlisting}
Predicted classification result: [0 1 0 1 1 1 0 1 1 1 1 1 1 0 0 0 0 0 0 0]
Actual classification result: [0 1 0 1 1 1 0 1 1 1 1 1 1 0 0 0 0 0 0 0]
{'Acc': 1.0}
\end{lstlisting}

As seen in the print above, the model's prediction accuracy has reached $100\%$, meaning that the anticipated classification results are perfectly consistent with the actual classification results.

\subsection{Quantum Approximate Optimization Algorithm}


View demo code of this section: \democode{05}{5.2_Quantum_Approximate_Optimization_Algorithm} \ \democodegithub{05}{5.2_Quantum_Approximate_Optimization_Algorithm}

\subsubsection{Background}

In this section, we present an introduction to the Quantum Approximate Optimization Algorithm (QAOA) and demonstrate its implementation using \MindQuantum. Additionally, we explore advanced applications in Section D, where we apply techniques STA to achieve improved convergence ansatz DC-QAOA \cite{PhysRevResearch.4.013141}.

QAOA is a powerful quantum algorithm designed to address combinatorial optimization problems by harnessing the capabilities of quantum computers. It was first proposed by Farhi et al. in 2014 \cite{farhi2014quantum}, and since then, numerous variants have been developed to overcome the limitations of the original algorithm.

Inspired by the trotterized version of the quantum adiabatic algorithm, QAOA is categorized as a variational quantum algorithm. The conventional QAOA comprises two main components: the quantum part, which involves a parameterized circuit ansatz, and the classical part, responsible for optimizing these parameters. To employ the QAOA algorithm effectively, we define the problem as finding the ground state of a problem Hamiltonian. The ansatz is constructed from two fundamental building blocks, namely the mixer layer and the problem layer:

\begin{equation}
    U(\gamma, \beta) = \prod_{l=1}^p e^{-i\beta_l \hat{H}_{m}}e^{-i\gamma_l\hat{H}{{p}}},
\end{equation}
where $\hat{H}_{{m}} = \sum_i \hat{\sigma}^x_i$ represents the mixer Hamiltonian, where $\hat{\sigma}^x$ corresponds to the Pauli $x$ operator. On the other hand, $\hat{H}_{{p}}$ denotes the problem Hamiltonian.

The quantum state begins from the ground state of $\hat{H}_{\text{mixer}}$, which is $|\psi_0\rangle=|+\rangle^{\otimes n}$. Then, the quantum state for the QAOA ansatz is expressed as:
\begin{equation}
    |\psi_p(\gamma, \beta)\rangle = e^{-i\beta_p\hat{H}_m}e^{-i\gamma_p\hat{H}_p}\cdots e^{-i\beta_1\hat{H}_{m}}e^{-i\gamma_1\hat{H}_{p}}|\psi_0\rangle.
\end{equation}
To determine the cost or objective function, we calculate the expectation value of the problem Hamiltonian $\hat{H}_{\text{prob}}$ with respect to the ansatz state $|\psi_p(\gamma, \beta)\rangle$. This involves performing repeated measurements of the final state in the computational basis:
\begin{equation}
    F_p(\gamma, \beta) =\langle\psi_p(\gamma, \beta)|\hat{H}_{p} |\psi_p(\gamma,\beta)\rangle.
\end{equation}
The primary objective of the QAOA is to find the optimal parameters $(\gamma^*, \beta^*)$ that minimize the expectation value $F_p(\gamma, \beta)$. Achieving this goal involves employing a classical optimization algorithm to iteratively update the parameters $\gamma$ and $\beta$:
\begin{equation}
    (\gamma^*, \beta^*) = \text{arg}\min_{\gamma, \beta} F_p(\gamma, \beta).
\end{equation}

By finding these optimal parameters, QAOA enables quantum computers to efficiently tackle complex optimization challenges, offering a promising avenue for various practical applications.

\subsubsection{Implementation}

\begin{figure}[H]
    \centering
    \includegraphics[width=0.49\textwidth]{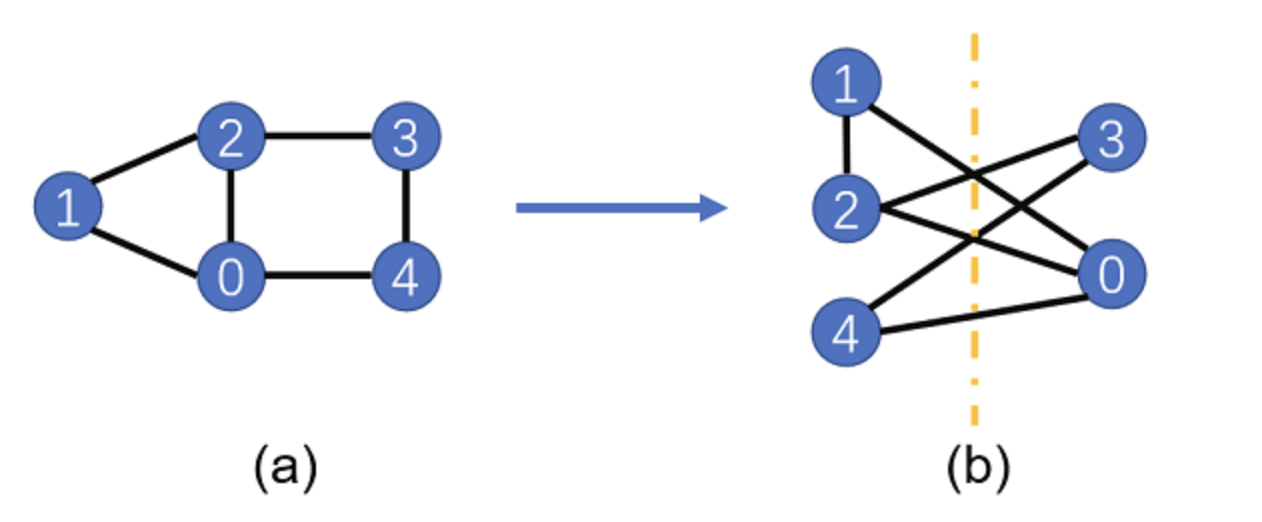}
    \caption{Max-cut graph}
    \label{5.1_QAOA}
\end{figure}

For this tutorial, we use QAOA algorithm to handle the Max-Cut problem,  which is an NP-complete problem in graph theory. As shown in figure \ref{5.1_QAOA}, it needs to divide the vertices of a graph into two parts and make the most edges be cut. But when the number of vertices in the graph increases, it is difficult for us to find an effective classical algorithm to solve the Max-Cut problem, because it is very likely that there is no polynomial time algorithm for this type of NP-complete problem.

Let us first generate a graph structure consisting of five vertices and six edges using \textit{NetworkX} \cite{SciPyProceedings_11}.
\begin{lstlisting}
g = nx.Graph()
nx.add_path(g, [0,1])
nx.add_path(g, [1,2])
nx.add_path(g, [2,3])
nx.add_path(g, [3,4])
nx.add_path(g, [0,4])
nx.add_path(g, [0,2])
nx.draw(g,with_labels=True, font_weight='bold')
\end{lstlisting}

For QAOA algorithm, first we need to convert the Max-Cut problem into a Hamiltonian, it's ground state energy is solution of the problem. The problem Hamiltonian of the Max-Cut problem can be built as
\begin{equation}
    \hat{H} = \sum_{(i,j)\in C}(\hat{Z}_i\hat{Z}_j-1)/2,
\end{equation}
where $C$ is the set of all edges. In the following function, \code{build\_ham}, the corresponding Hamiltonian can be constructed by inputting the graph.
\begin{lstlisting}
def build_ham(g):
    hc = QubitOperator()
    for i in g.edges:
        hc += QubitOperator(f'Z{i[0]} Z{i[1]}')
    return hc
\end{lstlisting}
Next, construct the QAOA ansatz, in which function \code{build\_hc} and \code{build\_hb} construct the quantum circuit according to the problem Hamiltonian and the mixer Hamiltonian. The ansatz can be directly built in function \code{build\_ansatz} by inputting the graph.

\begin{lstlisting}
def build_hc(g,para):
    hc = Circuit()
    for i in g.edges:
        hc += Rzz(para).on(i)
    return hc
def build_hb(g, para):
    hc = Circuit()
    for i in g.nodes:
        hc += RX(para).on(i)
    return hc
def build_ansatz(g, p):
    c = Circuit()
    for i in range(p):
        c += build_hc(g,f'g{i}')
        c += build_hb(g,f'b{i}')
    return c
\end{lstlisting}

In this example, p = 4 is selected, indicating that the four-layer QAOA quantum circuit is used.  And \code{init\_state\_circ} is a quantum circuit for preparing an initial state on a uniformly superposed state.
\begin{lstlisting}
p = 4
ham = Hamiltonian(build_ham(g))
ansatz = build_ansatz(g, p)
init_state_circ = UN(H, g.nodes)
\end{lstlisting}

The goal of QAOA is to find the optimal parameters that make the cost function or the expectation of Hamiltonian $\mathcal{E}=\langle \psi(\gamma, \beta)|\hat{H}|\psi(\gamma, \beta)\rangle$ minimize.
This problem does not require a coding-layer quantum circuit, so we use \MQAnsatzOnlyLayer as a quantum neural network to be trained and the Adam as the optimizer.
\begin{lstlisting}
import mindspore as ms
ms.set_context(mode=ms.PYNATIVE_MODE, device_target="CPU")

total_circuit = init_state_circ + ansatz
sim = Simulator('mqvector', total_circuit.n_qubits)
grad_ops = sim.get_expectation_with_grad(ham, total_circuit)
net = MQAnsatzOnlyLayer(grad_ops)
opti = nn.Adam(net.trainable_params(), learning_rate=0.05)
train_net = nn.TrainOneStepCell(net, opti)

for i in range(600):
    if i%10 == 0:
        print("train step:", i, ", cut:", (len(g.edges)-train_net())/2)
\end{lstlisting}
\begin{lstlisting}
train step: 0 ,   cut: [3.0001478]
train step: 10 ,  cut: [4.1718774]
train step: 20 ,  cut: [4.6871986]
train step: 30 ,  cut: [4.7258005]
train step: 40 ,  cut: [4.804503]
train step: 50 ,  cut: [4.8477592]
train step: 60 ,  cut: [4.8705964]
train step: 70 ,  cut: [4.9060946]
train step: 80 ,  cut: [4.933446]
train step: 90 ,  cut: [4.9356637]
train step: 100 , cut: [4.938308]
train step: 110 , cut: [4.9390197]
train step: 120 , cut: [4.939068]
train step: 130 , cut: [4.9392157]
train step: 140 , cut: [4.939249]
train step: 150 , cut: [4.939247]
train step: 160 , cut: [4.939255]
train step: 170 , cut: [4.939257]
train step: 180 , cut: [4.939257]
train step: 190 , cut: [4.939257]
\end{lstlisting}

According to the above training results, we found that the number of edge cuts corresponding to the ground state energy of the Hamiltonian of this problem approaches 5.

In \MindQuantum, we can directly import the function \MaxCutAnsatz to build ansatz for the max-cut problem.
\begin{lstlisting}
import numpy as np
from mindquantum.algorithm.nisq import MaxCutAnsatz
graph = [(0, 1), (1, 2), (0, 2)]
p = 1 # layer
maxcut = MaxCutAnsatz(graph, p)
maxcut.circuit.svg()
\end{lstlisting}
\begin{figure}[H]
    \centering
    \includegraphics[width=0.49\textwidth]{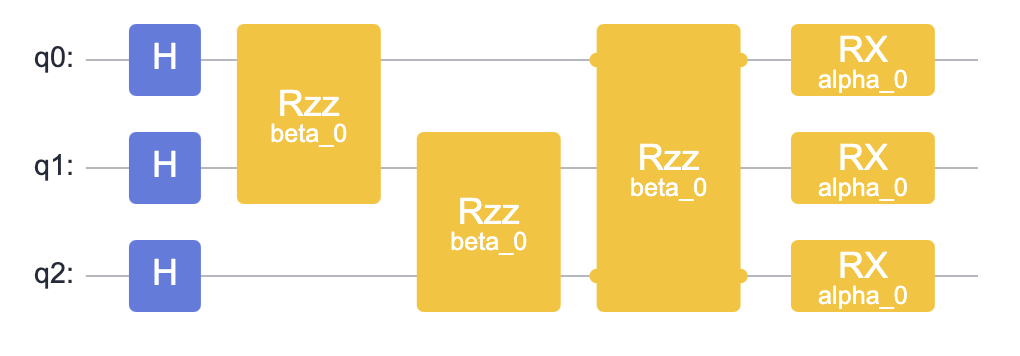}
    \caption{Max-cut problem Ansatz}
    \label{5.1_QAOA_ansatz}
\end{figure}
At the same time, the Hamiltonian can also be obtained
\begin{lstlisting}
print(maxcut.hamiltonian)
\end{lstlisting}
\begin{lstlisting}
3/2 [] +
-1/2 [Z0 Z1] +
-1/2 [Z1 Z2] +
-1/2 [Z0 Z2]
\end{lstlisting}
The cutting scheme of the Max-Cut problem and the number of cutting edges of the cutting scheme can be obtained through the function $get\_partition$ and $get\_cut\_value$. The cutting scheme is a list array consisting of two list arrays, and each list array contains the cut nodes.
\begin{lstlisting}
partitions = maxcut.get_partition(5, np.array([4, 1]))
for i in partitions:
    print(f'partition: left: {i[0]}, right: {i[1]}, cut value: {maxcut.get_cut_value(i)}')
\end{lstlisting}
\begin{lstlisting}
partition: left: [0, 2],   right: [1], cut value: 2
partition: left: [1, 2],   right: [0], cut value: 2
partition: left: [1],   right: [0, 2], cut value: 2
partition: left: [0, 1, 2], right: [], cut value: 0
partition: left: [], right: [0, 1, 2], cut value: 0
\end{lstlisting}

\subsection{Variational Quantum Eigensolver}

View demo code of this section: \democode{05}{5.3_Variational_Quantum_Eigensolver} \ \democodegithub{05}{5.3_Variational_Quantum_Eigensolver}

\subsubsection{Background}

In quantum computing, the variational quantum eigensolver (VQE) is a flagship algorithm for solving eigenproblems on near-term quantum computers, which was originally developed by Peruzzo et al, and its theoretical framework was extended and formalized by McClean et al \cite{Peruzzo2014Peruzzo2014}.
VQE is widely used to compute the energetic properties of molecules and materials, for example, the study of electronic structures which is a critical application for quantum chemistry and condensed matter physics.
Conventional computational chemistry provides efficient methods to approximate such properties for small systems, but it becomes intractable and expensive for very accurate calculations on increasingly large systems. In addition, the interactions between electrons formally require computation that scales exponentially in the size of the system, rendering exact quantum chemistry methods in general intractable with conventional computing.
This bottleneck is the motivation for investigating methods such as the VQE, with the anticipation that these could one day outperform the conventional computing paradigm for these problems.

In its most general description, VQE aims to approximate the ground energy $E_0$ (the smallest eigenvalue within a specified Hilbert space) and the corresponding ground state $|E_0\rangle$ of a given Hamiltonian $\hat{H}$ by the Ritz variational principle.
The Rayleigh quotient for a given complex Hermitian matrix $M$ and nonzero vector $x$ is defined as:
\begin{equation}
    R(M,x)=\frac{x^*Mx}{x^*x}.
\end{equation}
And $R(M,x)$ always satisfies $E_0\leq R(M,x)\leq E_N$ (the smallest and largest eigenvalues).

Namely, giving an interesting Hamiltonian $\hat{H}$, and a presupposed wavefunction $|\psi\rangle$, the ground state energy $E_{0}$ of $\hat{H}$ always satisfies
\begin{equation}
    E_0 \leq \frac{\langle \psi |\hat{H}| \psi \rangle}{\langle \psi |\psi\rangle}.
\end{equation}
Therefore, the main objective of VQE is to find an applicable $|\psi^*\rangle$, such that the expectation value of $\hat{H}$ is minimized. In the same time, the $|E_0\rangle = |\psi^*\rangle$ and $E_0 = \langle \psi^* |\hat{H}| \psi^* \rangle$, if the ground energy is non-degenerate. (If the ground energy is degenerate, the $|\psi^*\rangle$ is usually in the space spanned by all ground states.)

However, since directly finding $|\psi^*\rangle$ is an intractable problem, VQE starts by defining a so-called ansatz wavefunction $|\psi(\vec{\theta})\rangle$, which can be implemented on a quantum device as a series of quantum gates and whose behavior can be described by parameters $\vec{\theta}$.

The other feature of VQE is that it is a hybrid algorithm that uses both classical and quantum computers.
The quantum eigensolver aims to find eigenvalues and eigenstates of a given Hamiltonian $\hat{H}$.
The Ritz variational principle tells us that for a given $\hat{H}$, and arbitrary state $|\vec{\theta}\rangle$,
\begin{equation}
    \lambda_{\min}\leq \frac{\langle \vec{\theta} |\hat{H}| \vec{\theta}\rangle}{\langle \vec{\theta}|\vec{\theta}\rangle} \leq \lambda_{\max},
\end{equation}
where $\lambda_{\min, \max}$ is the minimum (maximum) eigenvalue of $\hat{H}$, and $\langle \vec{\theta}|\vec{\theta}\rangle = 1, \forall \vec{\theta}$ in quantum computing.
With the help of Ritz variational principle, VQE converts the eigenproblem into an iterative optimization problem whose cost function is $\min_{\vec{\theta}}\hat{E}= \langle \vec{\theta} |\hat{H}| \vec{\theta}\rangle$.
By optimizing the value of $\vec{\theta}$ by a classical optimizer to minimize the $\hat{E}$, we can get the $\hat{E}(\vec{\theta}^*)\approx E_{gs}$ and $|\langle \vec{\theta}^*|E_0\rangle| \approx 1$, if the ground state of $\hat{H}$ is non-degenerate.

VQE starts with an initialized qubit state. A parameterized quantum circuit is then applied to this state to model the physics and entanglement of the electronic wavefunction.

The VQE pipeline contains
\begin{itemize}
    \item Hamiltonian construction and representation: fermionic space to qubit space transformations, encoding of operators
    \item Ansatz selection and state preparation: Hardware efficient, QubitUCCAnsatz, UCCAnsatz, Hartreefock state
    \item Measurement: expectation
    \item Parameter optimization: optimizer parameter-shift rule, finite difference, automatic differentiation
\end{itemize}

\subsubsection{Implementation}

Before studying any physical model or molecule by VQE, we must transform them into a qubit Hamiltonian considering the present leading quantum platforms as a qubit basis.
Here, we show how \MindQuantum\ simulates a quantum chemistry example LiH by step and step.
First we need to import some packages we need.
\begin{lstlisting}
from openfermion.chem import MolecularData
from openfermionpyscf import run_pyscf
\end{lstlisting}
The \code{openfermion} and the \code{openfermionpyscf} are open resourced packages that can be used to construct the corresponding Hamiltonian of a given chemical molecular configuration. For example, we can define the LiH molecular geometry in this way
\begin{lstlisting}
    dist=1.5
    geometry=[
    ["Li",[0.0,0.0,0.0*dist]],
    ["H",[0.0,0.0,1.0*dist]]
    ]
    basis="sto3g"
    spin = 0
    molecule = MolecularData(geometry,basis,multiplicity=2*spin+1)
    molecule = run_pyscf(molecule,run_scf=1,run_ccsd=1,run_fci=1)
\end{lstlisting}
Based on \code{molecule}, we can use the function in \MindQuantum\ to obtain the corresponding qubit Hamiltonian formula,
\begin{lstlisting}
    from mindquantum.algorithm.nisq import get_qubit_hamiltonian
    hamiltonian_QubitOP = get_qubit_hamiltonian(molecule)
\end{lstlisting}
So far, we finish building the Hamiltonian formula of LiH easily.
The next step of VQE is to choose a good wavefunction ansatz. The Unitary Coupled-Cluster Singles and Doubles (UCCSD) method is inspired by the classical CCSD in quantum chemistry. In \MindQuantum, the UCCSD ansatz of LiH can be built by
\begin{lstlisting}
    from mindquantum.algorithm.nisq import uccsd_singlet_generator, uccsd_singlet_get_packed_amplitudes
    ucc_fermion_ops = uccsd_singlet_generator(
    molecule.n_qubits, molecule.n_electrons, anti_hermitian=True)
    ucc_qubit_ops = Transform(ucc_fermion_ops).jordan_wigner()
    ansatz_circuit = TimeEvolution(ucc_qubit_ops.imag, 1.0).circuit
ansatz_parameter_names = ansatz_circuit.params_name
\end{lstlisting}
In general, the initial state of VQE for this chemistry problem is the Hartree-Fock state. We apply the Pauli $X$ gate to make sure that the initial state contains $|1\rangle$ as much as the number of electrons in the molecule.
\begin{lstlisting}
    hartreefock_wfn_circuit = Circuit([X.on(i) for i in range(molecule.n_electrons)])
\end{lstlisting}
So the total quantum circuit of VQE to solve the ground energy of LiH is
\begin{lstlisting}
    total_circuit = hartreefock_wfn_circuit + ansatz_circuit
\end{lstlisting}

In VQE, gradient descent methods are usually chosen as the optimizer to get the optimal parameter such that it can output the minimal expectation value.
In \MindQuantum, we can easily utilize MindSpore Quantum to get the gradient operator.
The gradient operator is a special function, when we input the initial parameters, it can output the corresponding gradient values.
In \MindQuantum, the gradient operator of VQE can be obtained as
\begin{lstlisting}
    grad_ops = Simulator('mqvector', total_circuit.n_qubits).get_expectation_with_grad(
    Hamiltonian(hamiltonian_QubitOp.real),
    total_circuit)
\end{lstlisting}
In general, we need to provide appropriate initial parameter for circuit such that we can more quickly converge to the minimal value. For UCCSD ansatz, the CCSD value is a good choice, which can be obtained by
\begin{lstlisting}
    init_amplitudes_ccsd = uccsd_singlet_get_packed_amplitudes(
    molecule_of.ccsd_single_amps, molecule_of.ccsd_double_amps, molecule_of.n_qubits, molecule_of.n_electrons)
init_amplitudes_ccsd = [init_amplitudes_ccsd[param_i] for param_i in ansatz_parameter_names]
\end{lstlisting}
So far, we have prepared all the things needed in VQE.
The final step of VQE is to minimize the cost function to get the expected value of energy, which can be done by using the \textit{SciPy} package.
\begin{lstlisting}
    from scipy.optimize import minimize

    def fun(theta, grad_ops, energy_list):
        energy, grad = grad_ops(theta)
        energy_list.append(energy)
        return np.squeeze(energy.real), np.squeeze(grad.real)

    energy_list = []
    res = minimize(fun, p0, args=(grad_ops, energy_list), method='bfgs', jac=True)
\end{lstlisting}
If everything works well, we will get the result like this
\begin{lstlisting}
    Step: 5,        energy: -7.878223282730547
    Step: 10,       energy: -7.880288481438961
    Step: 15,       energy: -7.882035668304055
    Step: 20,       energy: -7.882302370885741
    Step: 25,       energy: -7.882349803534313
    Step: 30,       energy: -7.882352702053751
    Step: 35,       energy: -7.8823527077335065
    Step: 40,       energy: -7.882352708347106
\end{lstlisting}
So far, we have used the \MindQuantum\ to find the approximate value of ground energy. In addition, we have also observed a new quantum eigen solver quantum, Full Quantum Eigensolver (FQE) algorithm \cite{wei2020full}, which eliminates the need for a classical optimizer and conducts all computations directly on the quantum computer through the utilization of quantum gradient descent.

\subsection{DC-QAOA}

View demo code of this section: \demonotebook{05}{5.4_DC_QAOA} \ \demonotebookgithub{05}{5.4_DC_QAOA}

\subsubsection{Background}
In the field of quantum computing, researchers are continuously seeking efficient and robust methods to tackle complex problems. Adiabatic Quantum Computation (AQC) is one such approach, where a quantum system evolves gradually from an initial Hamiltonian to a final Hamiltonian that encodes the desired solution. According to the adiabatic theorem, if this evolution is slow enough, the system will remain in the ground state of the Hamiltonian throughout the process:
\begin{equation}
    H(t) = (1-\lambda(t))H_{\text{initial}} + \lambda(t)H_{\text{final}},
\end{equation}
where \(\lambda(t) \in [0, 1]\). However, conducting AQC in practical scenarios faces challenges arising from external noise, imperfections, and the need for long evolution times. To overcome these limitations and enhance the efficiency of adiabatic evolution, one promising approach is the Quantum Approximate Optimization Algorithm (QAOA).

QAOA employs a parameterized quantum circuit to prepare a trial state that approximates solutions for optimization problems. By optimizing the ansatz parameters, QAOA seeks to find the optimal values that minimize the objective function, which encodes the problem's energy landscape. This method allows for shorter evolution times and can be more resilient to noise and imperfections, making it a viable alternative to AQC for practical quantum computing applications.

However, hardware implementations of QAOA face challenges due to noise, decoherence, and the complexity of parameter optimization. To enhance its performance and robustness, the Digitized Counterdiabatic Quantum Approximate Optimization Algorithm (DC-QAOA) introduces the counterdiabatic (CD) term:

\begin{equation}
    H_{\text{DC-QAOA}} = H_{\text{initial}}(\gamma) + H_{\text{final}}(\beta) + H_{\text{CD}}(\alpha),
\end{equation}

This approach combines ideas from both Adiabatic Quantum Computation (AQC) and QAOA by introducing an additional parameter per layer compared to QAOA, resulting in shallower quantum circuits without compromising accuracy:
\begin{equation}
    \label{eq:evo operator}
    U(\gamma, \beta) \to U(\gamma, \beta, \alpha), \quad F(\gamma, \beta) \to F(\gamma, \beta, \alpha),
\end{equation}
where \(\alpha\), \(\beta\), and \(\gamma\) represent the digitized parameters, \(F\) is the cost function, and \(U\) is the evolution operator.

The CD term acts as an additional component in the Hamiltonian, designed based on the instantaneous eigenstates and eigenenergies of the evolving Hamiltonian. It effectively suppresses errors and reduces the required evolution time, leading to enhanced efficiency and reliability in quantum computations.

To derive the precise form of the counterdiabatic term, various techniques, such as Lewis-Riesenfeld invariants and transitionless quantum driving~\cite{PhysRevA.83.062116}, have been applied. The implementation of the CD term is tailored to the specific problem and dynamics of the quantum system, utilizing available control techniques. When obtaining exact CD terms is challenging, an approximate CD driving approach is proposed, utilizing the nested commutator method  with the adiabatic gauge potential~\cite{PhysRevResearch.4.013141}:

\begin{equation}
    A_{\lambda}^{(l)} = i \sum_{k=1}^{l} \alpha_{k}(t) [H_{a}, [H_{a}, \ldots, [H_{a}, \partial_{\lambda} H_{a}]]].
\end{equation}

DC-QAOA holds promise for solving a wide range of optimization problems, from combinatorial optimization to protein folding~\cite{PhysRevApplied.20.014024} and financial portfolio optimization~\cite{PhysRevResearch.4.043204}. As quantum computing technology continues to advance, DC-QAOA represents an exciting avenue for leveraging the counterdiabatic term and parameter space digitization to unlock new frontiers in quantum optimization.

\textbf{Target problem}
\begin{itemize}
    \item[1.] Learn how to find the ground state of problem Hamiltonian
    \item[2.] Utilize quantum circuits to find optimized parameters
\end{itemize}
\textbf{Required Mindquantum functionalities}
\begin{itemize}
    \item[1.] gradient-based optimizer
    \item[2.] Batch quantum circuit simulator

\end{itemize}

\subsubsection{Implement}
As a mature variant of variational algorithm, this method requires two steps, which is building the ansatz circuit and optimization. Now let's look at how to implement DC-QAOA into the Ising model~\cite{PhysRevResearch.4.013141}. The general form of the Hamiltonian is given by the following equation:

\begin{equation}
    H_{final} = H_{prob} = -J_{ij}\sum_{i=1}^{n-1}\sigma^z_i\sigma^z_{i+1}-h_{i}\sum_{i=1}^{n}{\sigma_i^z} + k_{i}\sum_{i=1}^{n}\sigma_{i}^{x}.
\end{equation}

The initial state is chosen as $H_{initial} = \sum_{i}\sigma_{i}^{x}$, which should drives transitions between different states, facilitating the exploration of the solution space. To be specific, we choose longitudinal field Ising model (LFIM) Hamiltonian, in which $J_{ij}=1,~h_{i}=1,~k_{i}=0$.
\begin{lstlisting}
# Generate H_{initial} as H_{mixer}.
# n_qubits is the number of qubits
def generate_h_mixer(n_qubits:int):
    h_mixer = QubitOperator()
    for i in range(n_qubits):
        h_mixer += QubitOperator(f'X{i}')
    return h_mixer

# Generate H_{final} as H_{prob}.
def generate_h_prob(n_qubits:int, J:float, h:float):
    h_prob = QubitOperator()
    for i in range(n_qubits-1):
        h_prob += QubitOperator(f'Z{i} Z{i+1}', -J)
    for i in range(n_qubits):
        h_prob += QubitOperator(f'Z{i}', -h)
    return h_prob
\end{lstlisting}

In our case,  a pool of CD operators is defined using the nested commutator approach of the adiabatic gauge potential~\cite{PhysRevResearch.6.013147}
$$A = \{\sigma^{y},\sigma^{z}\sigma^{y}, \sigma^{y}\sigma^{z}, \sigma^{x}\sigma^{y}, \sigma^{y}\sigma^{x} \}.$$
Here we present a 12 qubits system and we consider CD term to be  $A^{*} = \sum_{i=1}^{n}\sigma_{i}^{y}$. Following we build an ansatz layer for DC-QAOA.
\begin{lstlisting}
def generate_h_cd(n_qubits:int):
    h_cd = QubitOperator()
    for i in range(n_qubits):
        h_cd += QubitOperator(f'Y{i}')
    return h_cd
\end{lstlisting}

\begin{lstlisting}
n_qubits = 12
J, h = 1, 1

h_prob = generate_h_prob(n_qubits, J, h)
h_mixer = generate_h_mixer(n_qubits)
h_cd = generate_h_cd(n_qubits)
\end{lstlisting}

In order to prepare the eigenstates for $H_{initial}$, we first put ${\rm prep\_circ}$, and then apply the evolution operator on it.
\begin{lstlisting}
# Prepare the eigenstates for $H_{mixer}$.
prep_circ = UN(H, n_qubits)
# Choose the number of layers p.
p = 1

ansatz_template =   u_p + u_m  + u_cd

ansatz = Circuit() + prep_circ

for i in range(p):
    ansatz += add_suffix(ansatz_template, str(i)) + BarrierGate()
\end{lstlisting}

As long as we have set up the ansatz circuit, we can change the number of layers or type of classical optimizer in order to find the optimal parameter set. We define the cost function using
\begin{equation}
    F(\alpha,~\beta,~\gamma) = \left<\psi_0\right|U^{\dagger}\ H_{prob} U\left|\psi_0\right>,
\end{equation}
where $U = \Pi_{i=0}^p U_{cd}(\alpha_i)U_m(\beta_i)U_p(\gamma_i)$ and this is just the evolution operator in Eq.~\ref{eq:evo operator}.
\begin{lstlisting}
def train(ham, ansatze, iteration):

    # bulid the quantum circuit

    circ = ansatze
    sim = Simulator('mqvector', ansatze.n_qubits)

    grad_ops = sim.get_expectation_with_grad(ham, circ)
    net = MQAnsatzOnlyLayer(grad_ops)
    opti = nn.Adagrad(net.trainable_params(), learning_rate=0.05)
    train_net = nn.TrainOneStepCell(net, opti)

    result = []
    for i in range(iteration):
        train_net()
        result.append(np.array(train_net()[0]))
    pr = dict(zip(ansatze.params_name, net.weight.asnumpy()))

    return result
\end{lstlisting}

\begin{figure}
    \centering
    \includegraphics[width = 1 \linewidth]{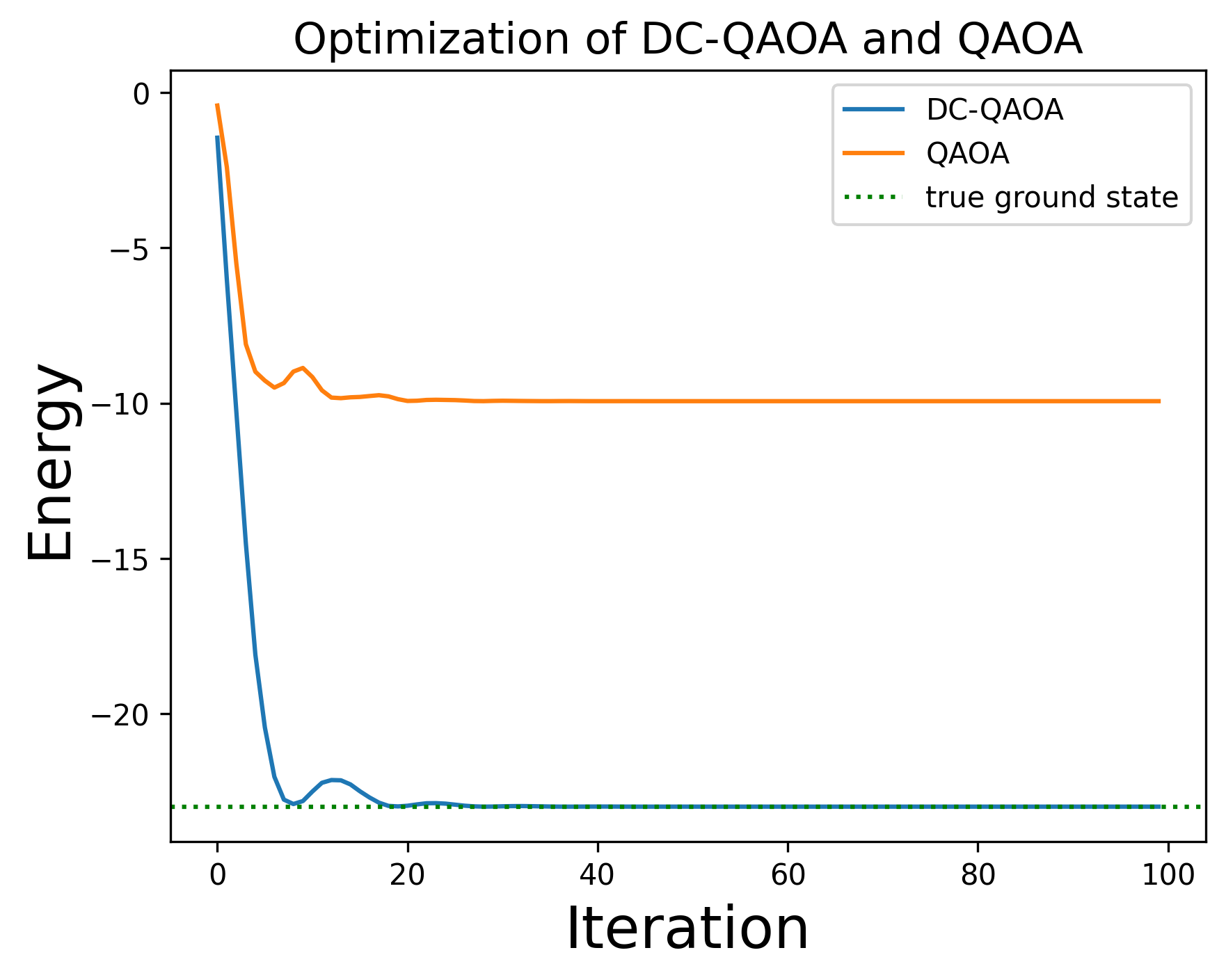}
    \caption{Energy as the function of iterations.}
    \label{fig:dc}
\end{figure}

In this tutorial, we will optimize the gradient using Adam optimizer~\cite{PhysRevResearch.6.013147}, which will have good enough results. If you plot the figure of mean energy, you will have Fig.~\ref{fig:dc}. Here, the dashed green line is the exact eigenvalue of $H_{prob}$, the solid blue line is the mean value of one-layer-DC-QAOA during the optimizing process, and the solid orange line is QAOA with one layer. We can notice that DC-QAOA has converged to the exact value in 20 steps while QAOA is trapped in the local minimal. Comparing these two algorithms from the ability to estimate the ground state, DC-QAOA has an advantage over QAOA.

\subsection{Symmetry enhanced VQE}
View demo code of this section: \demonotebook{05}{5.5_symmetry_enhanced_VQE} \ \demonotebookgithub{05}{5.5_symmetry_enhanced_VQE}

\subsubsection{Background}
Variational quantum algorithms have been primarily focused on identifying the ground state of complex many-body systems. In this context, the Variational Quantum Eigensolver (VQE) algorithm, designed to find the ground state of a many-body system  by minimizing its average energy, has emerged as a powerful tool. However, several critical phenomena in physics, including topological phases, necessitate knowledge of several low-energy eigenstates, not just the ground state. Consequently, extending VQE to higher energy eigenstates is of significant importance.

The weighted SSVQE method offers an alternative approach to generate all the $k$ lowest energy eigenstates of a given Hamiltonian $H$~\cite{nakanishi2019subspace}. This method utilizes a set of $k$ orthogonal initial states, denoted as $\{|\phi_{i}\rangle\}_{i=1}^{k}$ (where $\langle \phi_{i} | \phi_{j} \rangle = \delta_{ij}$), as inputs for a single parameterized quantum circuit, described by the unitary operator $U(\vec{\theta})$. Given that the initial states are orthogonal, the outputs $U(\vec{\theta})| \phi_{j} \rangle$, generated by the same circuit, maintain orthogonality. In the weighted SSVQE, the objective is to minimize the cost function
\begin{equation}
    \mathrm{cost} = \sum_{i=1}^{k} w_{i} \langle \phi_{i}| U^{\dagger}(\vec{\theta}) H U(\vec{\theta}) | \phi_{i} \rangle,
    \label{ssvqe_cost}
\end{equation}
where $w_1 > w_2 > \cdots > w_k$ are real positive numbers. Minimizing the cost function in Eq.~\eqref{ssvqe_cost} produces all the $k$ lowest energy eigenstates such that $|E_{i}\rangle = U(\vec{\theta}^{*})|\phi_{i}\rangle$.
A notable advantage of the weighted SSVQE is that it delivers all the $k$ lowest energy eigenstates through a single optimization process, without requiring any evaluation of quantum state overlaps. However, the algorithm becomes more resource-demanding as the number of target eigenstates increases.

Symmetry stands as one of the most fundamental concepts in physics, particularly in quantum mechanics. A majority of physical systems exhibit various types of symmetries that can be accurately described mathematically. The VQE algorithm can also significantly benefit from the integration of these symmetries. There are two ways to incorporate symmetries in the VQE algorithms: (i) designing the circuit to naturally generate the quantum states with the relevant symmetry~\cite{Lyu2020accelerated,Gard2020,barron2021preserving}, and (ii) adding extra terms to the cost function to penalize the quantum states without the relevant symmetry~\cite{mcclean2016theory,ryabinkin2018constrained}.

\subsubsection{Implementation}
Using the adaptable and comprehensive features of \MindQuantum, researchers can address various optimization targets by defining custom objective functions. In this section, we demonstrate the implementation of SSVQE using \MindQuantum\ to precisely determine the four lowest state energies of the Heisenberg Hamiltonian, while incorporating various symmetry strategies.

Initially, the construction of a symmetry-preserving ansatz is necessary, as detailed in Ref.~\cite{Lyu2023symmetryenhanced}.
\begin{lstlisting}
def entangling_gate(parameter, qubits):
    circuit = Circuit()
    circuit += RZ(-np.pi / 2).on(qubits[1])
    circuit += CNOT.on(qubits[0], qubits[1])
    circuit += RZ({parameter: -2}).on(qubits[0])
    circuit += RZ(np.pi / 2).on(qubits[0])
    circuit += RY({parameter: 2}).on(qubits[1])
    circuit += RY(-np.pi / 2).on(qubits[1])
    circuit += CNOT.on(qubits[1], qubits[0])
    circuit += RY({parameter: -2}).on(qubits[1])
    circuit += RY(np.pi / 2).on(qubits[1])
    circuit += CNOT.on(qubits[0], qubits[1])
    circuit += RZ(np.pi / 2).on(qubits[0])
    return circuit

def ansatz(N, layers, local_rot=True):
    circuit = Circuit()
    params_index = 0
    for layer_index in range(layers):
        for i in range(2):
            for j in range(i, N - 1, 2):
                circuit += entangling_gate(str(params_index), [j, j + 1])
                params_index += 1
        if local_rot:
            for i in range(N):
                circuit += PhaseShift(params_index).on(i)
                params_index += 1
    return circuit
\end{lstlisting}

SSVQE can be implemented as a combination of several VQEs, each with distinct initialization circuits but sharing a common parameterized ansatz.
For most VQE instance, specific tasks are defined by their corresponding measurement functions.
Using the \getexpectationwithgrad function, one can easily obtain the expectation value and the corresponding gradient with respect to the trainable variables. Here, as an example, we demonstrate the code implementation for obtaining the second- and the third-lowest state energies of the Heisenberg Hamiltonian. In this particular instance,
we use a $S_z$-conserving ansatz, which preserves the $z$ component of the total spin, and the total spin operator is added to the cost function as a penalty term. The cost function is then modified into a form of
\begin{equation}
    \mathrm{cost} = \sum_{i=1}^{k} w_{i} \langle \phi_{i}| U^{\dagger}(\vec{\theta}) [H + \beta(\hat{O} - c)^2] U(\vec{\theta}) | \phi_{i} \rangle,
    \label{ssvqe_cost_pen}
\end{equation}
where $c$ is a constant indicating the target subspace, and $\beta$ is a positive constant which is taken to be sufficiently large according to the form of penalty terms. Here, we do the expansion $(\hat{O} - c)^2 {=} \hat{O}^2 {-} c\hat{O} {+} c^2$. Together with the Hamiltonian, now we have three different measurement operators. The corresponding measurement operators in the case of Heisenberg Hamiltonian and the total spin operator can be generated with the following codes.

\begin{lstlisting}
def s_alpha(N, alpha: str = 'X'):
    alpha = alpha.upper()
    out = QubitOperator()
    for i in range(N):
        out += QubitOperator(alpha + f'{i}')
    return out * 0.5

def s_tot_op(N):
    s_x_op = s_alpha(N, 'X')**2
    s_y_op = s_alpha(N, 'Y')**2
    s_z_op = s_alpha(N, 'Z')**2
    op = s_x_op + s_y_op + s_z_op
    op.compress()
    return op

def ham_heis(n_qubits, J):
    ham = QubitOperator()
    for i in range(1, n_qubits):
        ham += QubitOperator("X{} X{}".format(i - 1, i), J)
        ham += QubitOperator("Y{} Y{}".format(i - 1, i), J)
        ham += QubitOperator("Z{} Z{}".format(i - 1, i), J)
    ham.compress()
    return Hamiltonian(ham)
\end{lstlisting}

Having predefined all necessary modules, including the initialization circuits, shared parameter ansatz, measurement operators, and the constant $\beta$, we can proceed with implementing the SSVQE algorithm. The following code snippet shows the implementation of the SSVQE class.
\begin{lstlisting}
class SSVQE:
    def __init__(self, n_qubits, init_circuits, pqc, ops, beta):
        self.n_qubits = n_qubits
        self.circs = [circ + pqc for circ in init_circuits]
        self.sim = Simulator('mqvector', n_qubits)
        self.beta = beta
        self.grad_ops = [self.sim.get_expectation_with_grad(ops, circ) for circ in self.circs]

    def energy(self, params):
        if self.sim is None:
            self.sim = Simulator('mqvector', self.n_qubits)
        energies = []
        for i in range(len(self.grad_ops)):
            f, g = self.grad_ops[i](params)
            energies.append(f[0, 0].real)
        return energies

    def __call__(self, inputs):
        cost = 0
        cost_grad = 0
        for i in range(len(self.grad_ops)):
            f, g = self.grad_ops[i](inputs)
            f1, f2, f3 = f[0, 0].real, f[0, 1].real, f[0, 2].real
            g1, g2, g3 = np.array(g[0, 0, :].real), np.array(g[0, 1, :].real), np.array(g[0, 2, :].real)
            cost += (f1 + self.beta * (f2 + f3)) * (len(self.grad_ops) - i)
            cost_grad += (g1 + self.beta * (g2 + g3)) * (len(self.grad_ops) - i)
        return cost, cost_grad
\end{lstlisting}

Upon completion of the algorithm implementation, the focus shifts to SSVQE optimization. The choice of classical optimizer remains flexible. For this instance, the \code{"L-BFGS-B"} optimizer from SciPy is selected.

\begin{lstlisting}
initial_parameter = (np.random.rand(len(ssvqe.circs[0].params_name)) - .5) * np.pi
res = scipy.optimize.minimize(ssvqe,
                    initial_parameter,
                    method='l-bfgs-b',
                    jac=True,
                    options={'disp': False})
\end{lstlisting}

\begin{figure}
	\centering
	\includegraphics[width=1\linewidth]{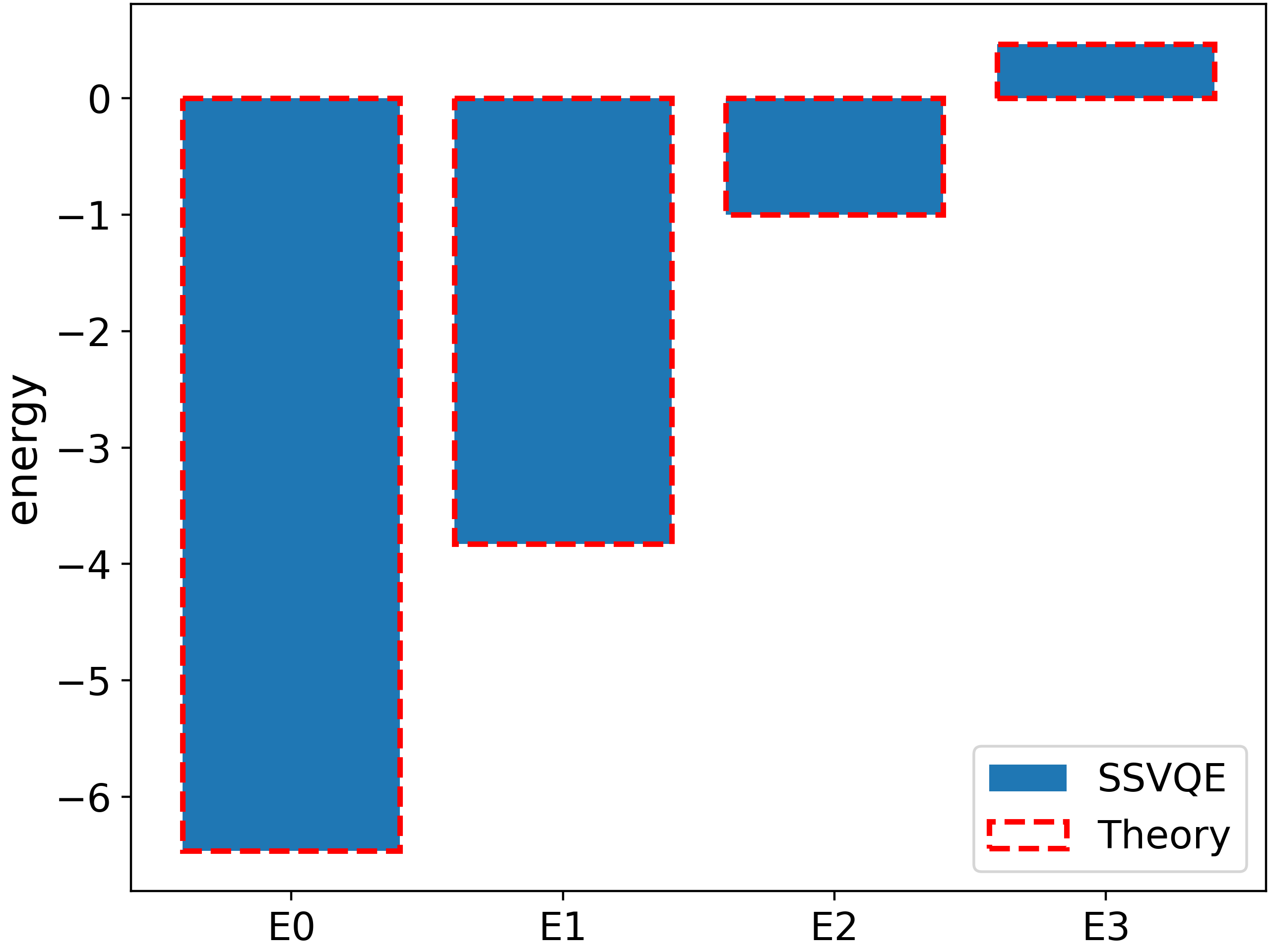}
	\caption{Simulation results of the SSVQE incorporating symmetries for 4-qubit Heisenberg Hamiltonian}
	\label{fig:SSVQE_result}
\end{figure}

This implementation completes training for the 4-qubit Heisenberg Hamiltonian in just seconds. One can take a look at the final results and generate a plot to compare the approximation results with the theoretical energies. In Fig.~\ref{fig:SSVQE_result}, we show the simulation results for the four lowest state energies of the $4$-qubit Heisenberg Hamiltonian. Clearly, by incorporating inherent symmetries in SSVQE, it is possible to precisely approximate both ground and excited state energies in a resource-efficient manner.

\subsection{Data-reuploading Classifier}
View demo code of this section: \demonotebook{05}{5.6_Data-reuploading_classifier} \ \demonotebookgithub{05}{5.6_Data-reuploading_classifier}

\subsubsection{Background}


Data classification is not only an important application of classical neural networks but also of quantum information and quantum computation in the future. The structure of a quantum classifier can be divided into three main components: the sub-circuit responsible for encoding data, the sub-circuit used for information processing and the measurements applied for extracting classification result from the final state.
The key to achieving accurate data classification lies in capturing the nonlinear features of the data. Many approaches carefully design quantum circuits to introduce this capability, such as the strategies inspired in artificial neural networks or kernel methods that widely used in classical machine learning \cite{PhysRevLett.122.040504, 2019Supervised_nature, Wan_2017_npjqi, hur2022quantum, chalumuri2021hybrid, oh2020tutorial, farhi2018classification, wrobel2021application}. Among them, the schemes that require fewer resources such as the number of qubits and quantum operations have received widespread attention during the current NISQ devices era.
Therefore, a natural motivation is to explore the minimum resources required to construct an effective quantum classifier.

It has been shown in Ref.~\cite{PerezSalinas2020datareuploading} that a single qubit, with the assistance of a classical subroutine, can provide sufficient computational capabilities to perform multi-classification tasks for multidimensional input data, similar to a single hidden-layered neural network. This may sound surprising, especially considering that a single qubit provides only a superposition of two sample states and operations on it merely involve rotations around the Bloch sphere.
The pivotal point of this scheme is to combine the data uploading unit and information processing unit and repeat them many times, i.e., \textit{Data re-uploading} to introduce the non-linearity necessary for a supervised classification task.

\begin{figure}
	\centering
	\begin{subfigure}{0.3\textwidth}
		\centering
		\includegraphics[width=\textwidth]{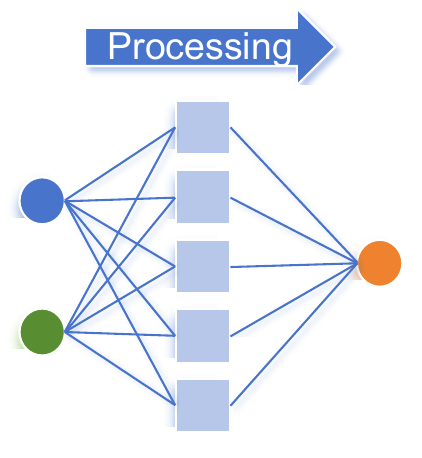}
		\caption{Physical topology graph.}
	\end{subfigure}
	\begin{subfigure}{0.3\textwidth}
		\centering
		\includegraphics[width=0.9\textwidth]{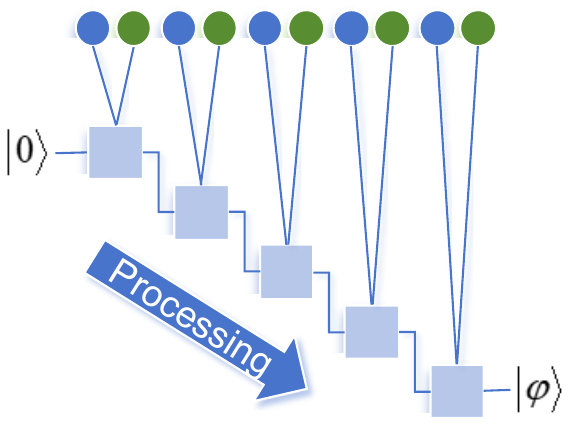}
		\caption{Logical circuit.}
	\end{subfigure}
    \captionsetup{justification=raggedright,singlelinecheck=false}
    \caption{Simplified workflows of (a) a single-hidden-layered classical neural network and (b) a single-qubit quantum classifier with structure of data re-uploading.
    In the classical neural network, each neuron receives all input data. Conversely, the qubit receives both the input data and the result of previous operations. It processes the combined information, and the final output is a quantum state that encodes multiple repetitions of input uploads and processing parameters.}
    \label{diagram_data_reuploading}
\end{figure}

The authors proved in their work that the equivalence between the data re-uploading scheme and the well-known Universal Approximation Theorem \cite{HORNIK1991251}, which argues that a single-layer network composed of enough neurons along with an activation unit can approximate any continuous function.

We precede the construction of the single-qubit classifier with an insight into the workflow of a classical single-layer neural network.
As it is shown in Fig.~\ref{diagram_data_reuploading} (a), the original data is fed into and processed by each neuron, or in other words, the original data is re-uploaded multiple times. And then in conjunction with the activation function, this feed-forward neural network captures non-linearity.

In order to introduce this capacity to the single-qubit classifier, we have also to re-upload the data into the qubit repeatedly, and without violating the principle of quantum no-cloning theorem. The data re-uploading scheme \cite{PerezSalinas2020datareuploading} is specifically designed to address this problem, and its workflow is depicted in Fig.~\ref{diagram_data_reuploading} (b). Similarly to the classical neural network, data points are introduced simultaneously into each processing unit, represented by a unitary rotation, of single-qubit quantum classifier.
However, the processing units are not only affected by the input data but also by the previous processing units.
The resulting output is a quantum state that will be measured to extract the classification result.

\begin{figure}[h]
    \begin{center}
        \includegraphics[width=0.9\linewidth]{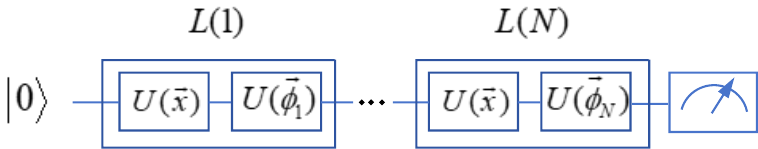}
    \end{center}
    \captionsetup{justification=raggedright,singlelinecheck=false}
    \caption{The single-qubit classifier with data re-uploading is comprised of layers $L(i)$, which serve as the building blocks of the quantum circuit. Each of these layers consists of a $U(x)$ gate, responsible for data uploading, and a trainable parameterized unitary $U(\phi)$. We repeat this building block $N$ times and calculate a cost function, which is related to the fidelity between the final state of the circuit and the corresponding target state of its class.}
    \label{data_reuploading_circuit}
\end{figure}

The explicit form of this single-qubit classifier is
shown in Fig.~\ref{data_reuploading_circuit}. Classical data is repeatedly re-uploaded sequentially, with processing units interspersed between them. The introduced data can be represented as qubit rotations. A data point from 3-dimensional space, $\overrightarrow{x}$, can be re-uploaded by a rotation of the qubit $U(\overrightarrow{x})$. Subsequent processing units are also rotations, whose angles will be optimized as trainable parameters to minimize the cost value $C(F)$ for performance improvement. The fidelity $F$ quantifies the closeness between the final state of the circuit and the target state corresponding to the label.

Then, the overall function of this quantum circuit is
$$\mathcal{U}(\overrightarrow{\phi},\overrightarrow{x})=U(\overrightarrow{\phi}_N)U(\overrightarrow{x})\cdots U(\overrightarrow{\phi}_1)U(\overrightarrow{x}),$$
which acts as
$$|\varphi\rangle=\mathcal{U}(\overrightarrow{\phi},\overrightarrow{x})|0\rangle.$$
The final classification of the data point will be determined by the measurement result on $|\phi\rangle$. Each pair of data uploading unit and processing unit can form a processing layer together:
$$L(i)\equiv U(\overrightarrow{\phi}_i)U(\overrightarrow{x}),$$
thus the classifier corresponds to
$$\mathcal{U}(\overrightarrow{\phi},\overrightarrow{x})=L(N)\cdots L(1).$$

The authors note that increasing the number of layers can provide the circuit with stronger expressivity, and leading to a higher classification accuracy.

\textbf{\textit{Target Problem:}}

Constructing and training a single-qubit data re-uploading quantum circuit to classify classical data.

\textbf{\textit{Required MindSpore Quantum functionalities:}}

1. Allowing the construction of interleaved and repetitive encoder and ansatz circuits, as well as enabling data re-uploading to the encoders.

2. Batching training over input data (multiple data and labels).

\subsubsection{Implementation}

In this example, we focus on a typical classification task: identifying points in different areas of a plane, and implement it based on \MindQuantum.

We create a dataset comprising numerous points with coordinates $\overrightarrow{x}=(x_1, x_2)$ randomly distributed on a unit area plane ($x\in[-1,1]$).
The task of the classifier is to determine, based on the coordinates of these data points, whether they fall within a circle of radius $r$, i.e., $x_1^2+x_2^2<r^2$.
And the radius of the circle is intentionally set to $r=\sqrt{\frac{2}{\pi}}$ to ensure an equal probability of points being inside (labeled ``0'' and corresponding to the target quantum state $|0\rangle$) or outside (labeled ``1'' and corresponding to the target quantum state $|1\rangle$) the circle.

\begin{lstlisting}
import numpy as np

def data_gen(n_samples=1000):
    center=[0.0, 0.0]
    rad=np.sqrt(2/np.pi)
    data, labels = [], []
    for i in range(n_samples):
        x = 2*(np.random.rand(2))-1
        if np.linalg.norm(x-center) <= rad:
            y = 0
        else:
            y = 1
        data.append(x)
        labels.append(y)
    return np.array(data), np.array(labels)

train_data, train_labels = data_gen(1000)
eval_data, eval_labels = data_gen(500)
test_data, test_labels = data_gen(5000)
\end{lstlisting}

The training set, validation set, and test set we created consist of 1000, 500, and 5000 random samples, respectively.

To encode the data onto the qubit, we use angles of rotation around the $x$ and $z$ axes to represent the coordinate information of the samples, i.e. $U(\overrightarrow{x})=U(x_1,x_2,0)=RZ(x_2)RX(x_1)$. Since our samples to be classified are only 2-dimensional, here we have padded 0.
Input data ($[x_1, x_2, x_3, \cdots x_{N-2}, x_{N-1},x_{N}]$) with $N$ dimensions can be split into multiple sets of three-dimensional components for encoding $U(\overrightarrow{x})=U(x_{N-2},x_{N-1},x_{N})\cdots U(x_1,x_2,x_3)$.

Next, we construct the circuit for a single-qubit classifier. We begin by applying an $H$ gate to the initial quantum state to generate a superposition state $(|0\rangle+|1\rangle)/\sqrt{2}$.
Afterwards, we add $RZ(x_1)$ and $RX(x_2)$ gates to upload the sample's coordinates $x_1$ and $x_2$ onto the qubit.
The subsequent processing unit is represented by a $U3$ gate with three independent trainable parameters to act as an arbitrary rotation.
These two sub-circuits are also named as ``encoder'' and ``ansatz'' respectively to indicate the parameters of the rotation gates will be directly fed into or gradually updated during training.
The encoder and ansatz together form a layer, and multiple such layers can be repeated to achieve the functionality of multiple data re-uploading and processing, thereby enabling points classification.

\begin{lstlisting}

from mindquantum import H, RZ, RX, U3, BarrierGate, Circuit, Simulator, Hamiltonian, QubitOperator, MQLayer

def circ(depth=1):
    circ = Circuit(H.on(0))
    circ += BarrierGate()
    for layer in range(depth):
        circ += Circuit([RZ('x1').on(0), RX('x2').on(0)]).as_encoder()
        circ += Circuit(U3(f'theta_{layer}', f'phi_{layer}',f'lambda_{layer}').on(0)).as_ansatz()
        circ += BarrierGate()
    return circ

example_circ = circ(depth=1)
example_circ.svg()
\end{lstlisting}

\begin{figure}[H]
\centering
\includegraphics[width=0.49\textwidth]
{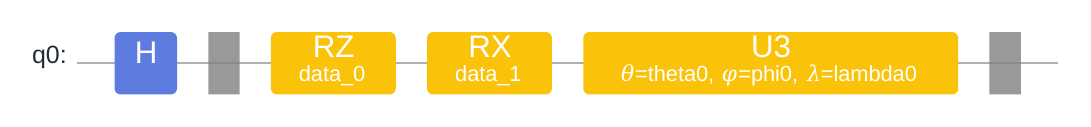}
\caption{The schematic diagram depicts a single qubit being subjected to an
$H$ gate to generate a uniform superposition state, along with an encoding and information processing layer for handling the classification task.}
\label{data_reuploading_circ0}
\end{figure}

The overall quantum circuit will transform the quantum state from $|0\rangle$ to $\varphi$, which requires to be measured.
As mentioned before, the classification result ``0'' (inside the circle) or ``1'' (outside the circle) can be determined according to the final measurement of the single qubit, e.g., a sampled point can be classified into ``0'' if the output probabilities $P(|0\rangle)>P(|1\rangle)$ and into ``1'' otherwise.
This approach can also be represented by the energy expectation of the final quantum state along the $Z$ direction, i.e., classifies the sampled point into ``0'' if $E(|\varphi\rangle)>0$ and else into ``1''. Hence, the loss function can be defined as $Loss=[E(|\varphi\rangle)-(1-2\cdot\mathrm{label})]^2$.

In the following code, we utilize the Adam optimizer with a learning rate of 0.1 as a classical subroutine to train the trainable parameters in the quantum circuit. The performance of classification will be enhanced by minimizing the loss function.

\begin{lstlisting}
import mindspore as ms
from mindspore import Tensor, ops
from mindspore.nn import Adam, TrainOneStepCell, LossBase,WithLossCell
ms.set_context(mode=ms.PYNATIVE_MODE, device_target='CPU')
import copy

circ = circ(depth=4)
ham = Hamiltonian(QubitOperator('Z0'))
sim = Simulator('mqvector', 1)
grad_ops = sim.get_expectation_with_grad(ham, circ)
qnet = MQLayer(grad_ops)

class MyLoss(LossBase):
    def __init__(self, reduction='mean'):
        super(MyLoss, self).__init__(reduction)
        self.square = ops.Square()

    def construct(self, logits, label):
        out =self.square(logits - (1-2*label))
        return self.get_loss(out)

loss = MyLoss()
opti = Adam(qnet.trainable_params(), learning_rate=0.02)
model = TrainOneStepCell(WithLossCell(qnet, loss), opti)

\end{lstlisting}
If the task to be solved with this single-qubit classifier involves many classes, one possible strategy consists on comparing the probability $P(|0\rangle)$ with several thresholds. For example, for a 4-categorized task, the determination of the classification result can be based on the interval in which the probability $P(|0\rangle)$ lies: $0\leqslant\lambda_1\leqslant\lambda_2\leqslant\lambda_3\leqslant1$.

Next, we train the parameters in the ansatz for 5000 iterations, recording and saving the highest validation accuracy achieved along with the corresponding trainable parameters. The batch size for training is 4.

\begin{lstlisting}
def eval(ansatz_params, eval_data, eval_labels):
    res = []
    for index in range(len(eval_label)):
        sim.reset()
        sim.apply_circuit(circ, pr=dict(zip(circ.params_name, np.hstack((eval_data[index], ansatz_params)))))
        exp = sim.get_expectation(ham).real
        clas = exp <= 0
        res.append(int(clas == eval_label[index]))
    return np.sum(res)/len(res)

acc_list = []
acc_max = 0.0

print('Training begins ...')

for i in range(2001):
    index = np.random.randint(0,len(train_data),4)
    loss = model(ms.Tensor(train_data[index]), ms.Tensor(train_labels[index]))
    acc = eval(qnet.weight.asnumpy(), eval_data, eval_labels)
    if acc > acc_max:
        acc_max = acc
        best_params = copy.deepcopy(qnet.weight.asnumpy())
    acc_list.append(acc)
    if i % 200 == 0:
        print(f'turn:{i}\t',f'\tacc_max:{acc_max}')

print('Training completes.\n')
print('Maximum validation accuracy in training:\n', acc_max)

test_acc = eval(best_params,test_data,test_labels)

print('Average test accuracy:\n', test_acc)
\end{lstlisting}

\begin{lstlisting}
Training begins ...
turn:0	 	  acc_max:0.55
turn:200	 	acc_max:0.566
turn:400	 	acc_max:0.622
turn:600	 	acc_max:0.622
turn:800	 	acc_max:0.7
turn:1000	 	acc_max:0.814
turn:1200	 	acc_max:0.814
turn:1400	 	acc_max:0.814
turn:1600	 	acc_max:0.814
turn:1800	 	acc_max:0.894
turn:2000	 	acc_max:0.894
Training completes.

Maximum validation accuracy in training:
0.894

Average test accuracy:
0.8876
\end{lstlisting}

\begin{figure}[H]
\centering
\includegraphics[width=0.49\textwidth]
{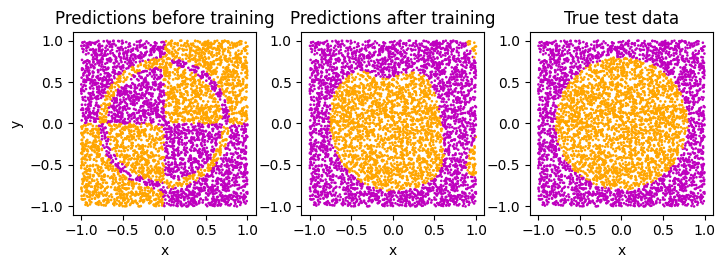}
\caption{Visualization Comparison of the data re-uploading single-qubit classifier's predictions before training, after training, and true data.}
\label{data_reuploading_performance}
\end{figure}

As seen in the information printed above, the model's prediction accuracy has reached $88.76\%$, meaning that the anticipated classification results are well consistent with the actual classification results.
Fig.~\ref{data_reuploading_performance} visualizes and compares the predictions before training, predictions after training, and the true classification data.
It can be observed that, after training, this single-qubit classifier is able to effectively classify the samples.

In addition, the authors also extended this scheme to a multiple-qubits version and achieved further improved performance, benefiting from the resulting more complex quantum superposition states and quantum entanglement.

\subsection{Q-GAN}
View demo code of this section: \href{https://gitee.com/mindspore/mindquantum/blob/research/paper_with_code/variational_quantum_circuits_enhanced_generative_adversarial_network/QGAN.py}{\color{ceruleanblue}{Gitee}} \ \href{https://github.com/donghufeng/mindquantum_tutorial_code/blob/master/whitepaper/code/chapt05/5.7_Q-GAN/QGAN.py}{\color{ceruleanblue}{GitHub}}

\subsubsection{Background}
Generative adversarial network (GAN) is a well-known machine learning model proposed by Goodfellow. It has outstanding performance in various challenging tasks, especially in image generation and video generation. The general GAN consists of a generator ${G}$ and a discriminator ${D}$. It accomplishes generation tasks through an adversarial game of ${G}$ and ${D}$. During the training, ${G}$ is trained to maximize the probability that ${D}$ misclassifies generated data as real data. Correspondingly, ${D}$ is trained to maximize the probability of successfully classifying real data, while minimizing the probability of misclassifying generated data. The objective function of GAN can be summarized as
\begin{equation}
  \begin{split}
    \mathop {\min }\limits_G \mathop {\max }\limits_D V(D,G) = {\rm{ }}{{\rm{E}}_{x\sim{p_r}(x)}}[\log D(x)]
    \\+ {{\rm{E}}_{z\sim{p_z}(z)}}[\log (1 - D(G(z)))]. \label{eq1}
  \end{split}
\end{equation}

QGAN, as a quantum version of GAN, has more advantages in sampling and generating discrete data than its classical counterpart. In addition, it has been proved the potential exponential speedup theoretically~\cite{lloyd2018quantum}. These lead to extensive research on QGAN. For example, the learning of the classical or quantum data~\cite{benedetti2019adversarial, zeng2019learning, zoufal2019quantum,ahmed2021quantum}, using the entangling power of quantum circuits to overcome issues of non-convexity and mode collapse~\cite{niu2022entangling}, discovering small molecular drugs~\cite{li2021quantum}, data enhancement~\cite{nakaji2021quantum} and anomaly detection with QGAN~\cite{herr2021anomaly}.

\subsubsection{Implementation}
\begin{figure}[ht]
  \centering
  \includegraphics[scale=0.2]{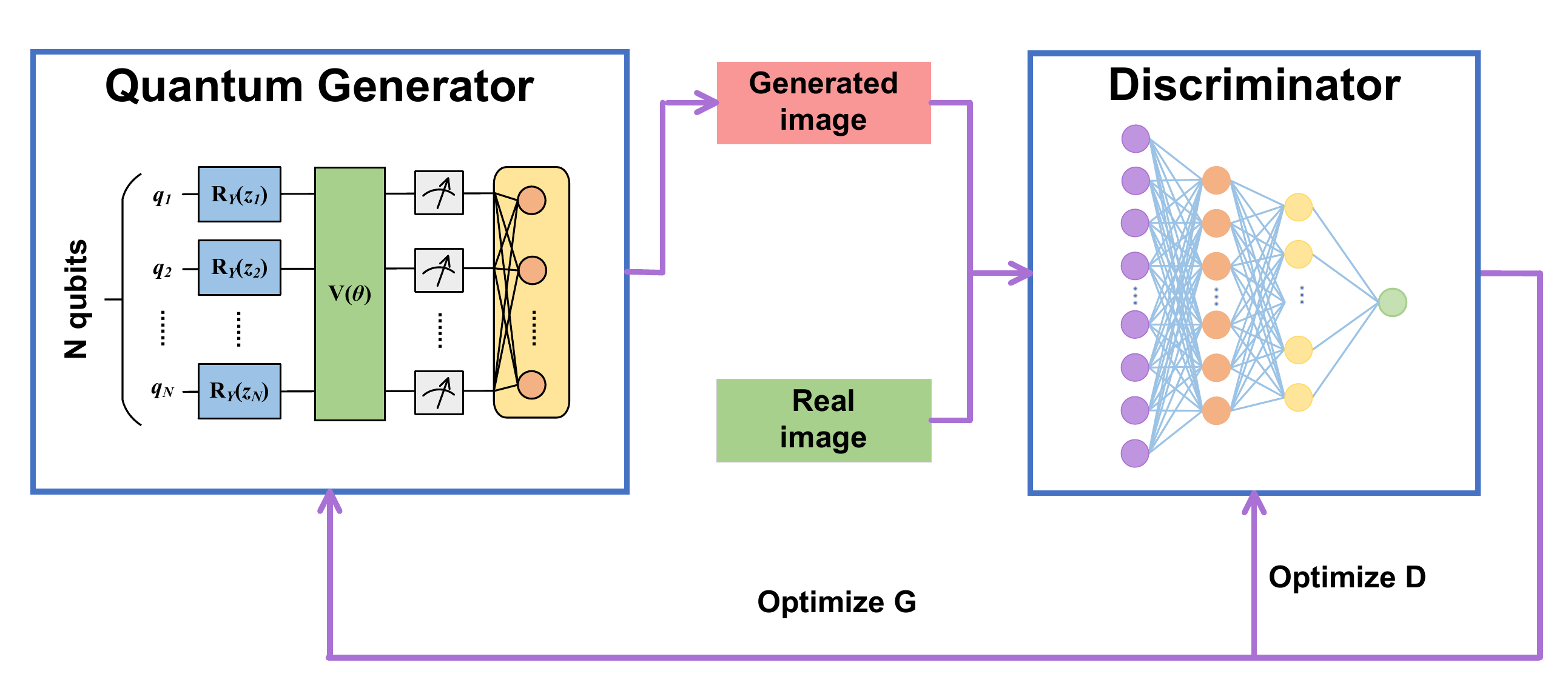}
  \caption{\label{qc-gan} The overall architecture and training process of QC-GAN.}
\end{figure}

We show an example of using \MindQuantum, building a hybrid quantum-classical generative adversarial network (QC-GAN) \cite{shu2024variational}, and then learning a probability distribution to generate handwritten digits through training. Specifically, the QC-GAN consists of a variational quantum circuit with one layer of neural network to form ${G}$ and a classical neural network to form ${D}$. It is trained similarly to the classical GAN with a noise vector input ${z}$ to the ${G}$. Then, the adversarial training of ${G}$ and ${D}$ realizes the generation of images.
Differently, the generated image is obtained by quantum circuit evolution, measurement, and nonlinear mapping of the neural network. The overall structure and training process of QC-GAN is shown in Fig.~\ref{qc-gan}.

The first step is loading the dataset. The experiment uses the handwritten digit set, which has a total of 70,000 handwritten digit images, including 60,000 training samples and 10,000 test samples, with an image size of 28*28 and a single channel. After loading the dataset, a series of preprocessing operations are performed to prepare for subsequent training. The model build-up consists of two networks, the generator ${G}$ and the discriminator ${D}$. For the discriminator part, it uses a general fully connected network structure. For the generator, it is divided into two parts, the quantum part and the classical part. Quantum circuits are built by calling the parameterized rotation gates of \MindQuantum. The quantum circuit is constructed as follows. The structure of the quantum circuit can be shown through \methodsummary method of \Circuit.
\begin{figure*}[htbp]
  \centering
  \includegraphics[scale=0.5]{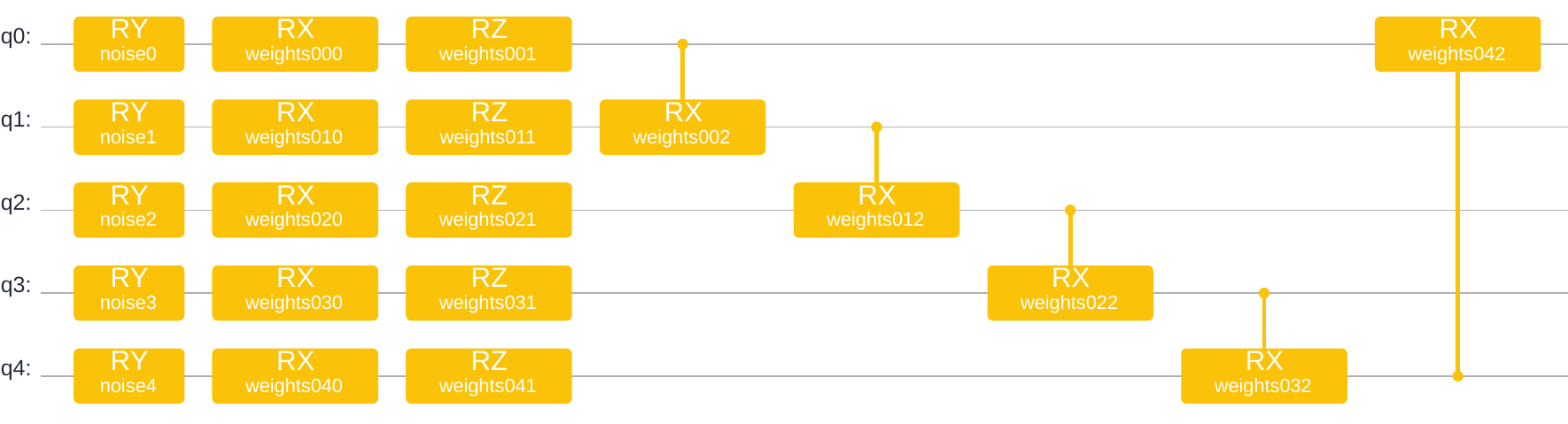}
  \caption{\label{quantum-circuit} The quantum circuit.}
\end{figure*}

\begin{lstlisting}
def quantum_circuit():
  encoder = Circuit()
  for i in range(qubits):
      encoder += RY(f'noise{i}').on(i)
  encoder = encoder.no_grad()
  ansatz = Circuit()
  for i in range(depth):
      for k in range(gate):
          for j in range(qubits):
              if k == 0:
                  ansatz += RX(f'weights{i}{j}{k}').on(j)
              elif k == 1:
                  ansatz += RZ(f'weights{i}{j}{k}').on(j)
              elif k == 2:
                  if j < 4:
                      ansatz += RX(f'weights{i}{j}{k}').on(j + 1, j)
                  else:
                      ansatz += RX(f'weights{i}{j}{k}').on(0, j)
  circuit = encoder.as_encoder() + ansatz.as_ansatz()
  return circuit
\end{lstlisting}
Stacking quantum circuits in layers and measuring the outputs constitutes a quantum layer.
\begin{lstlisting}
def quantum_layer(qubits):
    circuit = quantum_circuit()
    circ_l = Circuit()
    hams = quantum_measure(qubits)
    sim = Simulator('mqvector', circuit.n_qubits)
    sim_l = Simulator('mqvector', qubits)
    grad_ops = sim.get_expectation_with_grad(hams, circuit, circ_l, sim_l, parallel_worker=4)

    QuantumNet = MQLayer(grad_ops,
                         weight=ms.Tensor(np.random.uniform(-np.pi, np.pi, len(circuit.ansatz_params_name)),
                         dtype=ms.dtype.float32))
    return QuantumNet
\end{lstlisting}

Then, similar to the classical neural network construction, each layer of the network is defined and the forward propagation process is set to constitute the quantum generator model. The backward process automatically derives and updates the parameters.
\begin{lstlisting}
  class QuantumGenerator(nn.Cell):
    def __init__(self):
        super(QuantumGenerator, self).__init__()
        self.quantumlayer = quantum_layer(qubits)
        self.linear = nn.Dense(2**qubits, img_size * img_size, weight_init='uniform')
        self.relu = nn.ReLU()
        self.sigmoid = nn.Sigmoid()
        self.tanh = nn.Tanh()

    def construct(self, x):
        x = self.quantumlayer(x)
        x = x/(abs(x).max())
        x = self.linear(x)
        out = self.sigmoid(x)
        return out
\end{lstlisting}

The parameters of the quantum circuits are optimized according to the designed cost function. The generator needs to generate a probability distribution as real as possible, and the discriminator needs to distinguish between the real distribution and the generated distribution as much as possible. Their cost functions are designed separately as follows.
\begin{equation}
  {L_G} =  - {{\rm E}_{x\sim{p_f}(x)}}\left[ {\log D(x)} \right],
\end{equation}
\begin{equation}
  {L_D} =  - {{\rm E}_{x\sim{p_r}(x)}}\left[ {\log D(x)} \right] - {{\rm E}_{x\sim{p_f}(x)}}\left[ {\log D(x)} \right].
\end{equation}

\begin{figure}[htbp]
  \centering
  \includegraphics[scale=0.5]{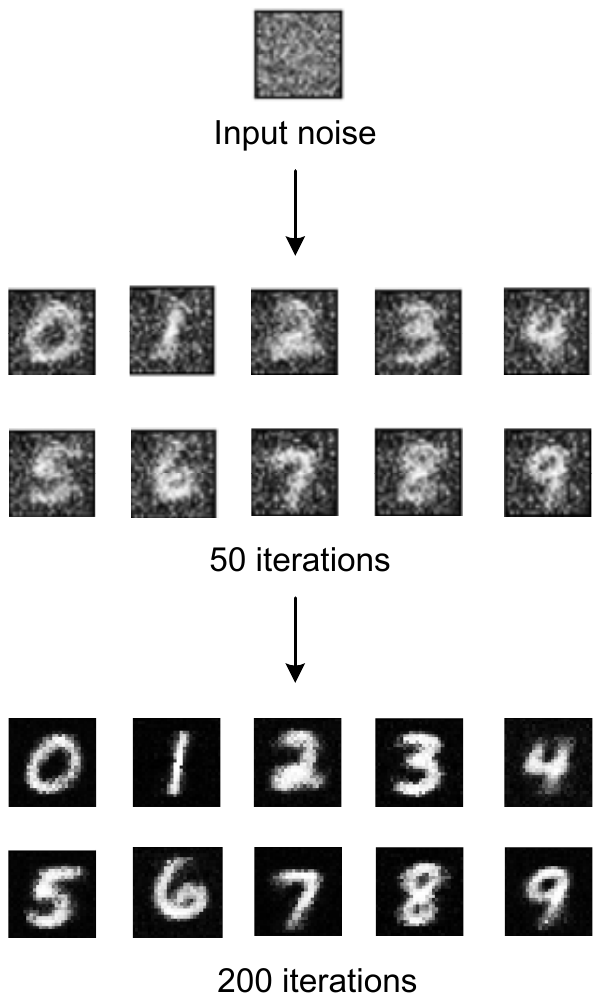}
  \caption{\label{qgan-result} The algorithms based on PQCs are optimized in classical and quantum systems.}
\end{figure}

According to the designed loss function and model, the generator and discriminator are trained and parameters are updated in turn. After training, the noise vector is used as input to the generator and the generated handwritten digits are output and the results are shown in Fig.~\ref{qgan-result}. It can be seen that as the number of iterations increases, the input noise is gradually able to produce a clear digital image. Notably, the number of iterations for training needs to be carefully designed and not more is better.

\subsection{Reinforcement Learning}
View demo code of this section: \democode{05}{5.8_Reinforcement_Learning} \ \democodegithub{05}{5.8_Reinforcement_Learning}

\subsubsection{Background}

\begin{figure}[ht]
  \centering
  \includegraphics[scale=0.4]{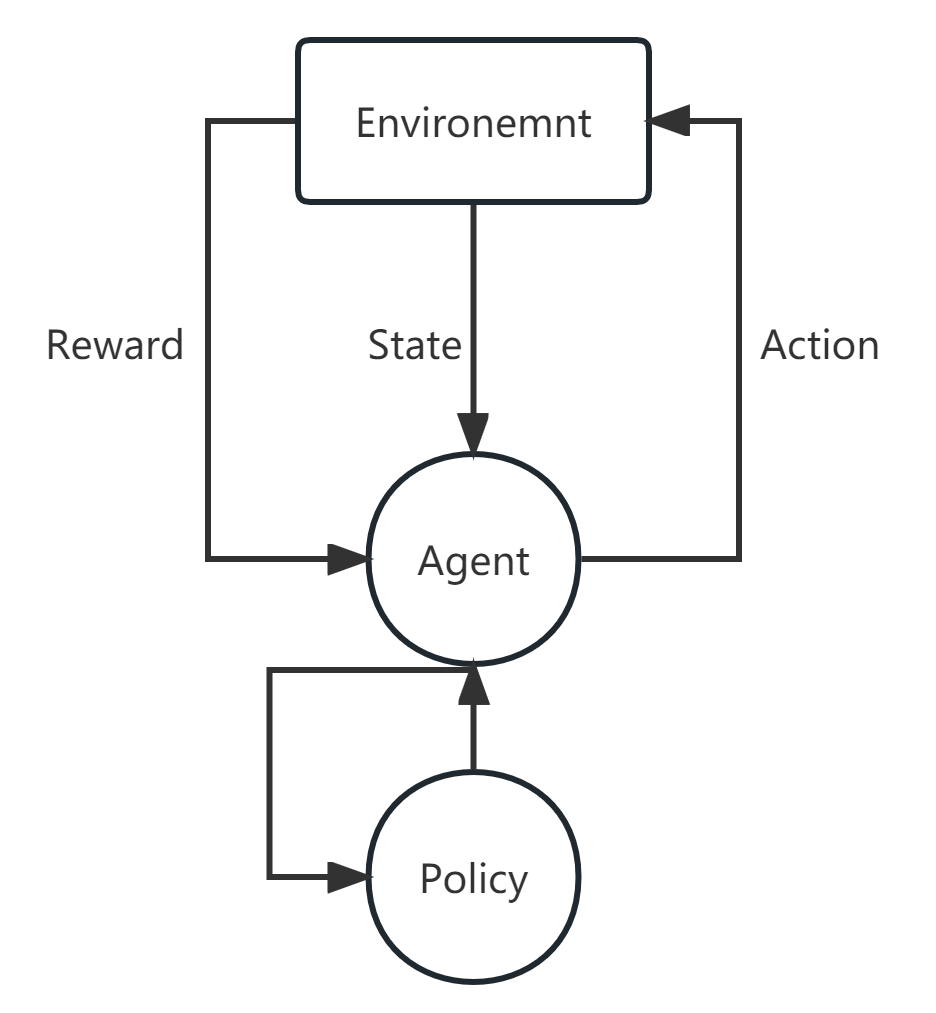}
  \caption{\label{RL_frame} The demonstration of RL working process.}
\end{figure}

An overall reinforcement learning algorithm(RL) is composed of Agent, Environment, State, Action, and Reward. After the Agent executes an action, the Environment switches to a new State according to the Action and gives a Reward according to certain rules. After that, the Agent updates its decision-making process according to the obtained new State and Reward and gives a new Action, and so on, until the entire control process is completed \cite{kaelbling1996reinforcement}. We denote $\pi_{\theta}$ as the policy function, $s$ as each state of the environment, $r$ as the reward in each state, and $a$ as the action of the agent. The architecture of RL is demonstrated in Fig.~\ref{RL_frame}.

We assumed an environment has $N$ different states. And we act action $a_{i-1}$ on state $s_{i-1}$ with the result $s_{i}$. The transition probability function from state $s_{i-1}$ to state $s_{i}$ can be $P(s_{i+1}|s_{i},a_{i})$. Meanwhile, there is a reward $r_i$ in each state calculated by reward function $R_i=R(a_i,s_i,s_{i-1})$. In the progress of environmental change, the reward expectation is only determined by the latest $state$, which is called the Markov decision process. Therefore, each $T$ step trajectory will be chosen by the agent with probability:

\begin{equation}
    \begin{split}
        &P(\tau|\pi_{\theta})=\\
        &\rho(s_0)\prod_{k=0}^{k=T-1}P(s_{k+1}|s_{k},a_{k})\pi_{\theta}(a_{k}|s_{k}),
    \end{split}
\end{equation}
where $\rho$ is the initial state distribution. Then the expectation of $T$ steps trajectory's reward is:
\begin{equation}
    \eta(\pi_{\theta})=\int_{\tau}P(\tau|\pi_{\theta})R(\tau)d\tau.
\end{equation}
In order to get the optimal policy function $\pi_{\theta}$, the agent is trained to maximize the expectation $\eta$, which is:
\begin{equation}
    \pi_{\theta}^{opt}=\mathrm{argmax}_{\theta}\eta(\pi_{\theta}).
\end{equation}
In following sections, we will demonstrate how to train the agent in quantum device.

\subsubsection{Hybrid Deep Q-Network}

\begin{figure*}[ht]
    \centering
    \includegraphics[width=0.9\linewidth]{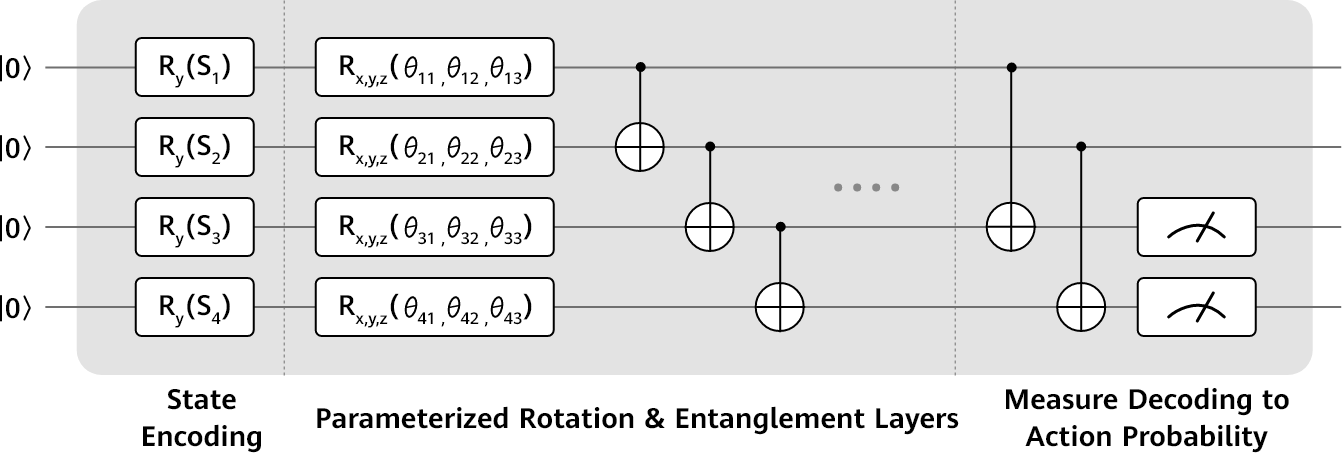}
    \caption{\label{HDQN} The architecture of HDQN.}
  \end{figure*}

Replace the neural network in the classical Deep Q-network's agent with the quantum variational circuit, we get the hybrid deep Q-network(HDQN)~\cite{Q_rl}. The architecture of HDQN is illustrated in FIG.~\ref{HDQN}.

We calculate the parameters $\theta$ by classical neural network and use them to construct the quantum variational circuit. Finally, the quantum part applies the action to the environment. Meanwhile, the environment responds with a reward $r_i$. We denote $e_i=(s_i,a_i,r_{i+1},s_{i+1})$ as an experience. In the training stage, we will use these experiences as training samples to adjust the parameters in both the classical part and quantum part according to loss function $L(\theta)$:
\begin{equation}
\begin{split}
    L(\theta_{j})=&E[R_{i+1}+\\
    &\gamma \max_{\alpha^{'}}Q(s_{i+1},a^{'},\theta_{j}^{-})-Q(s_{i},a_{i},\theta_{i})^2].
\end{split}
\label{loss func}
\end{equation}

\begin{figure*}[ht]
  \centering
  \includegraphics[scale=0.5]{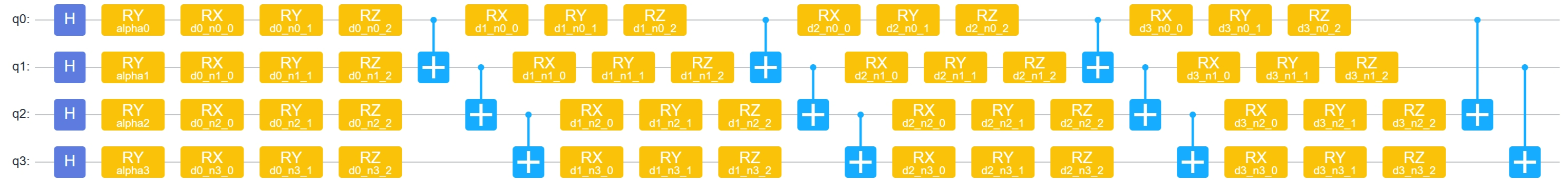}
  \caption{\label{Agent_qc} The overall quantum circuit of agent in RL.}
\end{figure*}

\textit{Variational quantum circuit} -- The variational quantum circuit is a quantum circuit using adjustable parameters that can be changed by the external environment. Here, according to the architecture in~\cite{Q_rl}, we construct the quantum circuit with the combination of encoder and ansatz. Here, we apply the Hardware Efficient circuit as the ansatz. The overall quantum circuit is illustrated in Fig.~\ref{Agent_qc}.

\textit{The implement in MindQuantum} -- Here, we give the example of realizing HDQN by \MindQuantum. The first step is to construct the quantum variational circuit including the encoder and the ansatz.
\begin{lstlisting}
def __get_encoder(self):
    encoder = Circuit()
    encoder += UN(H, 4)
    for i in range(4):
        encoder += RY(f'alpha{i}').on(i)
    encoder = encoder.no_grad()
    quantum circuit which has no gradient: no_grad()
    encoder.summary()
    return encoder
def __get_ansatz(self):
    ansatz = HardwareEfficientAnsatz\\
        (4, single_rot_gate_seq=[RX,RY,RZ],\\
        entangle_gate=X, depth=3).circuit
    ansatz += X.on(2,0)
    ansatz += X.on(3,1)
    ansatz.summary()
    return ansatz
self.circuit = self.encoder + self.ansatz
\end{lstlisting}

After the progress of the quantum circuit, we need to construct an observable Hamiltonian as a measurement basis. Here, we only measure the second and third qubits on $\sigma_y$ base. The Hamiltonian is as follows:
\begin{equation}
    H=I\otimes I\otimes\sigma_y\otimes\sigma_y.
\end{equation}
We can construct the Hamiltonian by \MindQuantum\ as follows:
\begin{lstlisting}
def __get_observable(self):
    hams=[Hamiltonian(QubitOperator(f'Y{i}'))\\
        for i in [2, 3]]
    return hams
\end{lstlisting}
Moreover, we can easily get the expectation and calculate the gradient of expectation by function \getexpectationwithgrad, which can be used to optimize parameters.

Next, we need to realize the classical part, which is a three-layers fully connected neural network with activation function rule.
\begin{lstlisting}
class Critic(nn.Cell):
    def __init__(self):
        super(Critic, self).__init__()
        self.relu = nn.ReLU()
        self.fc1 = nn.Dense(4, 64)
        self.fc2 = nn.Dense(64, 256)
        self.fc3 = nn.Dense(256, 1)

    def construct(self, x):
        x = self.relu(self.fc1(x))
        x = self.relu(self.fc2(x))
        x = self.fc3(x)
        return x
\end{lstlisting}

\textit{Environment} -- Here, we focus on the problem of the gym game \code{"CartPole-v0"}. It's a game that has an unbalanced cart on a track. This game has two inputs $0$ and $1$, which correspond to two directions left and right respectively. Users should control the trace moving toward these two directions and keep the cart balanced as long as possible.

The first step is to create the environment by gym.
\begin{lstlisting}
import gym
env = gym.make('CartPole-v0')
\end{lstlisting}
\textit{Memory buffer} -- In the progress of training, the training samples are randomly selected from the previous experience $e_i$. Therefore, we need to construct a buffer to store and sample the experience $e_i$.
\begin{lstlisting}
class Memory(object):
    ......
\end{lstlisting}

The full implementation can be found at the beginning of this section.

\subsubsection{Training}
The final step is model training. At first, we initialize HDQN with random parameters and let the actor interact with the environment. Next, we store the interaction result in the memory buffer as training samples. Finally, we sample from the memory buffer and adjust the parameters both in the classical part and quantum part with loss function (\ref{loss func}). Because of the space limitation, we will not provide the detailed code here.

\begin{figure}[ht]
  \centering
  \includegraphics[scale=0.3]{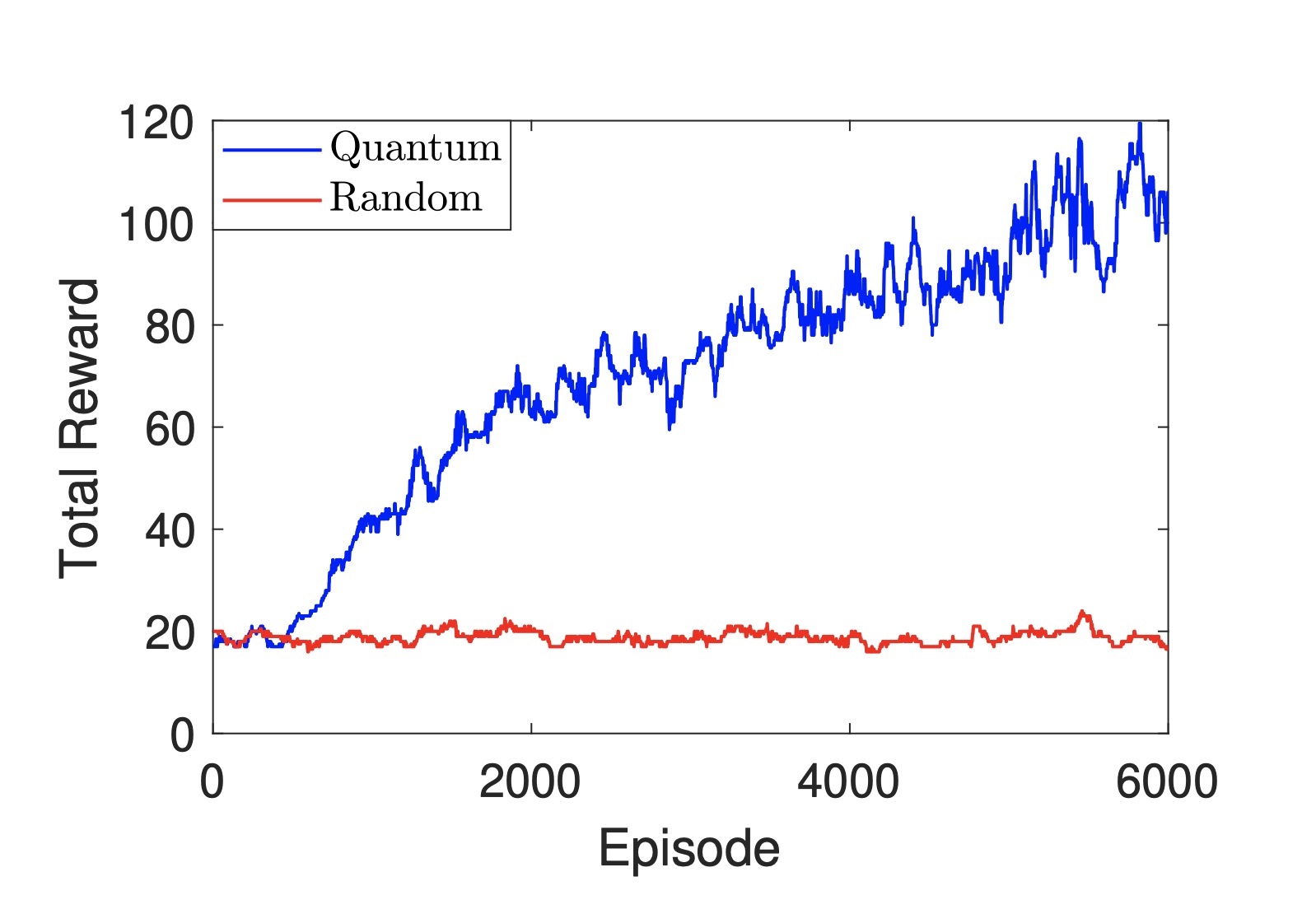}
  \caption{\label{stochastic} Comparison between RL and randomized strategy.}
\end{figure}

\textit{Result of numerical simulation} -- In order to evaluate the result of the HDQN. We do some experiments to evaluate the performance. As shown in Fig.~\ref{stochastic} We compare the average results from multiple experiments with randomized strategy experiments. The results show that reinforcement learning using quantum circuits as the agent has indeed learned the strategy.

\subsection{Quantum Singular Value Decomposition}
View demo code of this section: \democode{05}{5.9_Quantum_Singular_Value_Decomposition} \ \democodegithub{05}{5.9_Quantum_Singular_Value_Decomposition}

\subsubsection{Background}
Singular Value Decomposition (SVD) is an important matrix decomposition in linear algebra. As a generalization of eigendecomposition on arbitrary dimensional matrices, SVD is widely used in the field of machine learning, including matrix compression, recommendation systems, and natural language processing. It is defined as follows:
Given a complex matrix $M \in \mathbb{C}^{m \times n}$, define the SVD of the matrix $M$ as $M = UDV^\dagger$. Where $U$ is $m \times m$ unitary matrix, $V$ is $n \times n$ unitary matrix, that is, satisfy $UU^\dagger = I$, $VV^\dagger = I$. $D$ is a diagonal matrix $m \times n$, usually with the elements of the main diagonal arranged from largest to smallest, and each element is called a singular value of the matrix $M$.

Quantum algorithms for SVD have been proposed in \cite{kerenidis2016quantum, rebentrost2018quantum}, which leads to applications in solving linear systems of equations \cite{wossnig2018quantum} and developing quantum recommendation systems \cite{kerenidis2016quantum}. However, these algorithms are too costly to be convincingly validated for near-term quantum devices. The leading strategy to solve various problems using noisy intermediate-scale quantum (NISQ) devices are called variational quantum algorithms.

A variational quantum algorithm is proposed for singular value decomposition (VQSVD) \cite{wang2021variational}, by formulating the task of SVD as an optimization problem. The detailed VQSVD algorithm is as follows:
\begin{enumerate}
    \item Prepare the input of VQSVD algorithm:
        \begin{itemize}
            \item A decomposition of the matrix $M$ into a linear combination of $K$ unitary matrices of the form $M = \sum_{k=1}^K c_k A_k$ with real numbers $c_k$.
            \item Positive weights $q_1 > \cdots > q_r > 0$.
            \item Computational basis {$\ket{\psi_1}, \cdots, \ket{\psi_T}$}, where $T$ is the desired rank.
            \item Parameterized circuits $U(\theta)$, $V(\phi)$ with initial parameters of $\theta, \phi$.
        \end{itemize}
    \item Enter a hybrid quantum-classical optimization loop to train the parameters $\theta$ and $\phi$ in the parameterized quantum circuits $U(\theta)$ and $V(\phi)$, compute the singular values of $M$: $m_j = \text{Re}\langle\psi_j|U(\theta)^{\dagger} M V(\phi)\ket{\psi_j}$. The goal is to maximum the designed loss $L(\theta, \phi)$:
    \begin{equation}
        L(\theta, \phi) = \sum_{j=1}^T q_j \times \text{Re} \langle\psi_j| U(\theta)^{\dagger} M V(\phi)\ket{\psi_j},
    \end{equation}
    where weights are added to make the calculated singular values descending.
    \item Obtain optimal parameters $\alpha^ \star$ and $\beta^\star$ and compute $U(\alpha^\star)$ and $V(\beta^\star)$.
    \item Obtain approximate singular values ${m_1, \cdots, m_r}$ and  singular vectors $U(\alpha^\star)$ and $V(\beta^\star)$.
\end{enumerate}

\subsubsection{Implementation} -- In this section, we show how to decompose a randomly generated $8 \times 8$ complex matrix using \MindQuantum.

First we need to introduce the required packages, define the required constants and set the weights, then use the \code{numpy.random.randint} function to randomly generate an $8 \times 8$ complex matrix $M$:
\begin{lstlisting}
from mindquantum import Simulator, MQAnsatzOnlyLayer, add_prefix
from mindquantum import Hamiltonian, Circuit, RY, RZ, X
import mindspore as ms
import numpy as np
from scipy.sparse import csr_matrix
from scipy.linalg import norm
from matplotlib import pyplot
import tqdm

n_qubits = 3  # qbits number
cir_depth = 20  # circuit depth
N = 2**n_qubits
rank = 8  # learning rank
step = 3
ITR = 200  # iterations
LR = 0.02  # learning rate

# Set equal learning weights
if step == 0:
    weight = ms.Tensor(np.ones(rank))
else:
    weight = ms.Tensor(np.arange(rank * step, 0, -step))

# Define random seed
np.random.seed(42)

def mat_generator():
    '''
    Generate a random complex matrix
    '''
    matrix = np.random.randint(
        10, size=(N, N)) + 1j * np.random.randint(10, size=(N, N))
    return matrix

# Generate matrix M which will be decomposed
M = mat_generator()
# m_copy is generated to error analysis
m_copy = np.copy(M)
print('Random matrix M is: ')
print(M)
# Get SVD results
U, D, v_dagger = np.linalg.svd(M, full_matrices=True)
\end{lstlisting}

Next define a hardware-efficient ansatz used in the simulation for $U(\theta)$ and $V(\phi)$, the variational ansatz used in the paper \cite{wang2021variational}. Since only real matrices are involved, the combination of rotation gates and CNOT is sufficient. In the ansatz, each same block(denoted in the dashed-line box) consists of a column of single-qubit rotations about the y-axis and z-axis following by a layer of CNOT gates which only connects the adjacent qubits.
\begin{lstlisting}
class Ansatz:
    def __init__(self, n, depth):
        self.circ = Circuit()
        num = 0
        for _ in range(depth):
            for i in range(n):
                self.circ += RY('theta' + str(num)).on(i)
                num += 1
            for i in range(n):
                self.circ += RZ('theta' + str(num)).on(i)
                num += 1
            for i in range(n - 1):
                self.circ += X.on(i + 1, i)
            self.circ += X.on(0, n - 1)
\end{lstlisting}

Then define a quantum network using the given Hamiltonian. The \setqs function of the simulator is used to set the state of the simulator to the given computational basis, so that we can obtain the measurement results under the basis, that is, the corresponding singular values. The \getexpectationwithgrad function is used to compute the gradient of the parameters in the circuit and the value of the following expression: $E(\theta) = \langle\phi| U_l^{\dagger}(\theta) H U_r(\theta)\ket{\psi}$. Then we can use \MQAnsatzOnlyLayer to build a quantum network layer based on a given basis, and its output is: $\text{Re}\langle\psi_j|U(\theta)^{\dagger} M V(\phi)\ket{\psi_j}$.
\begin{lstlisting}
def quantnet(qubits_num, hams, circ_right, circ_left=None, base=None):
    sim = Simulator('mqvector', qubits_num)
    if base is None:
        pass
    else:
        sim.set_qs(base)
    grad_ops = sim.get_expectation_with_grad(hams, circ_right, circ_left)

    quantumnet = MQAnsatzOnlyLayer(grad_ops, 'ones')
    return quantumnet
\end{lstlisting}

After sparring the decomposed $8 \times 8$ matrix $M$ and generating the corresponding Hamiltonian $H$, we can then instantiate ansatz $U_\text{ansatz}$ and $V_\text{ansatz}$ and build a required quantum network layer, which can be used to compute $\{m_j\}_{j=1}^T$, the singular values of $M$.
\begin{lstlisting}
u_ansatz = add_prefix(Ansatz(n_qubits, cir_depth).circ, 'u')
v_ansatz = add_prefix(Ansatz(n_qubits, cir_depth).circ, 'v')
ham = Hamiltonian(csr_matrix(M))
i_matrix = np.identity(N)
quantum_models = dict()
quantum_models['net_0'] = quantnet(n_qubits, ham, v_ansatz, u_ansatz, i_matrix[0])
for s in range(1, rank):
    quantum_models["net_" + str(s)] = quantnet(n_qubits, ham, v_ansatz, u_ansatz, i_matrix[s])
    quantum_models["net_" + str(s)].weight = quantum_models['net_0'].weight
\end{lstlisting}

Additionally, we can use MindSpore to build a hybrid quantum-classical network to realize weighted summation of quantum network layers and compute $L(\theta,\phi) = \sum_{j=1}^T q_j \times \text{Re} \langle\psi_j| U(\theta)^{\dagger} M V(\phi)\ket{\psi_j}$.
\begin{lstlisting}
class MyNet(ms.nn.Cell):
    '''
    define quantum-classic net
    '''
    def __init__(self):
        super(MyNet, self).__init__()

        self.build_block = ms.nn.CellList()
        for j in range(rank):
            self.build_block.append(quantum_models["net_" + str(j)])

    def construct(self):
        x = self.build_block[0]() * weight[0]
        k = 1
        for layer in self.build_block[1:]:
            x += layer() * weight[k]
            k += 1
        return -x
\end{lstlisting}
Now we can instantiate the hybrid quantum-classical network and start training using MindSpore:
\begin{lstlisting}
net = MyNet()
# Define optimizer
opt = ms.nn.Adam(net.trainable_params(), learning_rate=LR)
# Simple gradient descent
train_net = ms.nn.TrainOneStepCell(net, opt)
# Start training
loss_list = list()
for itr in tqdm.tqdm(range(ITR)):
    res = train_net()
    loss_list.append(res.asnumpy().tolist())
\end{lstlisting}

Finally, read the training results (the singular values) and compare them with the results of the classical singular value decomposition:
\begin{lstlisting}
singular_value = list()
for _, qnet in quantum_models.items():
    singular_value.append(qnet().asnumpy()[0])

print('Predicted singular values from large to small:', singular_value)
print("True singular values from large to small:", D)
\end{lstlisting}
Output:
\begin{lstlisting}
Predicted singular values from large to small: [54.83174, 19.169168, 14.88653, 11.093878, 10.533753, 7.648352, 5.5560594, -0.3320913]
True singular values from large to small: [54.83484985 19.18141073 14.98866247 11.61419557 10.15927045  7.60223249 5.81040539  3.30116001]
\end{lstlisting}

Intuitively, we can see that the singular values learned by using the hybrid quantum-classical network are similar to the real singular values.

\section{QuPack: Acceleration Engine}
\label{sec:qupack}
Inspired by LINPACK\cite{dongarra1979linpack}, we aim to establish a quantum acceleration library in the field of quantum computing to enhance the development and validation of quantum algorithms. Presently, it is available for free usage on the HiQ quantum computing cloud platform and has been named \QuPack.

\QuPack\ focuses on accelerating quantum algorithms in specific domains, such as VQE, QAOA, Quantum Pulse Engineering, and high-performance tensor network simulators. Building upon the policy-based design pattern of \MindQuantum, \QuPack\ will further optimize these policies to enhance the performance of quantum simulators in different hardware devices. In this chapter, we will introduce some essential modules within \QuPack\ to facilitate developers in getting started quickly.

\begin{figure}[ht]
    \centering
    \includegraphics[scale=0.5]{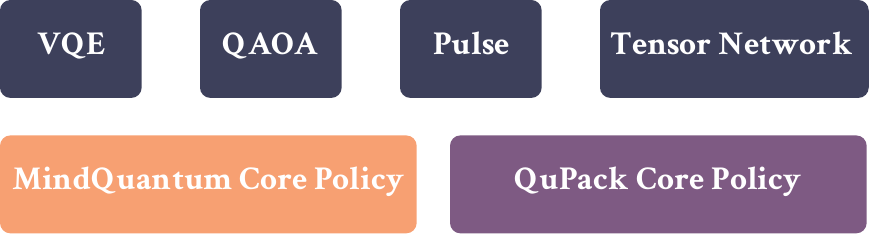}
    \captionsetup{justification=raggedright,singlelinecheck=false}
    \caption{\label{6_qupack_layer} The structure of \QuPack. Based on core simulation policy in \MindQuantum\ and specific designed policy in \QuPack, we are able to build high performance application like VQE, QAOA, Pulse Engineering and Tensor Network simulator.}
\end{figure}

\subsection{Variational Quantum Eigensolver}
\subsubsection{Background}
Simulating molecule system with high accuracy in VQE is both a time and memory consuming task. The memory used to simulate hydrogen chains in different sizes is shown in table~\ref{tab:memory_h_chain}

\begin{table}[ht]
    \begin{tabular}{ccc}
        \toprule
        Hydrogen chain & Qubit Number & Memory with Complex128 \\
        \midrule
        $H_2$          & 4            & 0.256kB                \\
        $H_6$          & 12           & 64kB                   \\
        $H_{10}$       & 20           & 16MB                   \\
        $H_{14}$       & 28           & 4GB                    \\
        $H_{18}$       & 36           & 1TB                    \\
        \bottomrule
    \end{tabular}
    \caption{Memory consumption for storing full amplitudes quantum state.}
    \label{tab:memory_h_chain}
\end{table}

\begin{figure}
    \centering
    \begin{subfigure}{0.3\textwidth}
        \centering
        \includegraphics[width=\textwidth]{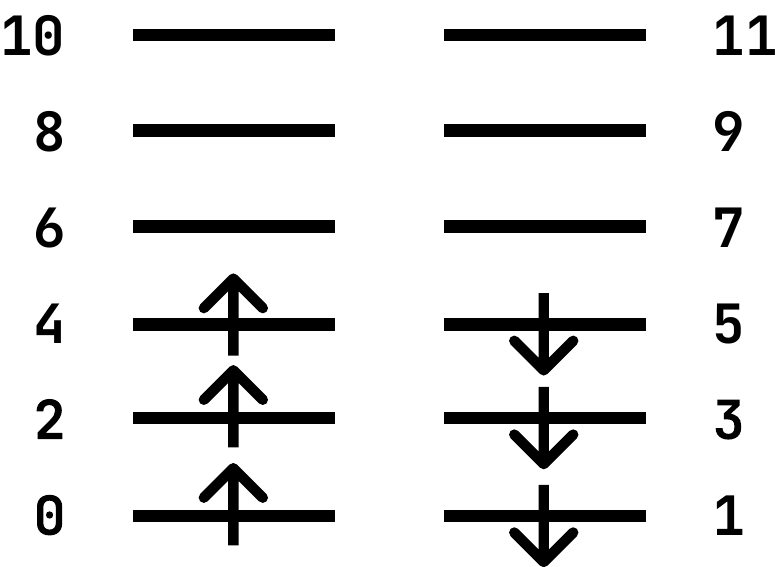}
        \caption{Hartree-Fork state of $H_6$.}
        \label{6_1_h6_a}
    \end{subfigure}
    \begin{subfigure}{0.3\textwidth}
        \centering
        \includegraphics[width=0.9\textwidth]{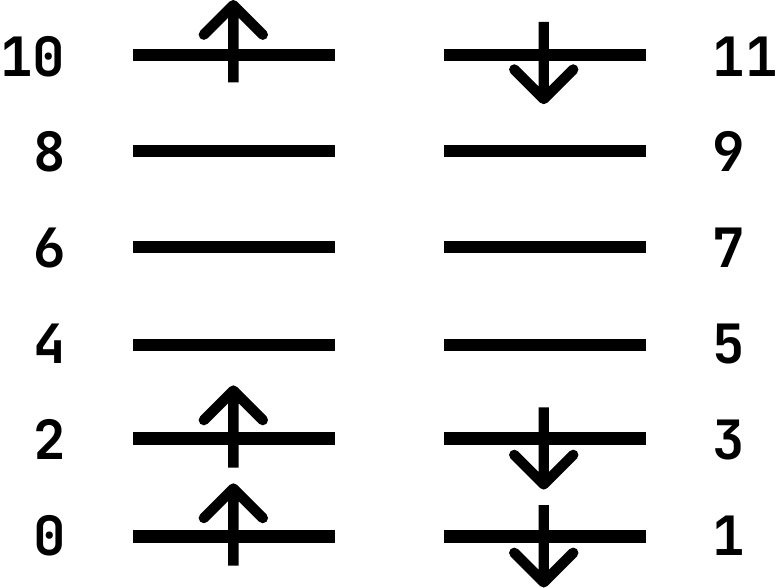}
        \caption{Spin-orbital after applying excitation operator $\hat{T}=\hat{a}_{11}^\dagger \hat{a}_{10}^\dagger \hat{a}_5 \hat{a}_4$.}
        \label{6_1_h6_b}
    \end{subfigure}
\end{figure}

Based on UCCSD (Unitary coupled-cluster with singles and doubles) theory, the ansatz is described as:
\begin{equation}
    |\Psi(\vec{\theta})\rangle = e^{(\hat{T}-\hat{T}^\dagger)}\ket{\psi_\text{HF}},
\end{equation}
where $\hat{T}$ is couple cluster operator, and $\ket{\psi_\text{HF}}$ is Hartree-Fork state. The couple cluster operator is written as:
\begin{equation}
    \hat{T}=\sum_{p\notin\text{occ},q\in\text{occ}}\theta_q^p \hat{a}_p^\dagger \hat{a}_q + \sum_{pq\notin\text{occ},rs\in\text{occ}}\theta_{rs}^{pq}\hat{a}_p^\dagger \hat{a}_q^\dagger \hat{a}_r \hat{a}_s,
\end{equation}
and under one order trotterization the ansatz can be decomposed to:
\begin{equation}
    |\Psi(\vec{\theta})\rangle\approx \prod_ie^{(\hat{T}_i-\hat{T}_i^\dagger)}\ket{\psi_\text{HF}}.
\end{equation}
Taking hydrogen chain $H_6$ for example, we need 12 qubits to simulate this system and there are 6 electrons with 3 spin up and 3 spin down. Fig.~\ref{6_1_h6_a} shows the Hartree-Fork state of $H_6$ and Fig.~\ref{6_1_h6_b} shows the spin-orbital under excitation operator $\hat{T}=\hat{a}_{11}^\dagger \hat{a}_{10}^\dagger \hat{a}_5 \hat{a}_4$.
\subsubsection{Electron and spin conservation}
In \QuPack, we impose the constraint that the total number of electrons and the total spin number of electrons remain constant throughout the evolution of the ansatz. This constraint significantly reduces the dimension of the Hilbert space that we need to simulate. Suppose the qubit number is $n_q$ and the electron number is $n_e$, the dimension of full amplitude state is $2^{n_q}$, but after electron and spin conservation, the dimension will be:
\begin{equation}
    \binom{n_q/2}{n_e/2}^2=\left(\frac{(n_q/2)!}{(n_e/2)!(n_q/2-n_e/2)!}\right)^2.
\end{equation}
The memory consumption will reduce from 1TB to $\sim$35GB.

Besides the memory reduction, we also optimized the evolution of coupled cluster operator. For a given coupled cluster operator $\hat{V}(\theta)=e^{\theta(\hat{T}-\hat{T}^\dagger)}$, it is easy to prove that $(\hat{T}-\hat{T}^\dagger)^3 = - (\hat{T}-\hat{T}^\dagger)$ and $(\hat{T}-\hat{T}^\dagger)^4 = -(\hat{T}-\hat{T}^\dagger)^2$. After Taylor expansion, we have:
\begin{equation}
    \hat{V}(\theta) = \mathbb{I} + ( 1-\cos\theta)(\hat{T}-\hat{T}^\dagger)^2 + \sin\theta (\hat{T}-\hat{T}^\dagger).
\end{equation}
Choose a calculation base in reduced Hilbert space $\ket{i}$, so that $\hat{T}\ket{i} = \ket{j}\neq0$, then we have $\hat{T}^\dagger\ket{i}=0$ and $\hat{T}^\dagger\ket{j}=\ket{i}$. The evolution of coupled cluster can be simplified to:
\begin{equation}
    e^{\theta(\hat{T}-\hat{T}^\dagger)}\begin{pmatrix}
        \ket{i} \\\ket{j}
    \end{pmatrix}=\begin{pmatrix}
        \cos\theta  & \sin\theta \\
        -\sin\theta & \cos\theta
    \end{pmatrix}\begin{pmatrix}
        \ket{i} \\\ket{j}
    \end{pmatrix}.
\end{equation}
And for $e^{i\theta(\hat{T}+\hat{T}^\dagger)}$, we have
\begin{equation}
    e^{i\theta(\hat{T}+\hat{T}^\dagger)}\begin{pmatrix}
        \ket{i} \\\ket{j}
    \end{pmatrix}=\begin{pmatrix}
        \cos\theta   & i\sin\theta \\
        -i\sin\theta & \cos\theta
    \end{pmatrix}\begin{pmatrix}
        \ket{i} \\\ket{j}
    \end{pmatrix}.
\end{equation}

When doing variational quantum algorithm on this system, we need to calculate the gradient of $\hat{V}(\theta)$, which is given as:
\begin{equation}
    \frac{\partial \hat{V}(\theta)}{\partial\theta}\begin{pmatrix}
        \ket{i} \\\ket{j}
    \end{pmatrix}=\begin{pmatrix}
        -\sin\theta & \cos\theta  \\
        -\cos\theta & -\sin\theta
    \end{pmatrix}\begin{pmatrix}
        \ket{i} \\\ket{j}
    \end{pmatrix}.
\end{equation}

Here is a demonstration code of how to build a quantum circuit of evolution of coupled cluster operator.

\begin{lstlisting}
from qupack.vqe import ExpExciteGate
from mindquantum.algorithm.nisq import uccsd_singlet_generator
from mindquantum.core.circuit import Circuit

ucc_fermion_ops = uccsd_singlet_generator(6, 4)
circ = Circuit()
for term in ucc_fermion_ops:
    circ += ExpExciteGate(term)
\end{lstlisting}

A specific simulator, known as \code{ESConservation}, has been designed to efficiently simulate the coupled cluster operators within a subspace of the Hilbert space. This specialized simulator, which shares the same API as the simulator in \MindQuantum, excels in accurately modeling the intricate dynamics of these operators.

\subsection{Quantum Approximate Optimization Algorithm}

\subsubsection{Introduction}
The Quantum Approximate Optimization Algorithm (QAOA) stands as one of the most promising Variational Quantum Algorithms (VQAs) that has garnered significant attention in recent years. QAOA is specifically designed for the purpose of finding approximate solutions to challenging combinatorial optimization problems using quantum computers. It achieves this by encoding the problem's associated Hamiltonian into a quantum circuit. QAOA then leverages adiabatic time evolution and layering techniques to optimize the variational parameters of the circuit. This optimization aims to construct an approximate solution to the problem, which can be obtained by measuring the QAOA circuit with the optimal parameter set.

\subsubsection{Basic Usage}
The QAOA class requires three input parameters:
\begin{itemize}
    \item n\_qubits: Corresponding to the number of qubits required for the QAOA circuit.
    \item depth: The number of layers in the ansatz circuit.
    \item h: The Hamiltonian associated with the problem to be solved.
\end{itemize}

Once these three parameters are provided, the simulator will generate the corresponding parameterized circuit.

Next, we'll illustrate how to use the QAOA simulator with a specific example of a max-cut problem. Firstly, we determine the circuit's size and generate the graph to be solved. In this case, we choose to compute a graph with 5 nodes (requiring 5 qubits), and set the depth of the ansatz circuit to 10 layers.

\begin{lstlisting}
from qupack.qaoa import QAOA
import networkx as nx

depth = 10
n_qubits = 5
graph = nx.complete_graph(n_qubits)
graph = list(graph.edges)
\end{lstlisting}

Next, we construct the Hamiltonian corresponding to the problem. It's important to note that in the QAOA problem, the Hamiltonian is diagonalized, meaning the Pauli operators in the input \code{QubitOperator} should only include Z. If the \code{QubitOperator} contains X or Y, the QAOA simulator will produce an error report.

\begin{lstlisting}
from mindquantum.core.operators import QubitOperator

ham = QubitOperator()
for node in graph:
    ham += QubitOperator('Z{} Z{}'.format(*node), 1)
\end{lstlisting}

Now, we can instantiate a QAOA simulator.
\begin{lstlisting}
sim = QAOA(n_qubits, depth, ham)
\end{lstlisting}

There are the following interfaces that can be called to implement various functions:

\begin{itemize}
    \item \code{get_expectation_with_grad()}: Based on the given quantum gate parameters, obtain the expectation value of the final state of the circuit with respect to the Hamiltonian, along with the gradients of each parameter.
    \item \code{get_expectation()}: Based on the given quantum gate parameters, obtain the expectation value of the evolved quantum state with respect to the Hamiltonian.
    \item \code{evolution()}: Run the circuit to evolve the quantum state based on the given quantum gate parameters.
    \item \code{get_qs()}: Get the current quantum state of the simulator.
    \item \code{set_qs()}: Set the current quantum state of the simulator.
\end{itemize}

The quantum gate parameters consist of two components: \textit{gamma} and \textit{beta}. \textit{Gamma} represents the parameters for Rzz gates, while \textit{beta} represents the parameters for RX gates. These parameters should be provided as one-dimensional lists or \code{numpy.ndarray}. Each entry in the array corresponds to the parameters for all Rzz (or RX) gates in one layer of the ansatz circuit. Therefore, the lengths of \textit{gamma} and \textit{beta} should match the depth.

Below, we will set all Rzz gate parameters to 1 and all RX gate parameters to 2, for the purpose of demonstrating these interfaces.

\begin{lstlisting}
gamma = [1] * depth
beta = [2] * depth
expectation, gamma_grad, beta_grad = sim.get_expectation_with_grad(gamma, beta)
print("expectation: ", expectation)
print("gradient with respect to gamma: ", gamma_grad)
print("gradient with respect to beta: ", beta_grad)
\end{lstlisting}
Output:
\begin{lstlisting}
expectation:  2.4222583315861312
gradient with respect to gamma:  [21.06869750223758, -46.18915014748717, 9.446358158199573, 11.021141208662032, -28.573493584250016, 35.674550037397935, -37.62868932163818, -13.136244193793484, 29.942571461897476, -21.026027385573684]
gradient with respect to beta:  [-3.264446786318293, 9.713011195857428, -11.622454620714056, 10.958548703603533, -2.4125637780460947, -4.314674707938554, 10.2451658203689, -9.335013624102858, 7.359235130728157, -1.431435776010227]
\end{lstlisting}

\subsection{Tensor Network Simulator}
\subsubsection{Introduction}
In recent years, NISQ devices have undergone great development. Different classes of simulators on classical computers have blossomed to address the challenging task of simulating large quantum circuits. Currently, simulators for quantum circuits can be categorized into two primary categories:
\begin{itemize}
    \item state vector simulation: The state vector approach simulates a direct evolution of the quantum state. In \MindQuantum, the \code{"mqvector"} is in this category. However, this approach stores the full state information, leading to an exponential growth in memory requirements.
    \item tensor network simulation: The tensor approach performs the simulation by contracting a tensor network \cite{Huggins_2019,PhysRevLett.128.030501}.
\end{itemize}

In the tensor network approach, the quantum circuit is represented as a tensor network, with one-qubit gates described as rank-2 tensors, two-qubit gates as rank-4 tensors, and $n$-qubit gates as rank-2$n$ tensors. Simulating the quantum circuit is consequently transformed into the task of contracting the associated tensor network.

A~\emph{tensor} of rank $r$ is a~multidimensional array $T[i_1,\dots,i_r] \equiv T[\mathbf{i}]$ with complex entries, where the~indices $(i_1,\dots,i_r) \equiv \mathbf{i}$ are usually called \emph{legs}, and the dimension of each leg is called its \emph{bond dimension}. The~\emph{shape} of the~tensor $T[i_1,\dots,i_r]$ is the~vector $(d_1,\dots,d_r)$, where each $d_j$ is the~bond dimension of the~tensor leg $i_j$;  $j=\overline{1,r}$. For example, a~vector $T[i_1]$ of length $n$ is an~order~$1$ tensor of shape~$(n)$, and an~$m\times n$ matrix $T[i_1,i_2]$ is an~order~$2$ tensor of shape~$(m, n)$.

Later, we would be interested in evaluating the summation of a collection of tensors $T_1[\mathbf{i}_1],\dots, T_m[\mathbf{i}_m]$ sharing some common legs,
\begin{equation}\label{eq:sum}
    {\rm{sum}}=\sum_{j_1,\dots,j_s} T_1[\mathbf{i}_1]\cdots T_m[\mathbf{i}_m],
\end{equation}
where the~sum is over all possible values of the~legs $j_1,\dots,j_s$, which we call the \emph{closed} legs. All the~rest legs of the~tensors $T_1,\dots,T_m$ are called \emph{open}.

\begin{figure}
    \centering
    \begin{subfigure}{0.5\textwidth}
        \centering
        \includegraphics[width=\textwidth]{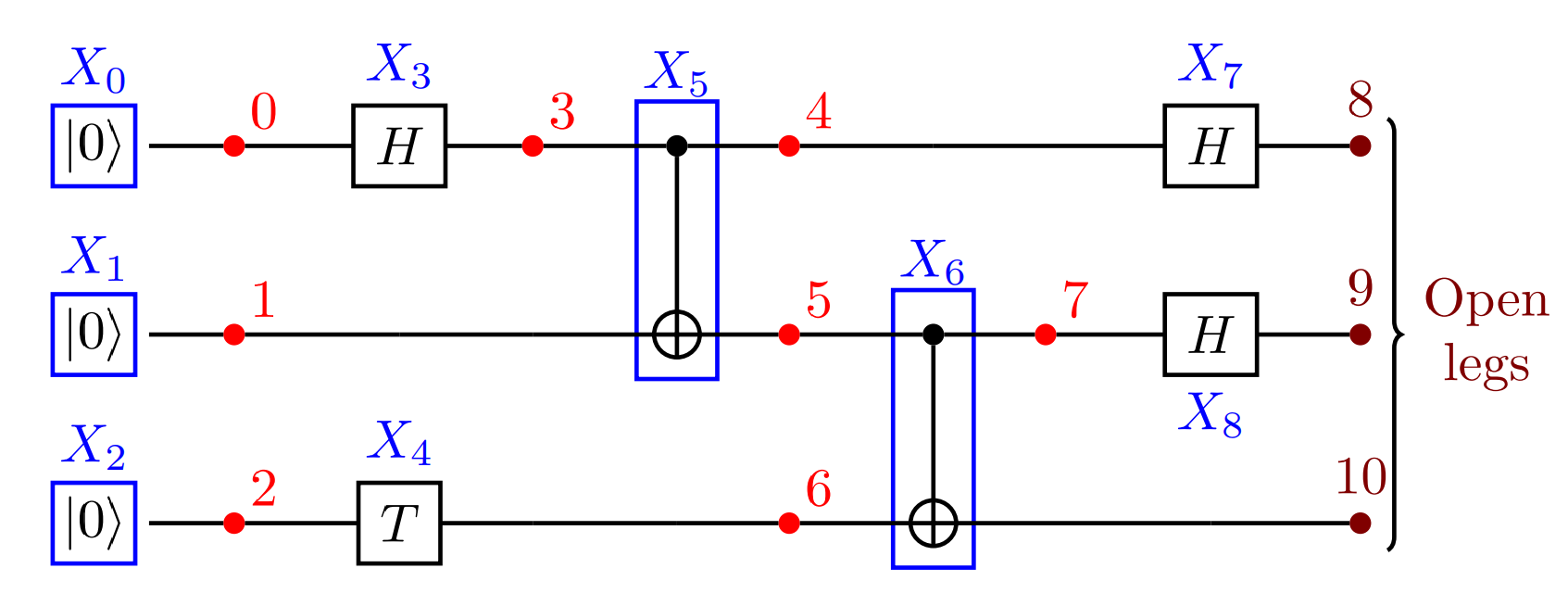}
        \caption{The quantum circuit.}
        \label{6_3_circ}
    \end{subfigure}
    \begin{subfigure}{0.4\textwidth}
        \centering
        \includegraphics[width=0.9\textwidth]{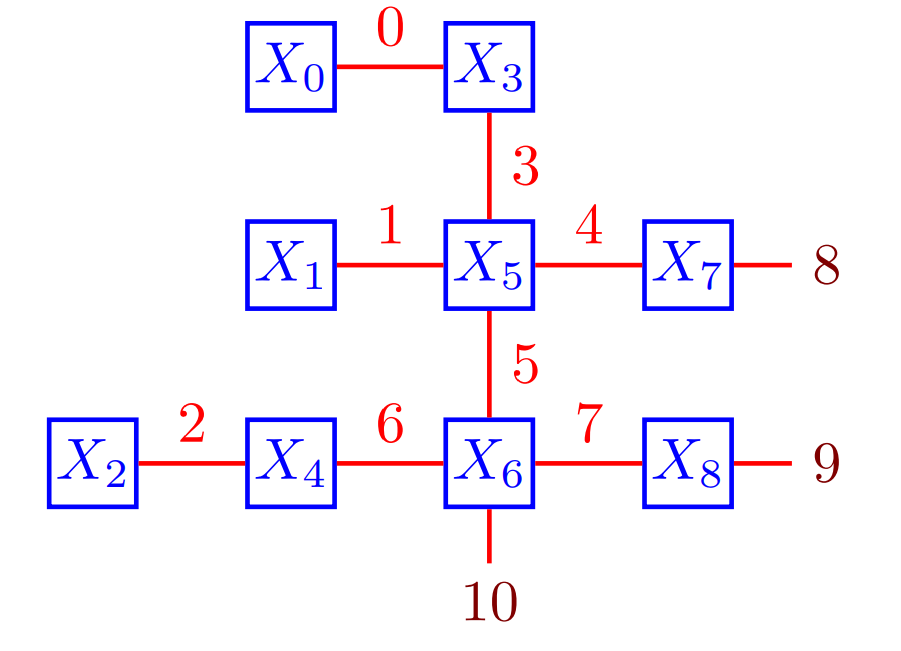}
        \caption{The tensor network.}
        \label{6_3_tn}
    \end{subfigure}
    \caption{An example of the quantum circuit and the corresponding tensor network.}
    \label{fig:6_3_tn_circ}
\end{figure}

Take the Fig.~\ref{fig:6_3_tn_circ} for example, $(X_{0}, \ldots, X_{8})$ are the tensors in this tensor network. We denoted the legs by the~numbers 0--7, and the~open legs by the~numbers 8--10.

A direct application of the tensor approach typically leads to an exponential increase in complexity as the number of qubits and circuit depth grows. This makes simulating large-scale circuits challenging to achieve within a reasonable timeframe. However, when the simulation is performed by calculating only one or a small batch of state amplitudes at the circuit's end, the tensor network's complexity is limited by the size of the largest tensor involved in the contraction process. This size, in turn, grows exponentially with the tree width of the graph corresponding to the tensor network \cite{Markov_2008}.

In \QuPack, we offer an ultra-fast tensor network simulation method for computing single amplitudes and multi-amplitudes \cite{kalachev2022multitensor}, and we also provide circuit sampling simulation capabilities for hundreds of qubits. In addition, we have also implemented quantum noise channels in tensor network simulation, enabling us to simulate the behavior of very large real-world quantum chips.

\subsubsection{Basic Usage}
In the following, we will provide a detailed explanation of the usage of this Tensor network simulator.
This module supports the following methods: Sampling,  expectation, and amplitudes calculation.

\textit{Sampling and expectation} -- The usage is exactly the same as the method in \MindQuantum. Actually we just wrap the tensor simulator 'MQ\_TNSim' as a backend for \MindQuantum \  simulator so that it supports all the functions as before.
\begin{lstlisting}
from qupack.tensor import MQ_TNSim
sim = Simulator(MQ_TNSim(n_qubits=n_qubits), n_qubits)
res = sim.sampling(circ, shots=shots)
\end{lstlisting}

\textit{Amplitudes calculation} -- The calculation of amplitudes constitutes a specific functionality within the tensor network simulator. While the computation of the full-amplitude obviates the necessity for this functionality due to its inclusion of all amplitudes, amplitude calculation plays a pivotal role in certain processes such as XEB calculation and tasks involving large-scale circuit sampling. Consequently, in this context, the tensor network simulator \code{TNSimulator} is employed exclusively for amplitude calculations.

\begin{lstlisting}
from qupack.tensor import TNSimulator
sim = TNSimulator(circuit_file='data/circuits/simple/circuit_n12_m14_s0_e0_pEFGH.qsim', backend='numpy')  # available backends : numpy, cupy, csim
n = sim.qubit_count
bitstring = [randint(0,1) for i in range(n)]
amp = sim.get_amplitude(bitstring)
\end{lstlisting}

When \code{TNSimulator} is initialized, the following backends can be specified for different tasks:
\begin{itemize}
    \item numpy --- Standard library for tensor operations on CPU. Fastest for simple circuits but very slow in multi-amplitude simulation mode.
    \item cupy --- GPU tensor contraction backend. Good for circuits of intermediate complexity.
    \item csim --- CUDA implementation of simulator. Fastest for long simulations (> 10 seconds), but has sufficient overhead for simpler cases.
\end{itemize}

In addition to single amplitude calculation, it also accommodates multiple amplitudes and batch amplitudes calculations:
\begin{lstlisting}
# batch-amplitudes
amps = sim.get_batch(bitstring, output_idx=[0,1,2])
# multi-amplitudes
amps = sim.get_amplitudes(bitstrings)
\end{lstlisting}

\subsection{Pulse Engineering}
\textit{Introduction} -- This section is about the open quantum system simulation including the dynamic of a quantum system and the quantum optimal control. In the following sections, we will demonstrate the implementation in each scenario.

\textit{Basic operations} -- Before we start realizing specific algorithms, we will illustrate some basic operations. In every question, we always need to construct a Hamiltonian to describe the quantum system. In the general case, a quantum system often contains two parts of Hamiltonians. One is the drift part which is time-independent, and the other is the control Hamiltonian, which changes with time. Therefore, we use the following way to construct drift Hamiltonian and control Hamiltonian. Firstly, let's construct the drift Hamiltonian:
\begin{equation}
    H_{d}=\sum_{i\in\{x,y,z\}}\prod_{j=1}^{n}\sigma_{j}^{i}.
\end{equation}
\begin{lstlisting}
from mindquantum import QubitOperator as Q
for i in range(qubit_num-1):
    H_d=H_d*(Q('X{}'.format(i+1))\
        +Q('Y{}'.format(i+1))\
        +Q('z{}'.format(i+1)))
H_d=H_d.matrix()
\end{lstlisting}
\textit{Unitary evolution} -- Based on this quantum system, We can study the characteristics of quantum systems in many ways. Firstly, we can study the dynamic of this quantum system in a closed environment. For a closed system, its evolution is governed by the Schr\"{o}dinger equation.
\begin{equation}
    i\hbar\frac{\partial}{\partial t}\Psi=\hat{H}\Psi,
\end{equation}
where the $\Psi$ is a wave function and $\hat{H}$ is the Hamiltonian of the quantum system. If we truncate the system into limited energy level $l$, the equation can be rewritten into:
\begin{equation}
    i\hbar\frac{d}{dt}\ket{\psi}=H\ket{\psi},
    \label{shro}
\end{equation}
where $\ket{\psi}$ is the state vector of $l$ energy level and $H$ is the matrix representation of the Hamiltonian. In \MindQuantum\ we can use a list to define the quantum state $\ket{\psi}$ as the initial state of the whole evolution progress.
\begin{lstlisting}
import numpy as np
psi0=np.zeros((1,2**qubit_num),dtype=complex)
psi0[0][0]=1+0j
\end{lstlisting}
Given a Hamiltonian, we can calculate the unitary (non-dissipative) time-evolution of an arbitrary initial state vector $\ket{\psi_{0}}$ using the \MindQuantum\ function MQsesolve, and we can get the result state in each time slot.
\begin{lstlisting}
t_list=np.linspace(0.0,10.0,100)
H=[[H_d]]
result=MQsesolve(H,psi0,t_list)
\end{lstlisting}
The function MQsesolve calculates states in each time slot during the schr\"{o}dinger unitary evolution. If we want to add control into the closed system, we could easily append the control Hamiltonian into the Hamiltonian list with the time function. The MQmesolve will identify scenarios by itself. We add a control Hamiltonian in the following form in the case of a two-qubit system.
\begin{equation}
    H_{c}=\sin(0.2\pi t)\sigma_{0}^{y}\sigma_{1}^{z},
\end{equation}
which we can easily realize in \MindQuantum.
\begin{lstlisting}
qubit_num=1
H_c=Q('Y0 Z1',qubit_num).matrix()
H=[[H_d],[H_c,np.sin(2*np.pu*0.1*t_list)]]
result=MQmeseolve(H,psi0,t_list,col_ops=[])
\end{lstlisting}
\textit{Non-unitary evolution} -- In more cases, we always have to consider the effect of the environment on an open quantum system. The effect from the bath will cause the energy leakage to a higher energy level or some uncertainty in the phase difference between states of the system. Therefore, we have to describe the state in such quantum systems in terms of density matrix. The density matrix is used to describe a probability distribution of a quantum state $\ket{\psi}$, which is $\rho=\sum_{k}p_{k}\ket{\psi_{k}}\bra{\psi_{k}}$. Here, $p_k$ is the probability of the state $\psi_k$.

A typical approach to describe an open quantum system is often the Lindblad Master equation \cite{lindblad1976generators}. It is derived from the von Neumann equation which expands the scope of the system to include the environment.
\begin{equation}
    \dot{\rho}_{tot}(t)=-\frac{i}{\hbar}\left[H_{tot},\rho_{tot}(t)\right],
    \label{sysAndBath}
\end{equation}
where the total Hamiltonian is the combination of the Hamiltonian of the quantum system $H_{sys}$, environment Hamiltonian $H_{env}$, and Hamiltonian $H_{int}$ used to describe the interaction between the environment and the quantum system.
\begin{equation}
    H_{tot}=H_{sys}+H_{env}+H_{int}.
\end{equation}
Because in quantum computing, most people only focus on the quantum system, we can partially trace the environment freedom degrees in Eq.~(\ref{sysAndBath}) to get a Lindblad master equation for the motion of the original system density matrix $\rho=\rm{Tr}_{env}(\rho_{tot})$.
\begin{equation}
    \begin{split}
        \dot{\rho}&=-\frac{i}{\hbar}\left[H(t),\rho(t)\right]\\
        &+\sum_{n}\frac{1}{2}\left[2C_{n}\rho(t)C^{\dagger}_{n}-\rho(t)C_{n}^{\dagger}C_{n}-C_{n}^{\dagger}C_{n}\rho(t)\right],
    \end{split}
\end{equation}
where $C_{n}=\sqrt{\gamma_{n}}A_{n}$ and $A_n$ are collapse operators through which the environment couples to the system in $H_{int}$ with corresponding rates $\gamma_{n}$. In \MindQuantum, we have no functions to derive collapse operators. Therefore, we have to design $C_{n}$ as followings:
\begin{lstlisting}
G=0.5
g=0.04
dephase1=csr_matrix([[1,0],[0,0]], dtype=complex)
dephase2=csr_matrix([[0,0],[0,1]], dtype=complex)
decay=csr_matrix([[0,1],[0,0]], dtype=complex)
cl_ops=[[decay,np.sqrt(g)],[dephase1,\
    np.sqrt(G-g/2)],[dephase2,np.sqrt(G-g/2)]]
\end{lstlisting}
For the non-unitary evolution of a quantum system, i.e., evolution that includes incoherent processes such as relaxation and dephasing, it is common to use master equations. In \MindQuantum, we use MQmesolve for both. Though these two are different, MQmesolve can automatically determine which one is proper according to the input parameters.
\begin{lstlisting}
result=MQmesolve(H,psi0,t_list,col_ops=cl_ops)
\end{lstlisting}

\textit{Quantum optimal control} -- In addition to quantum evolution, \MindQuantum\ can also do tasks about quantum optimal control. There are three algorithms in \MindQuantum\ that can be chosen: (1)grape \cite{GRAPE_0, GRAPE_2}, (2)crab \cite{CRAB_0, CRAB_1}, and (3)goat. The task of optimal control is always to minimize or maximize some functions that define how well a control pulse drives a quantum system to target in closed or open system. In the following sections.

\textit{Closed quantum system} -- We denote $U_{targ}$ as the target and $U_{T}$ as the actual unitary evolution operator. What we want to do is to maximize the overlap:
\begin{equation}
    \tau_{n}=\bra{n}\hat{U}_{targ}^{\dagger}\hat{U}(T)\ket{n}.
\end{equation}
Consider two following functionals:
\begin{equation}
    \begin{split}
        &J_{T,sm}=1-\frac{1}{N^{2}}\left|\sum_{n=1}^{N}\tau_{n}\right|^{2}=1-\frac{1}{N^{2}}\sum_{n=1}^{N}\sum_{m=1}^{N}\tau_{n}\tau_{m}^{\dagger}\\
        &J_{T,re}=1-\frac{1}{N}Re\left[\sum_{n=1}^{N}\tau_{n}\right].
    \end{split}
    \label{close}
\end{equation}
If ${\ket{n}}$ is a set of eignstates, $J_{T,sm}$ can be simply defined as $\left|\text{Tr}(\hat{U}_{targ}^{\dagger}\hat{U}(T))\right|^{2}$ and $J_{T,re}$ as $\text{Re}[\text{Tr}(\hat{U}^{\dagger}_{targ}\hat{U}(T))]$. Thus, $J_{T,sm}$ gets minimum when $\hat{U}(T)\ket{n}=e^{i\phi}\hat{U}_{targ}\ket{n}$ for all vectors $\ket{n}$ with arbitrary phase $\phi$. However, $J_{T,re}$ only gets minimal value when $\phi=0$. The functionals can be used in many scenarios, including gate optimization \cite{He_2024}, state preparation \cite{He_2021, refId0}, and optimal control for batch of state evolution.

\textit{Open quantum system} -- In an open quantum system, we have to use a mixed state because of the uncertainty caused by the bath environment. We have to adjust the Eq. (\ref{close}) to let it be able to characterize an evolution to a mixed state also. In this case, the function can be
\begin{equation}
    \begin{split}
        &J_{T,hs}=\frac{1}{2N}\sum_{n=1}^{N}\text{Tr}\left[(\hat{\rho}_{n,targ}-\hat{\rho}_{n}(T))^{2}\right]\\
        &\!=\!\frac{1}{2N}\sum_{n=1}^{N}\left(\text{Tr}(\hat{\rho}^{2}_{n,targ})\!+\!\text{Tr}(\hat{\rho}^{2}_{n}(T))\!-\!2\mathrm{Re}[\text{Tr}(\hat{\rho}_{n,targ}\hat{\rho}_{n}(T))]\right)\\
        &\!\!=\!\!\frac{1}{2N}\!\!\sum_{n=1}^{N}\!\!\braket{\hat{\rho}_{n,targ}|\hat{\rho}_{n,targ}}\!\!+\!\!\braket{\hat{\rho}_{n}(T)|\hat{\rho}_{n}(T)}\!\!-\!\!2\mathrm{Re}\!\braket{{\hat{\rho}_{n,targ}|\hat{\rho}_{n}(T)}}
    \end{split}
\end{equation}
on the basis of squared Hilbert-Schmidt distance:
\begin{equation}
    D_{hs}=\sqrt{\mathrm{Tr}(\hat{\sigma}-\hat{\rho})^{2}}.
\end{equation}
Moreover, the state can be rewritten as:
\begin{equation}
    \begin{split}
        &\ket{\hat{\rho}_{n,targ}}=\hat{\hat{P}}_{targ}\ket{\hat{\rho}(0)}\\
        &\ket{\hat{\rho}_{n}(T)}=\hat{\hat{P}}(T)\ket{\hat{\rho}_{n}(0)},
    \end{split}
\end{equation}
where $\hat{\hat{P}}_{targ}$ is the transformation operator and $\hat{\hat{P}}(T)$ is the open system evolution operator.

\textit{The gradients of functionals} -- Both algorithms GRAPE and GOAT methods need parameter derivatives to get results. Therefore, we have to induce the gradient of both closed and open quantum systems.

For gate optimization in closed system Eq.~(\ref{close}), the gradient is as following:
\begin{equation}
    \begin{split}
        &\frac{\partial}{\partial \alpha}J_{T,sm}=-\frac{2}{N^{2}}\mathrm{Re}\left[\sum_{n=1}^{N}\sum_{m=1}^{N}\frac{\partial \tau_{n}}{\partial \alpha}\tau^{*}_{m}\right]\\
        &\frac{\partial}{\partial \alpha}J_{T,re}=-\frac{1}{N}\mathrm{Re}\left[\sum_{n=1}^{N}\frac{\partial \tau_{n}}{\partial \alpha}\right],
    \end{split}
\end{equation}
where the $\alpha$ is a class of abstract parameters. The overlap between the target gate and the optimal result is:
\begin{equation}
    \begin{split}
        \frac{\partial \tau_{n}}{\partial \alpha}&=\left<n\left|\hat{U}^{\dagger}_{targ}\frac{\partial\hat{U}(T)}{\partial \alpha}\right|n\right>\\
        &=\left<n_{targ}\left|\frac{\partial \hat{U}(T)}{\partial \alpha}\right|n\right>,
    \end{split}
\end{equation}
and the trace of the density matrix can be:
\begin{equation}
    \begin{split}
        \frac{\partial}{\partial \alpha}\braket{\hat{\rho}_{n}(T)|\hat{\rho}_{n}(T)}=2\mathrm{Re}\left<\hat{\rho}_{n}(T)|\frac{\partial \hat{\rho}(T)}{\partial \alpha}\right>.
    \end{split}
    \label{overlap}
\end{equation}
Meanwhile, the gradient of the open system can be:
\begin{equation}
    \begin{split}
        \frac{\partial}{\partial \alpha}J_{T,hs}&=\frac{1}{N}\mathrm{Re}\sum_{n=1}^{N}\left<\hat{\rho}_{n}\left|\left(\hat{\hat{P}}^{\dagger}(T)+\hat{\hat{P}}^{\dagger}_{targ}\right)\frac{\partial \hat{\hat{P}}(T)}{\partial \alpha}\right|\hat{\rho}_{n}\right>\\
        &=\frac{1}{N}\mathrm{Re}\sum_{n=1}^{N}\left<\hat{\rho}_{n,targ}+\hat{\rho}_{n}(T)\left|\frac{\partial\hat{\hat{P}}(T)}{\partial \alpha}\right|\hat{\rho}_{n}\right>.
    \end{split}
\end{equation}
Here, the gradient is still available for vector $\bra{\hat{\rho}_{n,targ}+\hat{\rho}_{n}(T)}$.

\textit{Gradient ascent pulse engineering} -- Gradient ascent pulse engineering (GRAPE) algorithm is used in pulse optimal control. It divides the evolution time into many time slots. In each slot, the pulse is controlled by a time-independent Hamiltonian. The whole evolution can be regarded as a piecewise function. The reason is that solving the time-independent Schr\"{o}dinger equation costs less time complexity. We denote the control Hamiltonian as $H_{k}$, the unitary operator $\hat{U}_{m}$ in $m$'th slot can be:
\begin{equation}
    \hat{U}_{m}=\mathrm{exp}\left[-\frac{i}{\hbar}\Delta t\left(\hat{H}_{d}+\sum_{k=1}^{K}c_{k}^{(m)}\hat{H}_{k}\right)\right],
\end{equation}
where $c_{k}=c_{k}(t)$ is the constant coefficient in $m$'th slot for $k$'th control Hamiltonian. What we want is to adjust the parameter $c_{k}$ to find the optimal result of the control pulse, and we can rewrite the Eq.~(\ref{overlap}) as follows:
\begin{equation}
    \begin{aligned}
         & \frac{\partial}{\partial c_{k}^{m}}\braket{\psi_{T}|\psi(t_M)}                                                                             \\
         & \approx\bra{\psi_{T}}\hat{U}_{M}...\hat{U}_{m}(-i\Delta t\hat{H}_{k})\hat{U}_{m-1}...\hat{U}_{1}\ket{\psi(0)}                              \\
         & =(\hat{U}_{m}^{\dagger}...\hat{U}_{M}^{\dagger}\ket{\psi_{T}})^{\dagger}(-i\Delta t\hat{H}_{k})(\hat{U}_{m-1}...\hat{U}_{1}\ket{\psi(0)}),
    \end{aligned}
\end{equation}
and use the approximate formula of transformation derivative to calculate the gradient:
\begin{equation}
    \begin{aligned}
        \frac{\partial}{\partial c_{k}^{m}} & =\frac{\partial}{\partial c_{k}^{m}}e^{-\frac{i}{\hbar}\Delta t\left(\hat{H}_{d}+\sum_{k=1}^{K}c_{k}^{(m)}\hat{H}_{k}\right)} \\
                                            & \approx e^{-\frac{i}{\hbar}\Delta t\left(\hat{H}_{d}+\sum_{k=1}^{K}c_{k}^{(m)}\hat{H}_{k}\right)}(-i\Delta t\hat{H}_{k})      \\
                                            & =\hat{U}_{m}(-i\Delta t\hat{H}_{k}).
    \end{aligned}
\end{equation}
It was shown that higher-order gradient approximation, a vital method for second-order differential optimization(like L-BFGS-B in package), can lead to faster and more precise convergence of GRAPE. The general equation for the gradient is:
\begin{equation}
    \begin{split}
        \frac{\partial}{\partial c_{k}^{(m)}}\braket{\psi_{T}|\psi(t_{M})}&=(\hat{U}_{m}^{\dagger}...\hat{U}_{M}^{\dagger}\ket{\psi_{T}})^{\dagger}\times\\
        &\left(\sum_{s=0}^{S}\frac{(i\Delta t)^{s+1}}{(s+1)!}[\hat{H}(t_m),\hat{H}_{k}]_s\right)\times\\
        &(\hat{U}_{m-1}...\hat{U}_{1}\ket{\psi(0)}),
    \end{split}
\end{equation}
where
\begin{equation}
    \begin{split}
        &\left[\hat{H}(t_m),\hat{H}_{k}\right]_{0}=\hat{H}_{k}\\
        &\left[\hat{H}(t_m),\hat{H}_{k}\right]_{s}=\left[\hat{H}(t_m),\left[\hat{H}(t_m),\hat{H}_{k}\right]_{s-1}\right].
    \end{split}
\end{equation}
In \MindQuantum, the order of gradient function can be first order ($s=0$) or second order ($s=1$), which can fulfill most scenarios. Higher orders are computationally harder and have no significant improvement.

If we want to use GRAPE an open system, we can adjust the formulas by following:
\begin{equation}
    \begin{split}
        \ket{\psi(t)}&\to \ket{\hat{\rho}(t)}\\
        \hat{H}(t)&\to \hat{\hat{L}}(t)\\
        \hat{U}_{n}&\to \hat{\hat{P}}_{n}.\\
    \end{split}
\end{equation}

In \MindQuantum, we provide two functions for state preparation and optimal gate control for both open and closed systems. Here, we illustrate the realization of Hadamard gate as an example. First, we need to define the target.
\begin{lstlisting}
target=gen_target(qubit_num,eigen_states,hadmamard)
\end{lstlisting}
Next, we need to give parameters required for evolution including qubit number, fidelity error, evolution time, max iteration number in optimization, and number of time slots.
\begin{lstlisting}
evo_time=4
num_tslots=200
fid_err=1e-3
\end{lstlisting}
Then we can use $MQ\_opt\_unitary\_grape$ for unitary evolution and $MQ\_opt\_pulse\_grape$ for the open system as follows:
\begin{lstlisting}
result_unitary = MQ_opt_unitary_grape(
    drift=drift,
    ctrls=ctrls,
    target=target,
    num_tslots=num_tslots,
    evo_time=evo_time,
    max_iter=1000,
    fid_err=fid_err,
)

result_open = MQ_opt_pulse_grape(
    drift=drift,
    ctrls=ctrls,
    cl_op=cl_ops,
    target=target,
    num_tslots=num_tslots,
    evo_time=evo_time,
    max_iter=1000,
    fid_err=fid_err,
)
\end{lstlisting}
Other than the parameters above, we provide more parameters for users to meet various requirements in optimization problems. However, because of the space limitation, we won't show more details here.

\textit{Chopped random-basis} -- Because of the limitation that GRAPE's result can't be used in the experiment directly, chopped random-basis(CRAB) was proposed. It takes into account real hardware restrictions and the result can be used in a real experiment.

Since CRAB is a gradient-free algorithm, we use Nelder-Mead optimization method in this package. Moreover, we choose a Fourier basis with the possibility of Gaussian bounding to realize CRAB. In Fourier basis, $k$'th control function can be defined as:
\begin{equation}
    c_{k}(t)=\sum_{s=1}^{S}g_{k}(t)(A_{s,k}\mathrm{cos}(\omega_{s,k}t)+B_{s,k}\mathrm{sin}(\omega_{s,k}t)),
\end{equation}
where $g_{k}(t)=\exp\left[(t-t_{0})^{2}/\sigma^{2}\right]$ or $g_{k}(t)=1$ depends on the pulse shape limitation. Depending on the situation, $\omega_{s}$ also has two different options. Firstly, we have no idea of system resonance frequencies, the frequencies will be randomly chosen: $\omega_{s}=(1+r_s)(2\pi s/T)$. Here $2\pi s/T$ is the $s$'th principal frequency and $r_s$ is a uniform random number in the range of $[-0.5,0.5]$. Secondly, frequencies ${\omega_{s,k}}$ will be chosen according to system resonances and control limitations.

The function of CRAB is similar to the GRAPE:
\begin{lstlisting}
result_unitary = MQ_opt_unitary_grape(
    drift=drift,
    ctrls=ctrls,
    target=target,
    num_tslots=num_tslots,
    evo_time=evo_time,
    max_iter=1000,
    fid_err=fid_err,
)

result_open = MQ_opt_pulse_grape(
    drift=drift,
    ctrls=ctrls,
    cl_op=cl_ops,
    target=target,
    num_tslots=num_tslots,
    evo_time=evo_time,
    max_iter=1000,
    fid_err=fid_err,
)
\end{lstlisting}

\textit{Gradient optimization of analytic controls} -- Gradient optimization of analytic controls (GOAT) is the last method in \MindQuantum\ for optimal control. The Hamiltonian of a controllable system is:
\begin{equation}
    \vec{H}(\vec{a},t)=\hat{H}_{0}+\sum_{k=1}c_{k}(\vec{a},t)\hat{H}_{k}.
\end{equation}
Here, $\vec{a}$ is the parameter vector. We can get the following equations by calculating the gradient of Eq.~(\ref{shro}) and the order changing of derivatives.
\begin{equation}
    i\hbar\partial_{t}
    \begin{pmatrix}
        \ket{\psi} \\
        \ket{\partial_{a}\psi}
    \end{pmatrix}
    =
    \begin{pmatrix}
        \hat{H}(t)             & 0          \\
        \partial_{a}\hat{H}(t) & \hat{H}(t)
    \end{pmatrix}
    \begin{pmatrix}
        \ket{\psi} \\
        \ket{\partial_{a}\psi}
    \end{pmatrix}
    \label{Goat_diff}
\end{equation}
with initial value $\partial_{a}\psi(0)=0$. In this package, we use a modified version of GOAT. Eq.~(\ref{Goat_diff}) is split into two parts. The first part is the ordinary Schr\"{o}dinger equation and the second part is the inhomogeneous equation:
\begin{equation}
    i\hbar\frac{d}{dt}\ket{\partial_{\vec{a}}\psi(t)}=\hat{H}(t)\ket{\partial_{\vec{a}}\psi(t)}+\partial_{\vec{a}}\hat{H}(t)\ket{\psi(t)}.
\end{equation}
These two parts can be calculated in the size of Schr\"{o}dinger equation rather than twice the lager size. This form avoids background propagation but needs forward propagation for each element in parameter vector $\ket{a}$. So far, we only realized the GOAT method in the closed system, and we can implement as following:
\begin{lstlisting}
result_unitary = MQ_opt_unitary_goat(
    drift=drift,
    ctrls=ctrls,
    target=target,
    num_tslots=num_tslots,
    evo_time=evo_time,
    max_iter=1000,
    fid_err=fid_err,
)
\end{lstlisting}

\section{Benchmarking}
\label{sec:benchmark}

View benchmarking code: \href{https://gitee.com/mindspore/mindquantum/blob/research/whitepaper/code/benchmark}{\color{ceruleanblue}{CPU}} \href{https://gitee.com/mindspore/mindquantum/blob/research/whitepaper/code/benchmark_gpu}{\color{ceruleanblue}{GPU}}

\MindQuantum\ places great emphasis on the simulation efficiency of NISQ algorithms, particularly variational quantum algorithms. In this section, we benchmarked \MindQuantum\ against other quantum computing frameworks. Firstly, we used random circuit simulation to benchmark the fundamental performance of the framework. Secondly, we used Quantum Approximate Optimization Algorithm (QAOA) for benchmarking, demonstrating MindQuantum's ability to solve practical problems. The table below shows the frameworks that participated in the benchmark.

\begin{table}[ht]
    \begin{tabular}{cc}
        \toprule
        Framework          & Version    \\
        \midrule
        MindQuantum        & 0.9.0      \\
        Qiskit             & 0.45.0     \\
        Projectq           & 0.8.0      \\
        Pennylane          & 0.33.0     \\
        PyQpanda           & 3.8.0      \\
        Qulacs             & 0.6.2      \\
        Tensorflow Quantum & 0.7.2      \\
        Intel-QS           & 2.0.0-beta \\
        Cuquantum          & 23.10.0    \\
        \bottomrule
    \end{tabular}
    \caption{The software version of benchmarking.}
    \label{tab:software version}
\end{table}

The hardware platform used for the benchmark test is Intel$^\circledR$ Xeon$^\circledR$ CPU E5-2620 v3 @ 2.40GHz (16 threads used in benchmarking) with SIMD enabled and the GPU is a NVIDIA-V100. The benchmark framework is pytest-benchmark.

\begin{figure*}
    \centering
    \begin{subfigure}{0.45\textwidth}
        \centering
        \includegraphics[width=\textwidth]{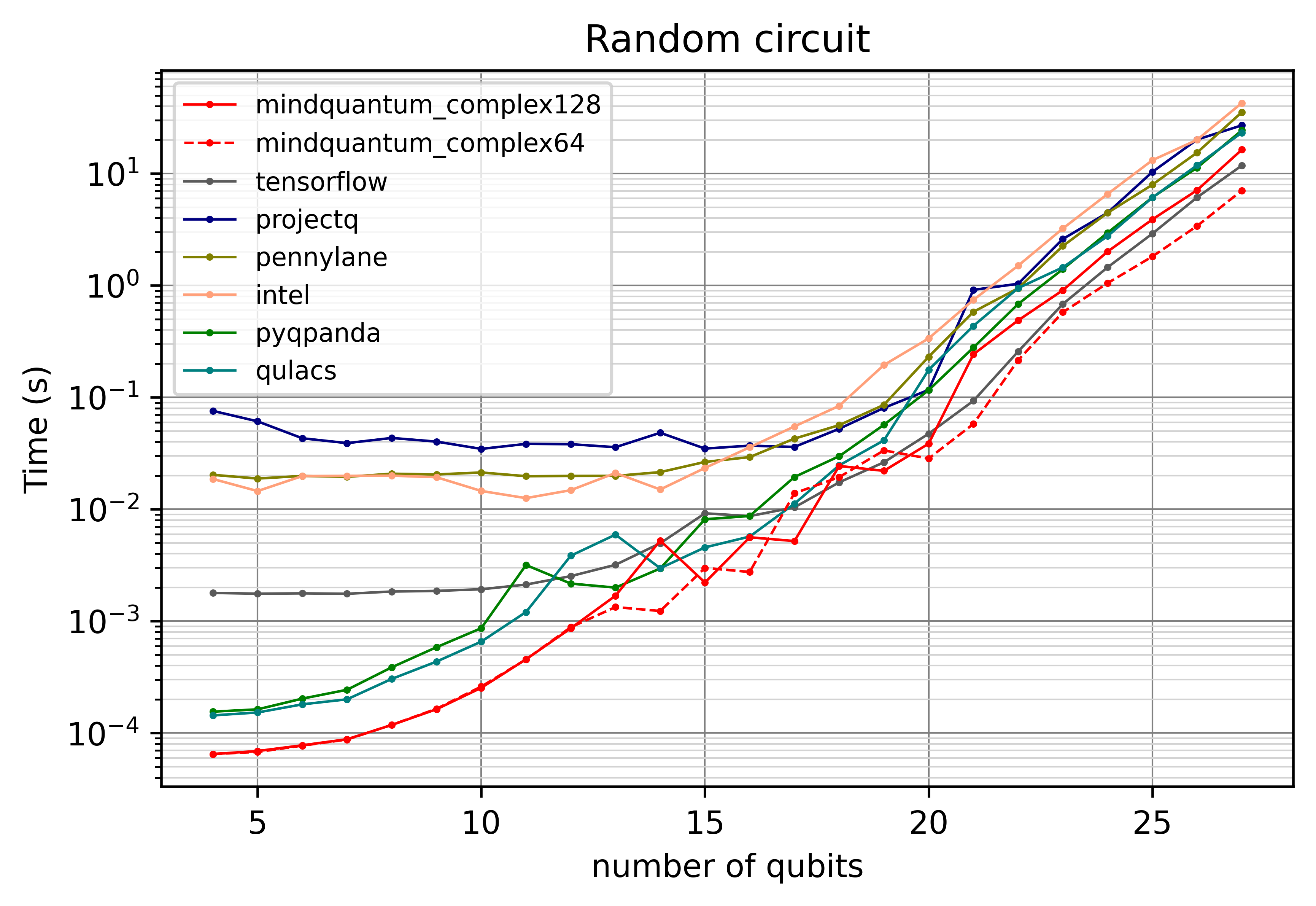}
        \caption{Random circuit benchmark with CPU backend.}
        \label{7_random_circ_cpu}
    \end{subfigure}
    \begin{subfigure}{0.45\textwidth}
        \centering
        \includegraphics[width=\textwidth]{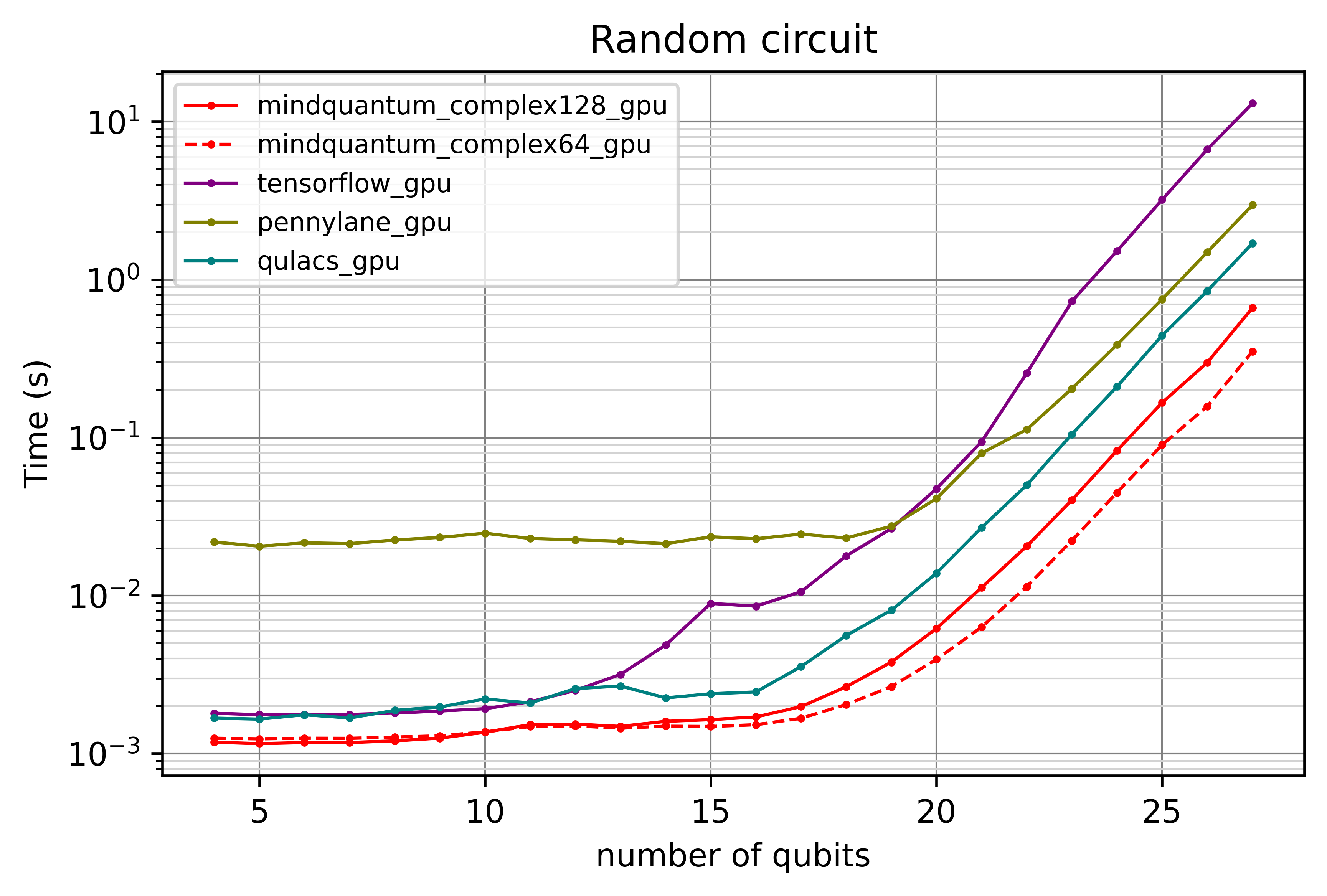}
        \caption{Random circuit benchmark with GPU backend.}
        \label{7_random_circ_gpu}
    \end{subfigure}
    \begin{subfigure}{0.45\textwidth}
        \centering
        \includegraphics[width=\textwidth]{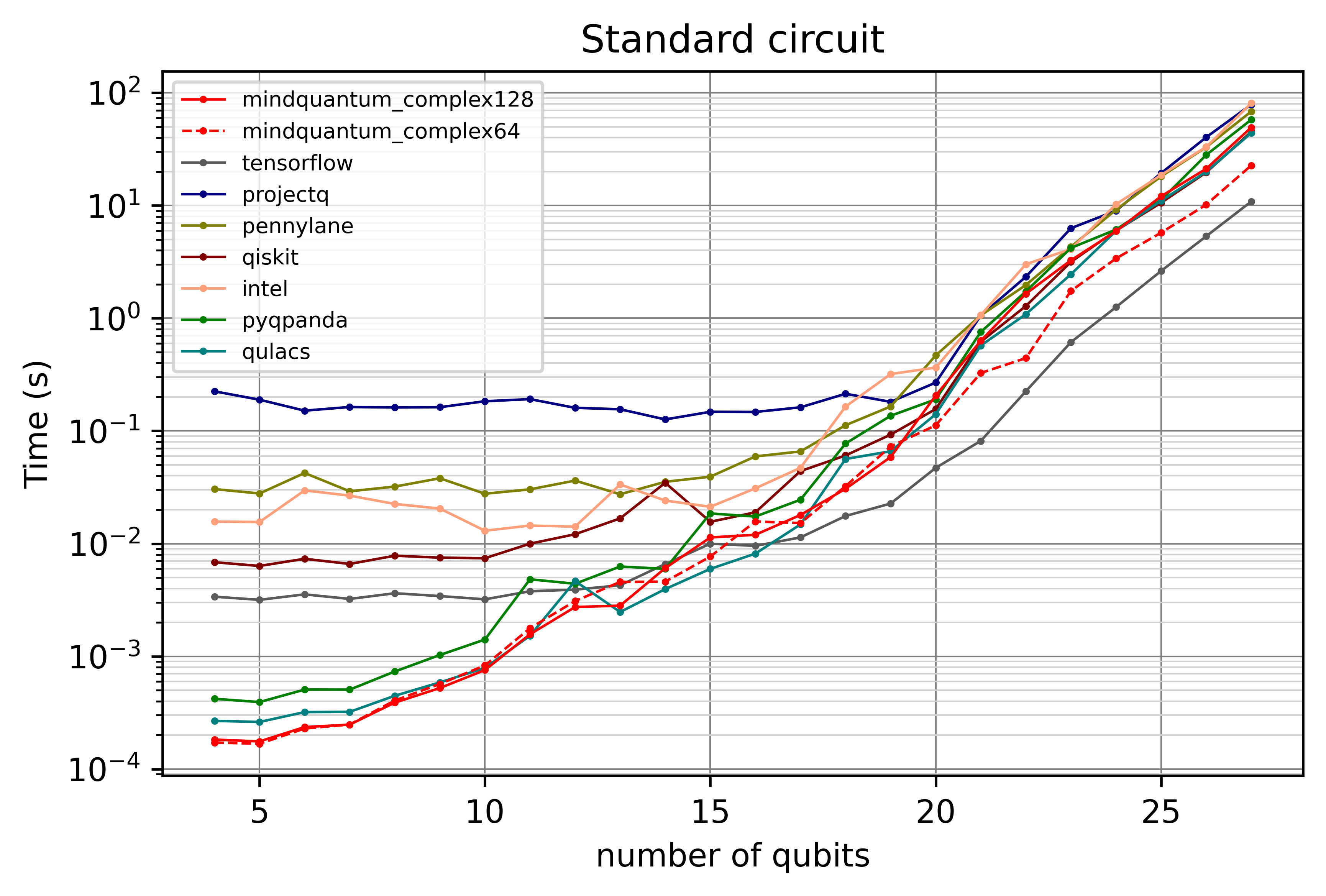}
        \caption{Standard circuit benchmark with CPU backend.}
        \label{7_random_circ_cpu}
    \end{subfigure}
    \begin{subfigure}{0.45\textwidth}
        \centering
        \includegraphics[width=\textwidth]{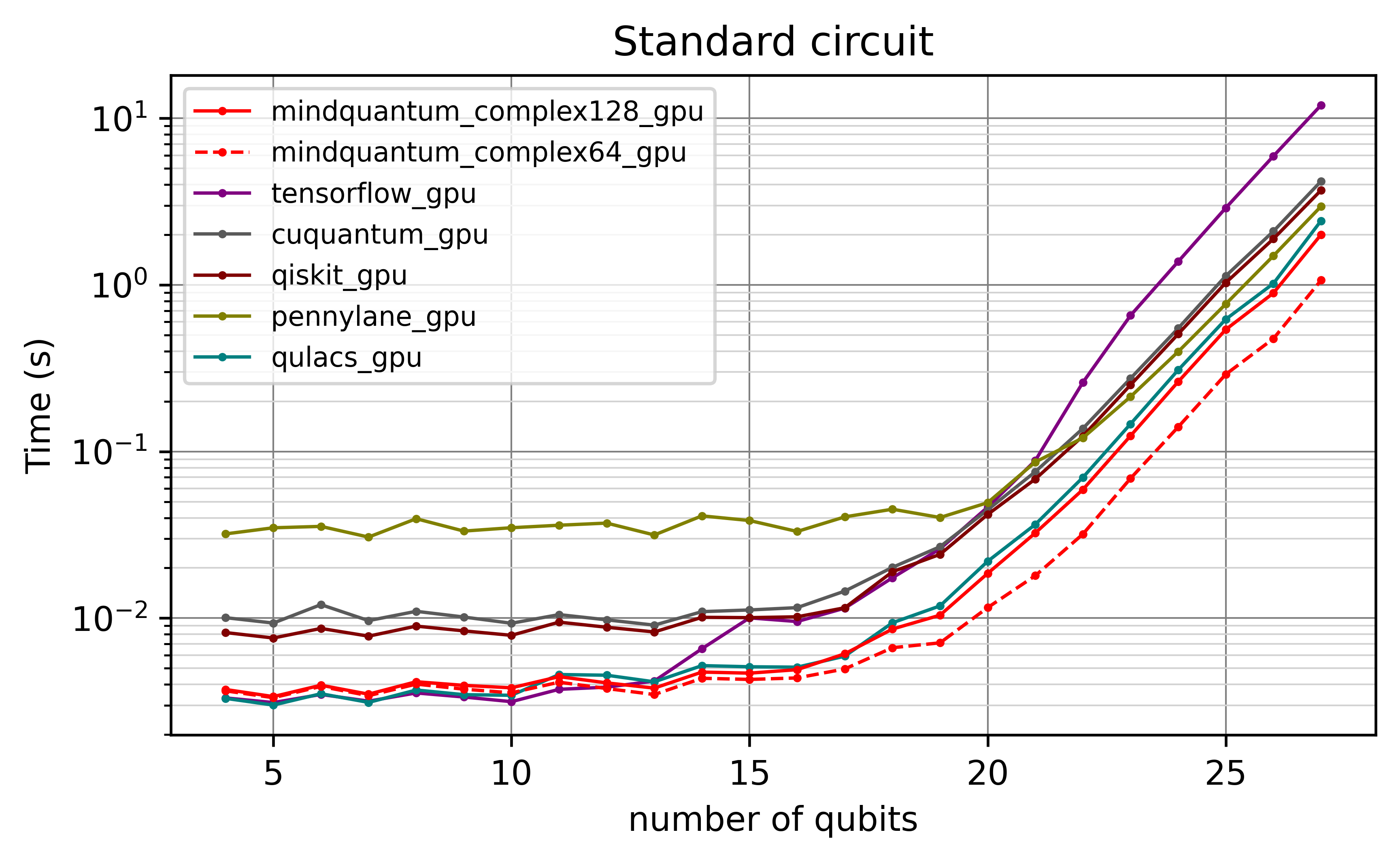}
        \caption{Standard circuit benchmark with GPU backend.}
        \label{7_random_circ_gpu}
    \end{subfigure}
    \caption{Benchmarking of randomly complex circuit and randomly simple circuit with different backends.}
    \label{fig:benchmark_circ}
\end{figure*}

\subsection{Running Random Circuit Task}

In random circuit benchmarking, we choose two different kinds of circuit. The first task is complex random circuit with qubit number start from 4 to 27. The complex random circuit contains X, Y, Z, H, CNOT, S, T, RX, RY, RZ, Rxx, Ryy, Rzz, SWAP gate and its control version. The second task is a simple random circuit that decomposed from the corresponded complex random circuit, and it only contains basic gate sets. The precision of simulator in benchmarking is double-precision except TensorFlow Quantum. The benchmark result is shown in Fig.~\ref{fig:benchmark_circ}.

In general, the simulation time increases exponentially with the circuit scale. However, when the number of qubits is small, \MindQuantum, Qulacs and Qpanda have a clear advantage in terms of time compared to other frameworks, which is due to the overhead of API call. It is worth noting that there is a slight dip at the 13-qubit position because \MindQuantum\ uses 13 qubits as a threshold, and when the number of qubits is greater than 13, it enables OpenMP to perform multi-threaded parallel computing, otherwise single thread simulation is used. This is because we found that multi-threaded computing would reduce the running speed when the number of qubits is small. As the circuit scale continues to increase, \MindQuantum\ and Qulacs maintain a good speed advantage, which can be judged that the two frameworks have been optimized to near the limit at the low-level implementation. When dealing with simple circuit with large qubit numbers, TensorFlow Quantum demonstrates advantages. There are two reasons for this. Firstly, it utilizes single-precision computation. Secondly, it incorporates more optimizations specifically targeted at fundamental quantum gates.

\begin{figure*}
    \centering
    \begin{subfigure}{0.9\textwidth}
        \centering
        \includegraphics[width=\textwidth]{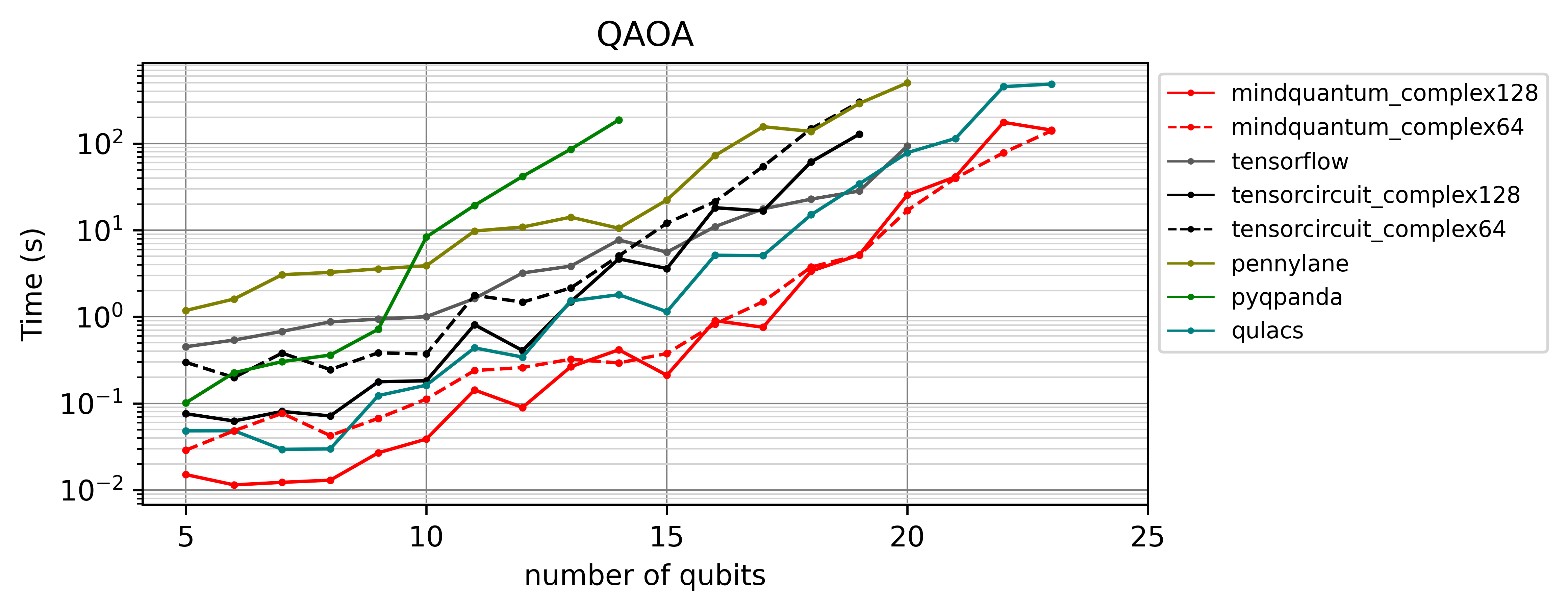}
        \caption{Benchmarking of QAOA end to end optimization with regular-4 graph in CPU.}
        \label{7_regular_4_cpu}
    \end{subfigure}
    \begin{subfigure}{0.9\textwidth}
        \centering
        \includegraphics[width=\textwidth]{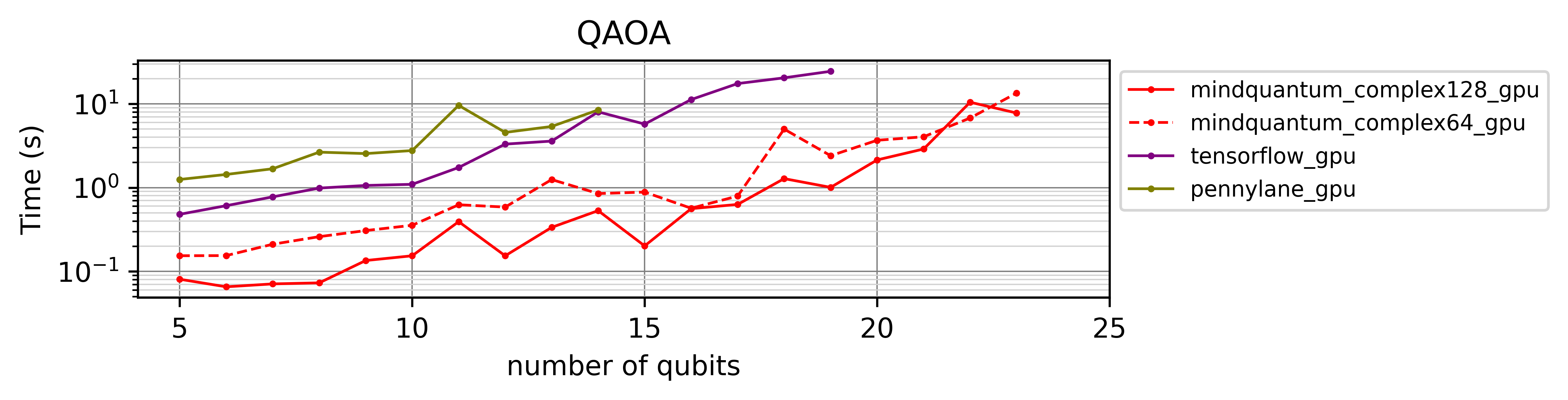}
        \caption{Benchmarking of QAOA end to end optimization with regular-4 graph in GPU}
        \label{7_regular_4_gpu}
    \end{subfigure}
    \caption{Benchmarking of QAOA with different backends.}
    \label{fig:benchmark_qaoa}
\end{figure*}

\subsection{QAOA Task}

Variational quantum algorithm is also a very important algorithm in NISQ. In this part, we choose Quantum Approximate Optimization Algorithm for benchmarking, which can further demonstrate the performance of the framework in solving practical problems. We apply QAOA to solve the max-cut problem of 4-regular graphs, and the size of the problem ranges from 5 to 23 nodes. The ansatz circuit is obtained by one-step Trotter decomposition. The optimizer is \code{scipy.optimize.minimize()} with the BFGS method. Since large-scale QAOA tasks are time-consuming, we stop the execution of each framework when the running time reaches a specified threshold. Therefore, in the result images, the maximum number of qubits that each framework data point can achieve is not consistent. The benchmark result is shown in Fig.~\ref{fig:benchmark_qaoa}.

Overall, as the problem size increases, the solution time of all frameworks grows exponentially. However, \MindQuantum\ is at least one orders of magnitude faster than other frameworks for problems. This is mainly due to MindQuantum's optimized adjoint method for getting gradient parameterized quantum circuit and efficient implementation of circuit evolution.

\section{Execution on Quantum Chip}
\label{sec:chip}
\subsection{Circuit Compilation and Optimization}
When simulating quantum circuits on a \Simulator, it is generally unnecessary to consider whether the quantum circuit can actually be executed on a quantum chip. Our focus lies solely on investigating the feasibility of the algorithm. However, when accounting for real quantum chips, the set of quantum gates executable on the chip is finite. Moreover, due to the existence of quantum noise, it is desirable to minimize the depth of the quantum circuit. This necessitates the compilation and optimization of the quantum circuit. In this section, we will elucidate how \MindQuantum\ performs the compilation and optimization of quantum circuit.

\begin{figure}
    \centering
    \begin{subfigure}{0.25\textwidth}
        \centering
        \includegraphics[width=\textwidth]{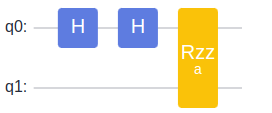}
        \caption{The original quantum circuit.}
        \label{fig:compiler_a}
    \end{subfigure}
    \begin{subfigure}{0.3\textwidth}
        \centering
        \begin{tikzpicture}[
                transparent/.style={circle, draw=black, fill=none, font=\sffamily, text centered, line width=1.5pt, minimum size=1cm}
            ]
            \node (q0) at (0, 0) [transparent] {$q_0$};
            \node (q1) at (3, 0) [transparent] {$q_1$};
            \node (H1) at (1.2, -1.5) [transparent] {$H$};
            \node (H2) at (1.2, -3) [transparent] {$H$};
            \node (Rzz) at (1.5, -4.5) [transparent] {$R_\text{zz}$};
            \draw [->, line width=1.5pt] (q0.south) to[out=270, in=180] (H1.west);
            \draw [->, line width=1.5pt] (H1.south) to[out=270, in=90] (H2.north);
            \draw [->, line width=1.5pt] (H2.south) to[out=270, in=135] (Rzz.north);
            \draw [->, line width=1.5pt] (q1.south) to[out=270, in=0] (Rzz.east);
        \end{tikzpicture}
        \caption{The DAG representation of quantum circuit.}
    \end{subfigure}
    \begin{subfigure}{0.45\textwidth}
        \centering
        \includegraphics[width=\textwidth]{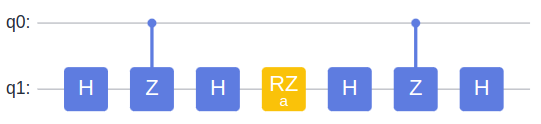}
        \caption{The circuit after compilation and optimization.}
        \label{fig:compiler_c}
    \end{subfigure}
    \caption{Converting a quantum circuit to DAG and compile it to with given compile rules.}
    \label{fig:compiler}
\end{figure}
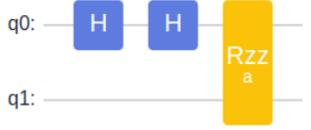
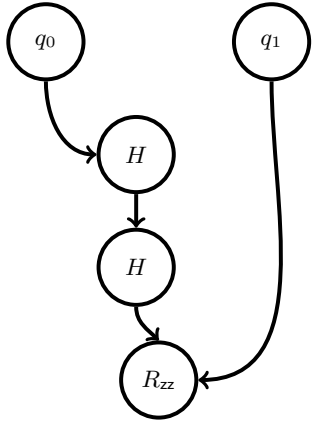
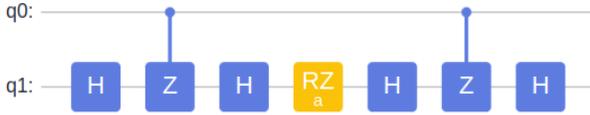

Here, we use DAG (Directed Acyclic Graph) as a tool to compile quantum circuit. As shown in Fig.~\ref{fig:compiler}, a quantum circuit is converted to a DAG. Based on the DAG we can apply built-in compilation rules, such as \BasicDecompose, \FullyNeighborCanceler and \GateReplacer, in \MindQuantum\ to finish quantum compilation and optimization. Furthermore, we have two robust managers in place for organizing these compilation rules, which are \SequentialCompiler and \KroneckerSeqCompiler. The rules in \SequentialCompiler will perform one by one in single time, but the rules in \KroneckerSeqCompiler will be executed until they can not compile anymore.

As an example, we will use different rules to compile and optimize the circuit in Fig.~\ref{fig:compiler_a}. First we build a \KroneckerSeqCompiler with \BasicDecompose and \FullyNeighborCanceler, so that we can decompose the $R_\text{zz}$ gate into basic quantum gate and canceled the two Hadamard gates. Next, we will replace the CNOT gate to CZ gate, assuming that the chip can only execute CZ gate.
\begin{lstlisting}
from mindquantum.core.circuit import Circuit
from mindquantum.algorithm.compiler import *

circ = Circuit().h(0).h(0).rzz('a', [0, 1])

compiler = SequentialCompiler([
    KroneckerSeqCompiler([
        BasicDecompose(),
        FullyNeighborCanceler()
    ]),
    CXToCZ()
])

new_circ = compile_circuit(compiler, circ)
\end{lstlisting}

\subsection{Qubit Mapping}

Due to some physical limitations in the realization of quantum hardware, today's quantum computers are qualified as noisy intermediate-scale quantum (NISQ) hardware. NISQ hardware is characterized by a few of qubits (50 to a few hundred) and noisy operations. Moreover, current realizations of superconducting quantum chips do not have the ideal all-to-all connectivity between qubits, but rather at most a nearest-neighbor connectivity. Due to these constraints, most quantum algorithms cannot be able to directly executed on the NISQ devices. Dynamically remapping logical qubits to physical qubits is needed to enable the two-qubit gates in the algorithm. This is called the qubit mapping problem \cite{10323857}.

The qubit mapping problem has been proved to be NP-Complete~\cite{siraichi2018qubit}. Considering the time-complexity, \MindQuantum\ chooses a swap-based bidirectional heuristic algorithm, named SABRE~\cite{li2019tackling}.

\begin{figure}
	\centering
	\begin{subfigure}{0.1\textwidth}
		\centering
		\includegraphics[width=\textwidth]{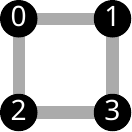}
		\caption{Physical topology graph.}
	\end{subfigure}
	\begin{subfigure}{0.3\textwidth}
		\centering
		\includegraphics[width=0.9\textwidth]{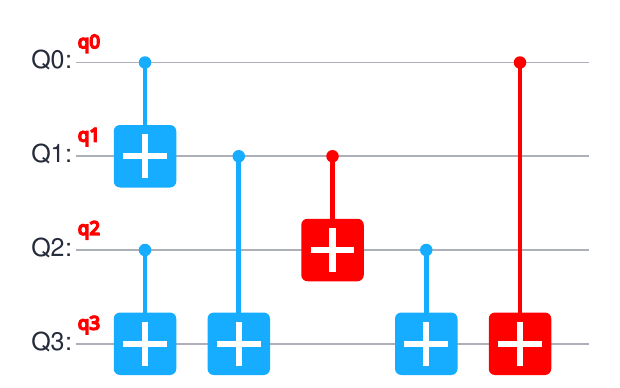}
		\caption{Logical circuit.}
	\end{subfigure}

	\begin{subfigure}{0.45\textwidth}
		\centering
		\includegraphics[width=\textwidth]{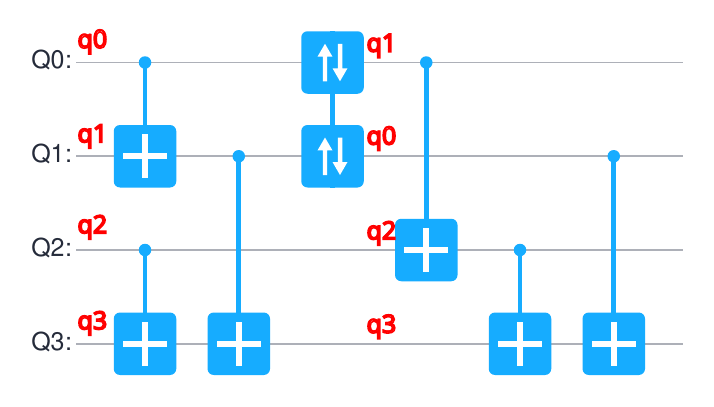}
		\caption{Fixed circuit.}
	\end{subfigure}

	\caption{An example of qubit mapping.}
	\label{fig:qubit-mapping-sample}
\end{figure}

Fig.~\ref{fig:qubit-mapping-sample} gives a sample of qubit mapping.
The physical hardware has its own physical connected graph. When we map the logical qubits to physical qubits, some two-qubit gates may not be able to realize, because two qubits are not connected in the physical connected graph. The trivial solution is using the swap gate. By some swap-operations, the logical circuit can be realized.

The goal of the algorithm is to reduce the number of swap gates and the circuit depth.

\begin{figure}
	\centering
	\begin{subfigure}{0.49\textwidth}
		\centering
		\includegraphics[width=\textwidth]{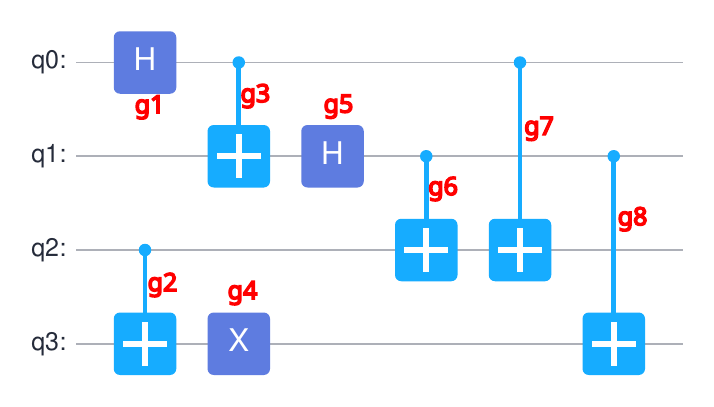}
		\caption{Quantum circuit.}
	\end{subfigure}
	\begin{subfigure}{0.49\textwidth}
		\centering
		\includegraphics[width=\textwidth]{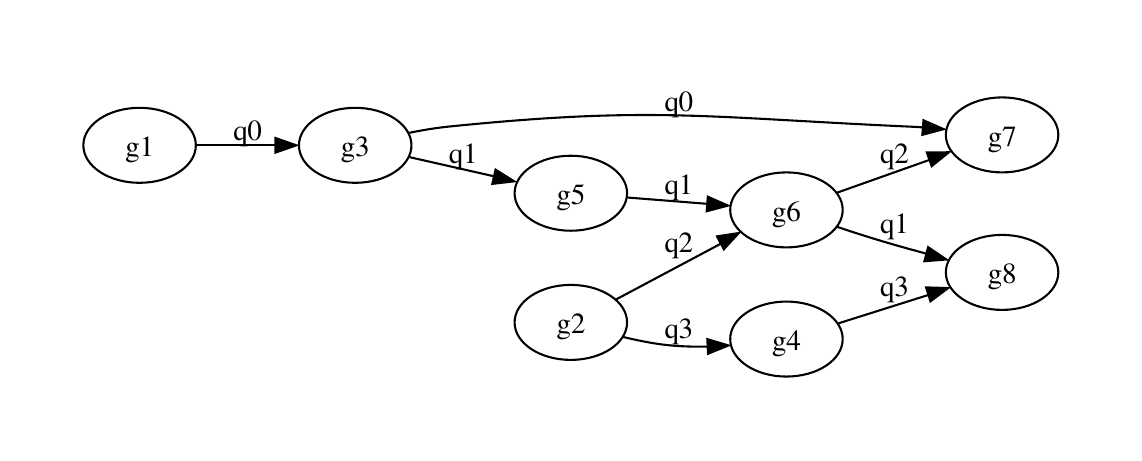}
		\caption{Generated DAG.}
	\end{subfigure}
	\caption{An example of DAG generation.}
	\label{fig:qubit-mapping-dag}
\end{figure}

Firstly, we can generate a directed acyclic graph (DAG) to represent the constraints between the two-qubit gates in the given quantum circuit. Single qubit gates can always be executed, therefore there is no need to consider them. We can label every two-qubit gate as $g_i(q_j, q_k)$, which means $g_i$ acts on $q_j$ and $q_k$. If $g_x(q_i, q_j)$ is followed by $g_y(q_j, q_k)$, then we add an edge $(g_x, g_y)$ to the DAG. Fig.~\ref{fig:qubit-mapping-dag} shows an example of DAG generation.

The core idea of the algorithm is trivial. Similar to topological sorting, we can construct a front layer which need to be executed now. If a gate can be executed under the current mapping, execute it and update the front layer. When there is no gate that can be executed, calculate available SWAPs. At last, we choose the SWAP which has the minimal heuristic function value.

The basic heuristic function is from the nearest neighbor cost function. For two physical qubit $Q_i$ and $Q_j$, $D[Q_i][Q_j]$ stands for the nearest distance in the physical connected graph. For a given front layer $F$ and current mapping $\pi$, the basic function is defined as
Eq.~\eqref{eq:qubit-mapping-h-basic}.
\begin{equation} \label{eq:qubit-mapping-h-basic}
	H_{\text{basic}} = \sum_{g \in F} D[\pi (g.q_1)][\pi (g.q_2)].
\end{equation}

$H_{\text{basic}}$ only considers the front layer. If we want to look ahead, we can add an extended set $E$, which contains some closet successors of the gates from $F$. The size of $E$ is flexible, depending on how much look-ahead ability we want. Also, the gates in $F$ should have a higher priority than $E$. So a weight parameter $0 \leq W < 1$ is added.

Up to now, the heuristic function tends to reduce the number of SWAP gates. We also want to reduce the depth of the circuit. There is a trade-off between them. We add a parameter $decay$ to each qubit. If a qubit already has many operations on it, it will have a bigger $decay$. Using $decay$, the algorithm will tend to choose those qubits that have fewer operations and increase the parallelism of the circuit. The fixed heuristic function is given by
Eq.~\eqref{eq:qubit-mapping-h}.
\begin{equation} \label{eq:qubit-mapping-h}
	\begin{aligned}
		H =    & \max \left( \rm{decay}(\rm{SWAP.q_1}), \rm{decay}(\rm{SWAP.q_2}) \right)                \\
		\times & \Big\lbrace \frac{1}{|F|} \sum_{g \in F} D[\pi(g.q_1)][\pi(g.q_2)]  \\
		+      & \frac{W}{|E|} \sum_{g \in E} D[\pi(g.q_1)][\pi(g.q_2)] \Big\rbrace.
	\end{aligned}
\end{equation}

\begin{figure}
	\centering
	\includegraphics[width=0.45\textwidth]{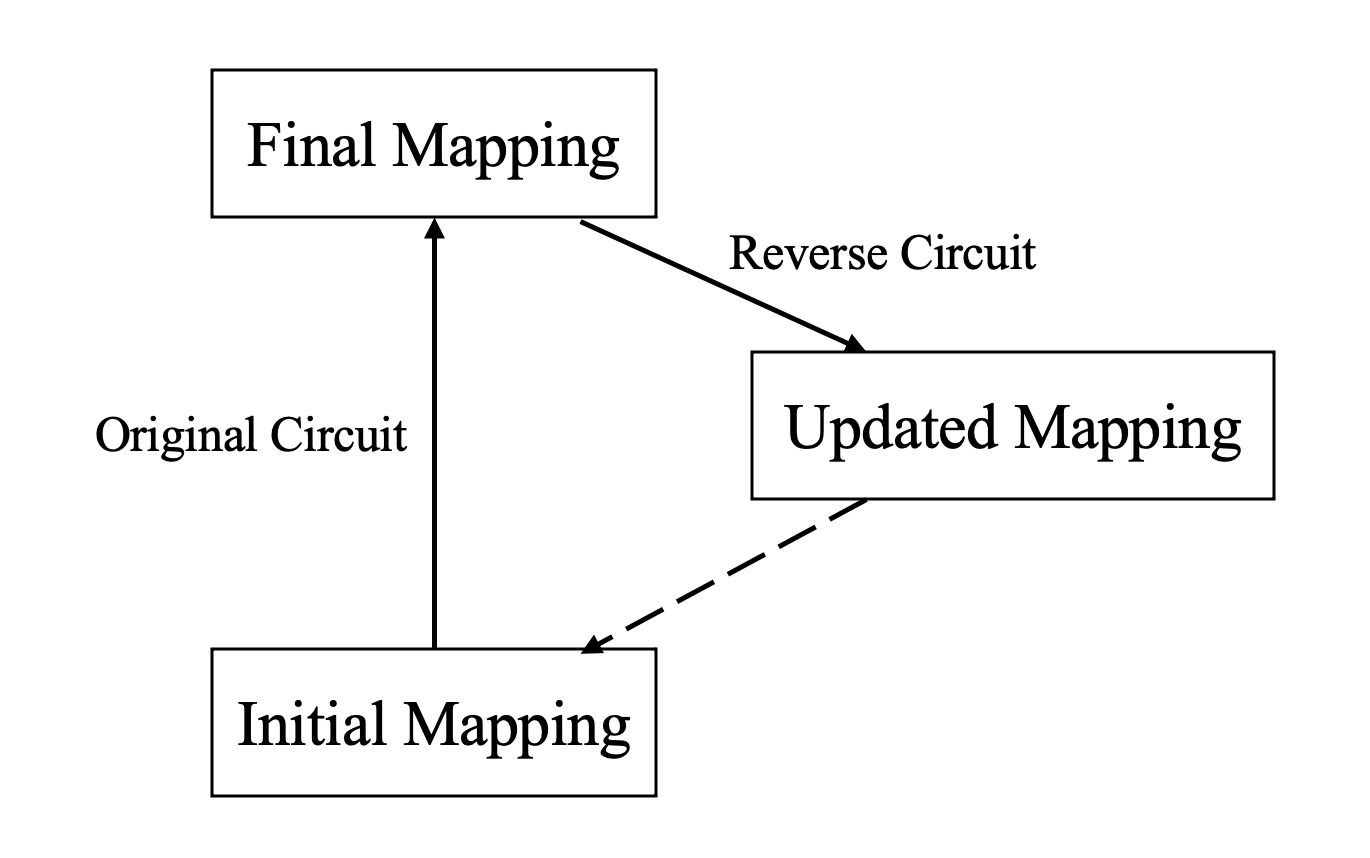}
	\caption{Initial mapping update procedure.}
	\label{fig:qubit-mapping-initial-mapping}
\end{figure}

Lastly, there is a problem of the initial mapping. It has been proved that initial mapping could have a huge impact on the final result. Here we use a method introduced in \cite{li2019tackling}.
Fig.~\ref{fig:qubit-mapping-initial-mapping}
shows the procedure of updating the initial mapping.

Here we introduce how to use SABRE in \MindQuantum. We can use \code{QubitNode} and \code{QubitsTopology} to generate physical topology graph.

\begin{lstlisting}
from mindquantum.device import QubitNode, QubitsTopology
from mindquantum.io.display import draw_topology

n: int = 5
topology = QubitsTopology([QubitNode(i, poi_x=i) for i in range(n)])
draw_topology(topology)
\end{lstlisting}

\begin{figure}
	\centering
	\begin{subfigure}{0.2\textwidth}
		\centering
		\includegraphics[width=0.8\textwidth]{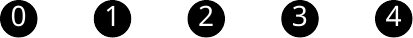}
		\caption{Disconnected graph.}
		\label{fig:qubit-mapping-topo1}
	\end{subfigure}
	\begin{subfigure}{0.2\textwidth}
		\centering
		\includegraphics[width=0.8\textwidth]{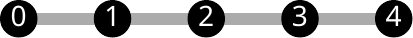}
		\caption{Connected graph.}
		\label{fig:qubit-mapping-topo2}
	\end{subfigure}
	\caption{Define a topology graph which contains 5 nodes.}
\end{figure}

The above code defines a topology graph which contains 5 nodes. \code{QubitNode(i, poi_x=i)} defines a \code{QubitNode} whose ID is \code{i} and x-axis coordinate is \code{i}. \code{draw_topology} is used to draw the SVG Fig.~\ref{fig:qubit-mapping-topo1}.

We can also use \code{q1 >> q2} or \code{q1 << q2} to connect different nodes, and use \code{q1 > q2} or \code{q1 < q2} to disconnect nodes. The following code connects Fig.~\ref{fig:qubit-mapping-topo1} and generate Fig.~\ref{fig:qubit-mapping-topo2}.

\begin{lstlisting}
for i in range(n-1):
topology[i] << topology[i+1]
draw_topology(topology)
\end{lstlisting}

\begin{figure}[H]
	\centering

	\begin{subfigure}{0.1\textwidth}
		\includegraphics[width=0.9\textwidth]{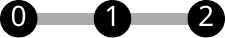}
		\caption{Linear.}
		\label{fig:qubit-mapping-linear}
	\end{subfigure}
	\begin{subfigure}{0.15\textwidth}
		\includegraphics[width=0.9\textwidth]{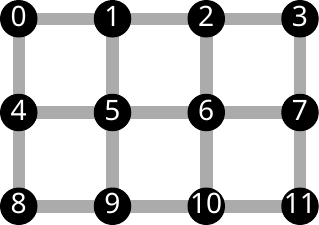}
		\caption{Grid.}
		\label{fig:qubit-mapping-grid1}
	\end{subfigure}
	\begin{subfigure}{0.2\textwidth}
		\includegraphics[width=0.9\textwidth]{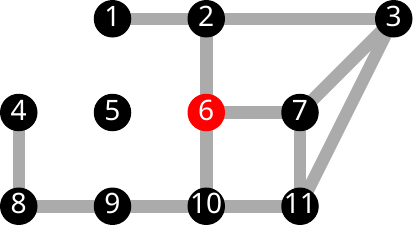}
		\caption{Modified.}
		\label{fig:qubit-mapping-grid2}
	\end{subfigure}

	\caption{Some useful structures and modified graph.}
\end{figure}

\MindQuantum\ also defines some useful structures. The following code defines a linear topology in Fig.~\ref{fig:qubit-mapping-linear} and a grid topology in Fig.~\ref{fig:qubit-mapping-grid1}.

\begin{lstlisting}
from mindquantum.device import LinearQubits, GridQubits

t1 = LinearQubits(3)
t2 = GridQubits(3,4)
draw_topology(t1)
draw_topology(t2)
\end{lstlisting}

We can also modify existing graph. The following codes remove a node, isolate a node, change a node's color, change a node's position and add a new edge. Fig.~\ref{fig:qubit-mapping-grid2} shows the modified graph.

\begin{lstlisting}
t2.remove_qubit_node(0)
t2.isolate_with_near(5)
t2.set_color(6, "#ff0000")
t2.set_position(3, 4, 0)
t2[3] << t2[11]
draw_topology(t2)
\end{lstlisting}

\begin{figure}
	\centering
	\begin{subfigure}{0.1\textwidth}
		\includegraphics[width=0.8\textwidth]{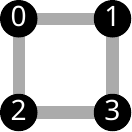}
		\caption{Physical topology graph.}
		\label{fig:qubit-mapping-physical-topo}
	\end{subfigure}
	\begin{subfigure}{0.35\textwidth}
		\includegraphics[width=\textwidth]{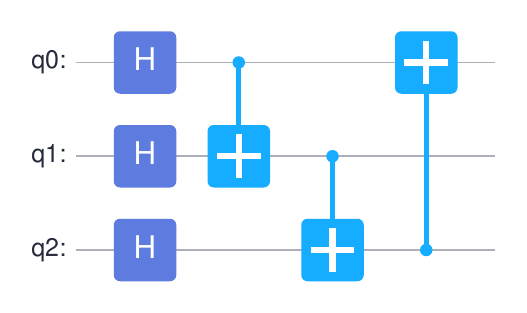}
		\caption{Logical circuit.}
		\label{fig:qubit-mapping-logical-circuit}
	\end{subfigure}
	\begin{subfigure}{0.45\textwidth}
		\includegraphics[width=0.9\textwidth]{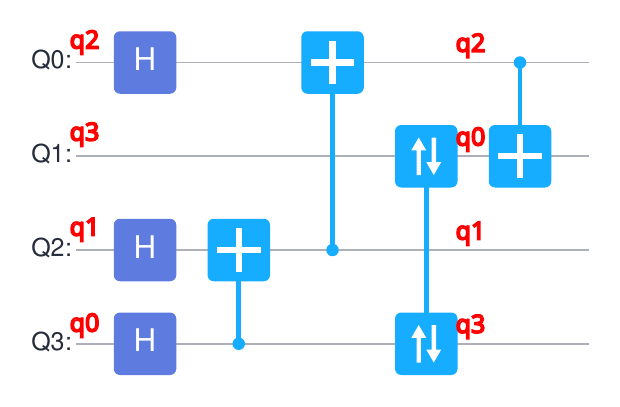}
		\caption{A possible physical circuit. The initial mapping is \code{[3, 2, 0, 1]} and final mapping is \code{[1, 2, 0, 3]}.}
		\label{fig:qubit-mapping-physical-circuit}
	\end{subfigure}

	\caption{An example of SABRE algorithm.}
\end{figure}

Combined with a given topology graph, we can use SABRE algorithm to generate a qubit mapping for any logical quantum circuit. The following codes define a topology graph \code{topo} and a logical circuit \code{circ}. SABRE solver needs 4 parameters:
\begin{itemize}
	\item \code{iter_num}: Number of iterations to generate initial mapping.
	\item \code{w}: $W$ parameter in Eq.~\eqref{eq:qubit-mapping-h}.
	\item \code{delta1}: Decay of single gates.
	\item \code{delta2}: Decay of CNOT gate.
\end{itemize}
The solver returns three parts:
\begin{itemize}
	\item \code{new_circ}: Physical circuit.
	\item \code{initial_mapping}: A list that represents the initial mapping. Here lower-case q stands for logical qubits and capital Q stands for physical qubits. For example, \code{[3, 2, 0, 1]} means logical $q_0$ stores in physical $Q_3$, $q_1$ in $Q_2$, $q_2$ in $Q_0$ and $q_3$ in $Q_1$.
	\item \code{final_mapping}: A list that represents the final mapping.
\end{itemize}
Fig.~\ref{fig:qubit-mapping-physical-circuit} shows a possible solution.

\begin{lstlisting}
topo = GridQubits(2, 2)
display_svg(draw_topology(topo))

from mindquantum.core.circuit import Circuit
from mindquantum.algorithm.mapping import SABRE

circ = Circuit().h(0).h(1).h(2).x(1, 0).x(2, 1).x(0, 2)
print("Logical circuit:")
display_svg(circ.svg())

solver = SABRE(circ, topo)
new_circ, init_mapping, final_mapping = solver.solve(5, 0.5, 0.3, 0.2)
print("Physical circuit:")
print(f"initial mapping: {init_mapping}")
print(f"  final mapping: {final_mapping}")
display_svg(new_circ.svg())
\end{lstlisting}

There are also some shortcomings. Current \MindQuantum\ only supports connected topology graphs. If physical topology is not connected, users have to allocate logical qubits to physical connected subgraphs manually.
Another one is gates' type. Current version only supports single gates and CNOT gate. Any quantum circuit can be realized by single gates and CNOT, therefore it's not a big problem.




\section{Summary}
\label{sec:summary}
We introduced \MindQuantum, a hybrid quantum-classical framework which focus on the design and implement of NISQ algorithm, especially Variational Quantum Algorithm. First, we showed the eDSL (embedded domain specific language) of how to manipulate quantum gate, quantum circuit and observable. \MindQuantum\ support arbitrary control qubits on any quantum gate, and based on JIT, high performance arbitrary costumed parameterized or non-parameterized gate is also supported. Then we showed how we implement state vector, density matrix and noisy quantum simulator and how we optimized the simulation with gate based optimization with SIMD, OpenMP, Kunpeng and GPU acceleration. Furthermore, we provide numerous application examples to help you quickly grasp the usage of \MindQuantum. These applications include VQE, QAOA and QNN. An acceleration engine call \QuPack\ was introduced, including how we speed up the simulation of VQE and QAOA algorithm. A benchmark on the performance compared with other quantum frameworks was demonstrated in Chapter ~\ref{sec:benchmark}, which showed advantages of \MindQuantum\ in several scenarios. Finally, we also showed the work that we have done about how to quickly compile and execute a quantum circuit on a real quantum chip.

\section{Acknowledgement}
\label{sec:acknowledgement}
We sincerely thank the authors of this paper and their respective contributions, as shown in table \ref{tab:contributions}.

\begin{table}[ht]
    \begin{tabular}{|p{3cm}|p{3cm}|}
        \toprule
        Author        & Section                                        \\
        \midrule
        Qingyu Li     & 2.1, 2.2, 5.3                                  \\
        Qiang Zheng   & 2.3                                            \\
        Jiale Liu     & 2.4                                            \\
        Yanling Lin   & 2.5, 5.9                                       \\
        Xusheng Xu    & 1, 2.6, 3.1, 3.2, 3.3, 4.1, 6.1, 7, 8.1, 9, 10 \\
        Zuoheng Zou   & 6.2, 6.3, 7                                    \\
        Junyuan Zhou  & 3.3, 4.2, 4.3, 7                               \\
        Xu Zhou       & 5.1                                            \\
        Jialiang Tang & 5.2                                            \\
        Ruoqian Xu    & 5.4                                            \\
        Chufan Lyu    & 5.5                                            \\
        Runhong He    & 5.6                                            \\
        Runqiu Shu    & 5.7                                            \\
        Zidong Cui    & 5.8, 6.4                                       \\
        Yikang Zhu    & 8.2                                            \\
        \bottomrule
    \end{tabular}
    \caption{Contributions of authors}
    \label{tab:contributions}
\end{table}

We are thankful to Shu Xu, Maolin Luo, Shijie Pan and Mosharev Pavel for their review of this paper. We also thank Damien Ngyuen, Kai Zhang, Qin Li, Jinlin Zeng, Qi Yan, FangFang Yi, Youkun Ren, Hui Yang, Jiwei Xie, Ting Wang, Haoran Yang, YiXing Feng, Ting Wang, Han Yu, Xiaoling Huan, Jianing Luo and many other people for the contribution of continuous integration and documentation and Shenyu Huang, Xiaopeng Cui, Bowen Liu, Yao Wang for the contribution in promoting. We would like to thank all the contributors and users of \MindQuantum. It is the collective effort and attention of you that keeps this project going.
This work is supported by the National Natural Science Foundation of China under the Grant No. 62273154, the China Postdoctoral Science Foundation under Grant No. 2023M740874, and the China Scholarship Council (CSC) under Grants No. 202206890003 and No. 202306890004.
\bibliography{lib.bib}

\end{document}